\documentclass[a4paper,11pt]{article}
\pdfoutput=1 
\usepackage{jcappub} 
\usepackage[T1]{fontenc} 
\usepackage{aas_macros}
\usepackage{xspace}
\usepackage{bm}
\usepackage{multirow}

\newcommand{\eg}{\mbox{e.\,g.,}\xspace}

\newcommand{\ie}{\mbox{i.\,e.,}\xspace}
\newcommand{\Ie}{\mbox{I.\,e.,}\xspace}
\renewcommand{\vec}[1]{\bm{#1}}

\definecolor{red}{rgb}{1,0,0}
\definecolor{green}{rgb}{0.25,0.75,0.25}
\definecolor{pblue}{rgb}{0.36,0.54,0.66}
\definecolor{byzantium}{rgb}{0.477,0.1875,0.422}

\title{Ultra-light axions and the \(S_8\) tension: joint constraints from the cosmic microwave background and galaxy clustering}

\author[a,1]{Keir K. Rogers,\note{Corresponding author.}}
\author[a, b]{Ren\'ee Hlo\v zek,}
\author[c, d, b, a]{Alex Lagu\"e,}
\author[e, f]{Mikhail M. Ivanov,}
\author[g, h, i, e]{Oliver H.\,E. Philcox,}
\author[e]{Giovanni Cabass,}
\author[e]{Kazuyuki Akitsu}
\author[j]{and David J.\,E. Marsh}

\affiliation[a]{Dunlap Institute for Astronomy \& Astrophysics, University of Toronto,\\50 St.\,George Street, Toronto, ON M5S 3H4, Canada}
\affiliation[b]{David A.\,Dunlap Department of Astronomy \& Astrophysics, University of Toronto,\\50 St.\,George Street, Toronto, ON M5S 3H4, Canada}
\affiliation[c]{Department of Physics and Astronomy, University of Pennsylvania,\\209 South 33rd Street, Philadelphia, PA 19104-6396, USA}
\affiliation[d]{Canadian Institute for Theoretical Astrophysics, University of Toronto,\\60 St.\,George Street, Toronto, ON M5S 3H8, Canada}
\affiliation[e]{School of Natural Sciences, Institute for Advanced Study,\\1 Einstein Drive, Princeton, NJ 08540, USA}
\affiliation[f]{NASA Hubble Fellowship Program Einstein Postdoctoral Fellow}
\affiliation[g]{Center for Theoretical Physics, Department of Physics, Columbia University,\\538 West 120th Street, New York, NY 10027, USA}
\affiliation[h]{Simons Society of Fellows, Simons Foundation, New York, NY 10010, USA}
\affiliation[i]{Department of Astrophysical Sciences, Princeton University,\\Peyton Hall, 4 Ivy Lane, Princeton, NJ 08544, USA}
\affiliation[j]{Theoretical Particle Physics and Cosmology, King's College London,\\Strand, London WC2R 2LS, UK}

\emailAdd{keir.rogers@utoronto.ca}
\emailAdd{hlozek@dunlap.utoronto.ca}
\emailAdd{alague@sas.upenn.edu}
\emailAdd{ivanov@ias.edu}
\emailAdd{ohep2@cantab.ac.uk}
\emailAdd{gcabass@ias.edu}
\emailAdd{kakitsu@ias.edu}
\emailAdd{david.j.marsh@kcl.ac.uk}

\abstract{We search for ultra-light axions as dark matter (DM) and dark energy particle candidates, for axion masses $10^{-32}\,\mathrm{eV} \leq m_\mathrm{a} \leq 10^{-24}\,\mathrm{eV}$, by a joint analysis of cosmic microwave background (CMB) and galaxy clustering data -- and consider if axions can resolve the tension in inferred values of the matter clustering parameter $S_8$. We give legacy constraints from \textit{Planck} 2018 CMB data, improving 2015 limits on the axion density $\Omega_\mathrm{a} h^2$ by up to a factor of three; CMB data from the Atacama Cosmology Telescope and the South Pole Telescope marginally weaken \textit{Planck} bounds at $m_\mathrm{a} = 10^{-25}\,\mathrm{eV}$, owing to lower (and theoretically-consistent) gravitational lensing signals. We jointly infer, from \textit{Planck} CMB and full-shape galaxy power spectrum and bispectrum data from the Baryon Oscillation Spectroscopic Survey (BOSS), that axions are, today, $< 10\%$ of the DM for $m_\mathrm{a} \leq 10^{-26}\,\mathrm{eV}$ and $< 1\%$ for $10^{-30}\,\mathrm{eV} \leq m_\mathrm{a} \leq 10^{-28}\,\mathrm{eV}$. BOSS data strengthen limits, in particular at higher $m_\mathrm{a}$ by probing high-wavenumber modes ($k < 0.4 h\,\mathrm{Mpc}^{-1}$). BOSS alone finds a preference for axions at $2.7 \sigma$, for $m_\mathrm{a} = 10^{-26}\,\mathrm{eV}$, but \textit{Planck} disfavours this result. Nonetheless, axions in a window $10^{-28}\,\mathrm{eV} \leq m_\mathrm{a} \leq 10^{-25}\,\mathrm{eV}$ can improve consistency between CMB and galaxy clustering data, \eg reducing the $S_8$ discrepancy from $2.7 \sigma$ to $1.6 \sigma$, since these axions suppress structure growth at the $8 h^{-1}\,\mathrm{Mpc}$ scales to which $S_8$ is sensitive. We expect improved constraints with upcoming high-resolution CMB and galaxy lensing and future galaxy clustering data, where we will further assess if axions can restore cosmic concordance.}

\begin{document}
\maketitle
\flushbottom

\section{Introduction}
\label{sec:intro}

While evidence for dark matter (DM) exists observationally \cite{2022ApJ...938..110B,2020A&A...641A...6P,2013ApJS..208...20B,2014ApJ...782...74H, 2014PhRvD..89j3508M, 2013JCAP...10..060S}, the fundamental nature of dark matter remains one of the greatest unsolved problems in science. Axions are a well-motivated particle candidate for DM \cite{1981PhRvL..47..402W,1981PhLB..104..199D,1982AIPC...93...66D,Dine:1982ah,1983rctm.proc..100A,1983PhLB..120..127P,1983PhLB..129...51S,1987PhR...150....1K,1992SvJNP..55.1063B}. Axions were proposed to solve the strong $CP$ problem \cite{pecceiquinn1977,weinberg1978,wilczek1978} and can arise in a string theory ``axiverse'' where many axions of different masses are produced, suggesting that no single axion, but rather a mixture, can dominate the dark sector \cite{1984PhLB..149..351W,2006JHEP...06..051S,pecceiquinn1977,1978PhRvL..40..223W,wilczek1978,Arvanitaki:2009fg}. Depending on their particle mass, axions behave either as DM or as a scalar field dark energy (DE) component \cite{2015PhRvD..91j3512H}. 
 Axions are proposed to resolve the so-called ``small-scale crisis'' in the clustering of matter \cite{2017ARA&A..55..343B,2015PNAS..11212249W,2017PhRvD..95d3541H}, as they suppress the growth of small-scale (sub-Mpc) structure depending on the mass of the axion. Improvements in our ability to model astrophysical effects in the formation of Galactic and sub-Galactic structure \citep[\eg][]{2014Natur.506..171P} and a more complete census of the Milky Way (MW) satellite galaxy population \citep[\eg][]{2020ApJ...893...47D} have demonstrated the viability of astrophysical solutions to the ``small-scale crisis.'' Further, complementary constraints from \eg the Lyman-alpha forest \citep{rogers/peiris:2021} and the MW sub-halo mass function \citep{2021PhRvL.126i1101N} have ruled out the axion mass \(\sim 10^{-22}\,\mathrm{eV}\) as being all the DM that is invoked to address small-scale issues. Nonetheless, advances in our understanding of axion structure formation including astrophysical effects \citep[\eg][]{lague/etal:2021, 2021PhRvD.103d3526R, dome/etal:2022, may/etal:2022, nori/etal:2022, vogt/etal:2022} now allows us to test empirically using cosmological data the existence of ultra-light axions, as motivated from fundamental theory, across the full mass range where their wavelength is astrophysically large (\(10^{-32}\,\mathrm{eV} \leq m_\mathrm{a} \leq 10^{-18}\,\mathrm{eV}\)).
 
 Observational constraints on the axion typically limit a combination of the axion mass and the axion energy density (or the fraction that axions make up of the total DM energy density). Multiple tracers have been used to constrain the allowed mass and density of axions including, \eg the cosmic microwave background \citep[CMB;][]{2015PhRvD..91j3512H,2018MNRAS.476.3063H,Farren_2022}, galaxy clustering \citep{2022JCAP...01..049L}, galaxy weak lensing \citep{dentler/etal:2022,kunkel/etal:2022}, the Lyman-alpha forest \citep{rogers/peiris:2021,2021PhRvD.103d3526R,2017arXiv170304683I,2017arXiv170800015K,2017arXiv170309126A,2016JCAP...08..012B}, dwarf galaxies \citep{2021PhRvL.126i1101N,dala/etal:2022,goldstein/etal:2022}, 21 cm observations \citep{hotinli/etal:2022,bauer/etal:2021, flitter/etal:2022}. A single axion species as the \textit{only} DM candidate is ruled out by the Lyman-alpha forest for masses less than $2 \times 10^{-20}$ eV (at 95\% c.l.) \citep{rogers/peiris:2021}. However, axions as a component of the DM or DE (either as a mixture of axions of different masses or in combination with other dark sector species) is still viable across the ultra-light mass range (\(10^{-32}\,\mathrm{eV} \leq m_\mathrm{a} \leq 10^{-18}\,\mathrm{eV}\)) and above (see, \eg Refs.~\cite{2022arXiv220314915A,2022arXiv220314923J} for reviews of searches for axion-like particles). In this work, we search for axions through their gravitational imprint in a compendium of CMB and large-scale structure data for \(m_\mathrm{a} \leq 10^{-25}\,\mathrm{eV}\) and allowing for sub-dominant axion energy densities \(\Omega_\mathrm{a} h^2\). We present legacy constraints from \textit{Planck} 2018 CMB data \citep{2016A&A...594A..13P}, the first study of high-resolution CMB data from the Atacama Cosmology Telescope \citep[ACT;][]{2020JCAP...12..047A} and the South Pole Telescope \citep[SPT;][]{2021PhRvD.104b2003D}, and a full joint analysis of \textit{Planck} CMB and full-shape galaxy power spectrum and bispectrum data from the Baryon Oscillation Spectroscopic Survey \citep[BOSS,][]{2022PhRvD.105d3517P,BOSS:2012dmf}.

We include high-resolution CMB data and we model axions in galaxy power spectrum data to smaller scales than previously considered (wavenumbers \(k < 0.4\,h\,\mathrm{Mpc}^{-1}\)) as probing smaller scales gains sensitivity to the scale-dependent suppression of heavier axions. The challenge is that exploiting smaller scales typically requires robust modelling of axion structure formation into the non-linear regime, moving beyond the well-established linear-order theory (\eg \texttt{axionCAMB} \citep{2015PhRvD..91j3512H,2022ascl.soft03026G}, \texttt{AxiCLASS} \citep{2011JCAP...07..034B}) that is sufficient for \textit{Planck} analyses\footnote{Refs.~\cite{Cookmeyer:2019rna,passaglia:2022bcr} discuss the robustness of the fluid approximations that are typically required to make linear-order axion perturbation calculations computationally tractable.}. Ref.~\cite{2022JCAP...01..049L} modelled axions into the mildly non-linear regime using the effective field theory of large-scale structure \citep[EFT of LSS;][]{2012JCAP...07..051B,2018JCAP...04..030S,Cabass:2022avo,DAmico:2020kxu,2020JCAP...05..005D,2020JCAP...06..001C,Ivanov:2019pdj,Ivanov:2019hqk,2020PhRvD.102f3533C,Ivanov:2022mrd}, finding that, similarly to massive neutrinos \citep{2017arXiv170704698S,Ivanov:2019hqk}, the galaxy bias and counterterm parameters capture non-linear effects after modifying the input linear matter power spectrum. Axion-induced wave effects are suppressed as they manifest on scales suppressed in the linear power spectrum.

In this work, we find that current high-resolution CMB data from ACT-DR4 and SPT-3G can be modelled by linear theory. However, upcoming and proposed future high-resolution CMB lensing data from, \eg ACT, SPT, Simons Observatory, CMB-S4, and galaxy weak lensing data from, \eg the Dark Energy Survey (DES), \textit{Rubin} Observatory, \textit{Euclid} will gain sensitivity to heavier axions (\(m_\mathrm{a} > 10^{-25}\,\mathrm{eV}\)) but will probe fully non-linear scales. Ref.~\cite{dentler/etal:2022} searched for axions as the only DM species in DES-Y1 galaxy shear data using an axion halo model that analytically captures the effect of axions on the formation and clustering of DM halos \citep{Marsh:2016vgj}. Ref.~\cite{2022arXiv220913445V} extended this model to the case of mixed axion and cold DM. A complementary approach is to capture non-linear modes using machine learning models called emulators which are trained on the outputs of cosmological simulations \citep[\eg][]{2009ApJ...705..156H,2014ApJ...780..111H,2017ApJ...847...50L,2019ApJ...874...95Z,2019MNRAS.484.5509E,2019JCAP...02..050B,2019JCAP...02..031R,2021JCAP...05..033P}. Emulators have been used successfully to set DM constraints, \eg with the Lyman-alpha forest \citep{rogers/peiris:2021,2022PhRvL.128q1301R,2021PhRvD.103d3526R}, where astrophysical effects can be captured in training simulations. Accurate emulator predictions rely on accurate input simulations. There is much progress in our ability to simulate axion structure formation using fluid approximations \citep{2018MNRAS.478.3935N} and by solving the full axion field equations \citep{2014NatPh..10..496S,2019PhRvL.123n1301M,2019PhRvD..99f3509L,2020PhRvD.102h3518S,dome/etal:2022,may/etal:2022,nori/etal:2022,kulkarni/etal:2022}.

There are discrepancies between CMB, galaxy clustering and galaxy shear inferences on the amplitude of matter density fluctuations \citep[see][for a recent review]{2022JHEAp..34...49A}. This is typically characterised by the matter clustering parameter \(\sigma_8\), the amplitude at redshift \(z = 0\) when averaged over \(8\,h^{-1}\,\mathrm{Mpc}\) scales, or by the degenerate combination \(S_8 \equiv \sqrt{\frac{\Omega_\mathrm{m}}{0.3}} \sigma_8\) (where \(\Omega_\mathrm{m}\) is the matter energy density), which is well constrained by large-scale structure experiments. The statistical significance of the so-called \(S_8\) tension ranges from 2 to 3 \(\sigma\) depending on the data considered; galaxy shear, in particular, drives the largest discrepancies with CMB data. Notwithstanding undetected systematic errors in the data, the \(S_8\) tension has proposed solutions based on physics beyond $\Lambda$CDM typically by introducing either a time-dependence or a scale-dependence in the DM dynamics. This can be achieved by, \eg coupling DM to DE \citep{2021PDU....3400899L,2022arXiv220906217P}, a complex dark sector (\eg atomic DM \citep{2010JCAP...05..021K,2013PhRvD..87j3515C,2022arXiv221202487B}), decaying DM \citep{2015JCAP...09..067E,2020JCAP...07..026P,2020arXiv200809615A,2021PhRvD.104l3533A,2022PhRvD.106b3516S}, or baryon-DM interactions \citep{2022PhRvD.106j3525D}; Ref.~\cite{2022MNRAS.516.5355A} more generally considers modifications to non-linear clustering including the effects of baryonic feedback.

Ultra-light axions form a component of the dark sector with a scale-dependent growth factor. We therefore hypothesise that axions could alleviate the $S_8$ tension, by behaving like standard cold DM at the scales probed by current CMB surveys, while suppressing the growth of structure at the smaller scales to which galaxy surveys are sensitive. We investigate this hypothesis by jointly analysing CMB and galaxy clustering data. The inclusion of galaxy shear measurements is left for future work. Another discrepancy in the $\Lambda$CDM model is the $H_0$ tension, the \(\sim 5 \sigma\) difference in the Hubble expansion rate today \(H_0\) as inferred from different direct and indirect distance ladders \citep[see, \eg][]{2022JHEAp..34...49A}. Many proposed solutions to the $H_0$ tension based on new physics, however, exacerbate the discrepancy in $S_8$ \citep[\eg][]{Hill:2020osr}. Models of ultra-light axions, with $m_\mathrm{a} \sim (10^{-27} - 10^{-26})$ eV, combined with modifications to the dynamics of the DE component \citep{2021arXiv210713391Y,2022arXiv220713086A,2021PhRvD.104h1303A} are invoked to alleviate simultaneously both parameter tensions. In this work, we stress the importance of assessing tension in the full parameter space. In testing the extent to which ultra-light axions can improve consistency between CMB and large-scale structure data, we therefore use metrics of tension that account for the full non-Gaussian posterior distribution.

In \S~\ref{sec:model}, we introduce our model for axion structure formation: the linear theory in \S~\ref{sec:linear} and the EFT of LSS that we use as our non-linear theory in \S~\ref{sec:eft}. We discuss our data in \S~\ref{sec:data}: CMB in \S~\ref{sec:cmb}, baryon acoustic oscillations (BAO) and supernovae in \S~\ref{sec:bao}, full-shape BOSS galaxy clustering in \S~\ref{sec:boss} and our parameter inference methods in \S~\ref{sec:inference}. We present results from the CMB, BAO and supernovae in \S~\ref{sec:cmb_results} and from BOSS galaxy clustering in \S~\ref{sec:boss_results}. In \S~\ref{sec:discussion}, we discuss these results and draw conclusions in \S~\ref{sec:conclusions}.

\section{Axion structure formation model}
\label{sec:model}

\subsection{Linear theory}
\label{sec:linear}

\subsubsection{Axion cosmology}
\label{sec:linear_cosmology}

In order to model the effect of ultra-light axions (ULAs) on the cosmic microwave background (CMB), we calculate linear-order perturbations using the Einstein-Boltzmann solver \texttt{axionCAMB}\footnote{\url{https://github.com/dgrin1/axionCAMB}.} \citep{2015PhRvD..91j3512H,2022ascl.soft03026G}. The fundamental equation governing the axion field \(\phi\) is the Klein-Gordon equation:
\begin{equation}
\label{eq:klein_gordon}
    \Box\phi-m_\mathrm{a}^2\phi=0,
\end{equation}
where \(\Box\) is the d'Alembert operator. We consider a temperature-independent axion mass, which is appropriate for string theory axions, where the mass switches on at a high energy scale (typically the geometric mean of the supersymmetry scale and the Planck scale \citep{2006JHEP...06..051S}). We ignore self-interactions of the axion (valid for initial field misalignment angles that are not tuned close to $\pi$ \citep{Zhang:2017dpp,Cedeno:2017sou,Leong:2018opi,Arvanitaki:2019rax}). The axion-photon coupling can affect CMB polarisation if it is large (see \eg Refs.~\cite{Pospelov:2008gg,Sigl:2018fba,Fedderke:2019ajk,Obata:2021nql}), but does not back-react significantly on the axion DM density (although see Ref.~\cite{Levkov:2020txo}). Cosmologically, all other axion couplings can lead only to a small thermal population of axions, which is negligible for couplings consistent with astrophysical limits, current constraints on the effective number of relativistic species $N_{\rm eff}$, and in the mass range that we consider (for related discussion, see, \eg Refs.~\cite{Baumann:2016wac,DEramo:2021psx}). Thus, we set the axion couplings to zero and consider only gravitational effects. The gravitational couplings of the axion are contained in the metric dependence of $\Box$.

Under the above approximations, the relic density of axions arises from the solution of Eq.~\eqref{eq:klein_gordon} for a spatially-homogeneous axion field $\phi=\phi(t)$ coupled to the Friedmann equation governing the evolution of the scale factor $a(t)$. The field is assumed to have an initial value $\phi_\mathrm{i}=\theta_\mathrm{i} f_\mathrm{a}$, where $\theta_\mathrm{i}$ is the initial field misalignment angle and $f_a$ is the decay constant. \texttt{axionCAMB} solves a shooting problem to fix $\phi_\mathrm{i}$ given $\Omega_\mathrm{a} h^2$, \(m_\mathrm{a}\) and cosmological parameters. In the mass range that we probe in this work, the relic density can be approximated as
\begin{equation}
\label{eq:axion_density}
\Omega_\mathrm{a} h^2 \approx \left\{
\begin{array}{ll}
5.23\times 10^{-2} \left(\frac{\Omega_\mathrm{m} h^2}{0.143}\right)^\frac{3}{4} \left(\frac{\phi_\mathrm{i}}{ 10^{17}\,\mathrm{GeV}}\right)^2 \left(\frac{1+z_{\rm eq}}{3400}\right)^{-\frac{3}{4}} \left(\frac{m_\mathrm{a}}{10^{-23}\,\mathrm{eV}}\right)^\frac{1}{2} \mbox{if $m_\mathrm{a}\gtrsim H(z_{\rm eq})$},\\
3.62\times 10^{-3}\left(\frac{\Omega_\mathrm{m} h^2}{0.143}\right) \left(\frac{\phi_\mathrm{i}}{10^{17}\,\mathrm{GeV}}\right)^2 \mbox{if $H_0\lesssim m_\mathrm{a}\lesssim H(z_{\rm eq})$},
\end{array}
\right.
\end{equation}
where \(\Omega_\mathrm{m} h^2\) is the total matter energy density, \(z_\mathrm{eq}\) is the redshift of matter -- radiation equality, \(H (z)\) is the Hubble expansion rate and \(H_0\) is its value today. The initial misalignment angle is expected to be determined by a random process, while the decay constant is a physical constant (possibly itself determined by a vacuum expectation value of some other scalar field). Taking the random $\theta_\mathrm{i}$ to be of order unity, $\Omega_a h^2$, for given axion mass and cosmological parameters, is a function of $f_a$. For $f_a$ near the Grand Unified Theory scale (between $10^{16}$ GeV and $10^{17}$ GeV), $\Omega_a h^2$ is large enough to contribute significantly to the total energy density (as shown in Eq.~\eqref{eq:axion_density}).

At early times, defined as when $H \gg m_\mathrm{a}$, \texttt{axionCAMB} solves fluid equations, equivalent to the full Klein-Gordon equation (Eq.~\ref{eq:klein_gordon}) at linear order in spatial fluctuations of $\phi$ and metric perturbations, in the synchronous gauge. The homogeneous axion field begins to oscillate when $H \approx m_\mathrm{a}$. After this time, \texttt{axionCAMB} uses the WKB approximation to adopt an effective fluid description \citep{Hu:1998kj,Amendola:2005ad,Hwang:2009js,Marsh:2015xka}. The fluid model is equivalent to the Madelung formulation \citep{1926NW.....14.1004M} and is accurate up to shell crossing. For further discussion of the accuracy of the adopted approximations, see Refs.~\cite{2015PhRvD..91j3512H,Cookmeyer:2019rna,passaglia:2022bcr}.

The main physical features that distinguish ULAs from standard $\Lambda$CDM components, as pertains to cosmological observables, are two-fold \citep{2015PhRvD..91j3512H}. First, the slow roll of the axion field when $H \gg m_\mathrm{a}$ leads to a distinctive background evolution equivalent to a fraction of the matter component behaving like an early form of dark energy. This leads to differences in the diffusion damping and Sachs-Wolfe contributions to the CMB, changes the sound horizon, and changes the distance to the surface of last scattering (if axions begin their oscillation after matter-radiation equality, \ie for \(m_\mathrm{a} \lesssim 10^{-28}\,\mathrm{eV}\)). 

Second, the gradient terms in the Klein-Gordon equation appear as an effective pressure opposing gravitational collapse, leading to a Jeans scale for the ULAs and, consequently, a suppression in the amplitude of density perturbations on small scales \citep{Khlopov:1985jw,Amendola:2005ad,Arvanitaki:2009fg,Marsh:2010wq}. If axions contribute non-negligibly to the total DM density, the Jeans scale can be approximated as
\begin{equation}
\label{eq:axion_Jeans}
\lambda_\mathrm{J}\approx 12.3\,(1+z)^\frac{1}{4}\left(\frac{\Omega_\mathrm{a} h^2}{0.01}\right)^{-\frac{1}{4}} \left(\frac{m_\mathrm{a}}{10^{-26}\,\mathrm{eV}}\right)^{-\frac{1}{2}} h^{-1}\,\mathrm{Mpc}.
\end{equation}
If ULAs begin to oscillate prior to matter -- radiation equality (\(m_\mathrm{a} \gtrsim 10^{-28}\,\mathrm{eV}\)), there is strong suppression of the matter power spectrum, compared to a pure cold DM model, for wavelengths smaller than the Jeans scale evaluated at matter -- radiation equality \citep{2000PhRvL..85.1158H}. If $m_\mathrm{a}\lesssim 10^{-28}\,\mathrm{eV}$, the physics of the Jeans scale (suppression of density perturbations relative to pure cold DM) is related to the horizon size at the time when the axion becomes non-relativistic. The magnitude of the relative suppression of the power spectrum, with respect to the cold DM limit, increases with the fraction of the DM composed of axions.

\subsubsection{\(S_8\) and the linear matter power spectrum}
\label{sec:linear_s8}

\begin{figure}[tbp]
\centering 
\includegraphics[width=\textwidth]{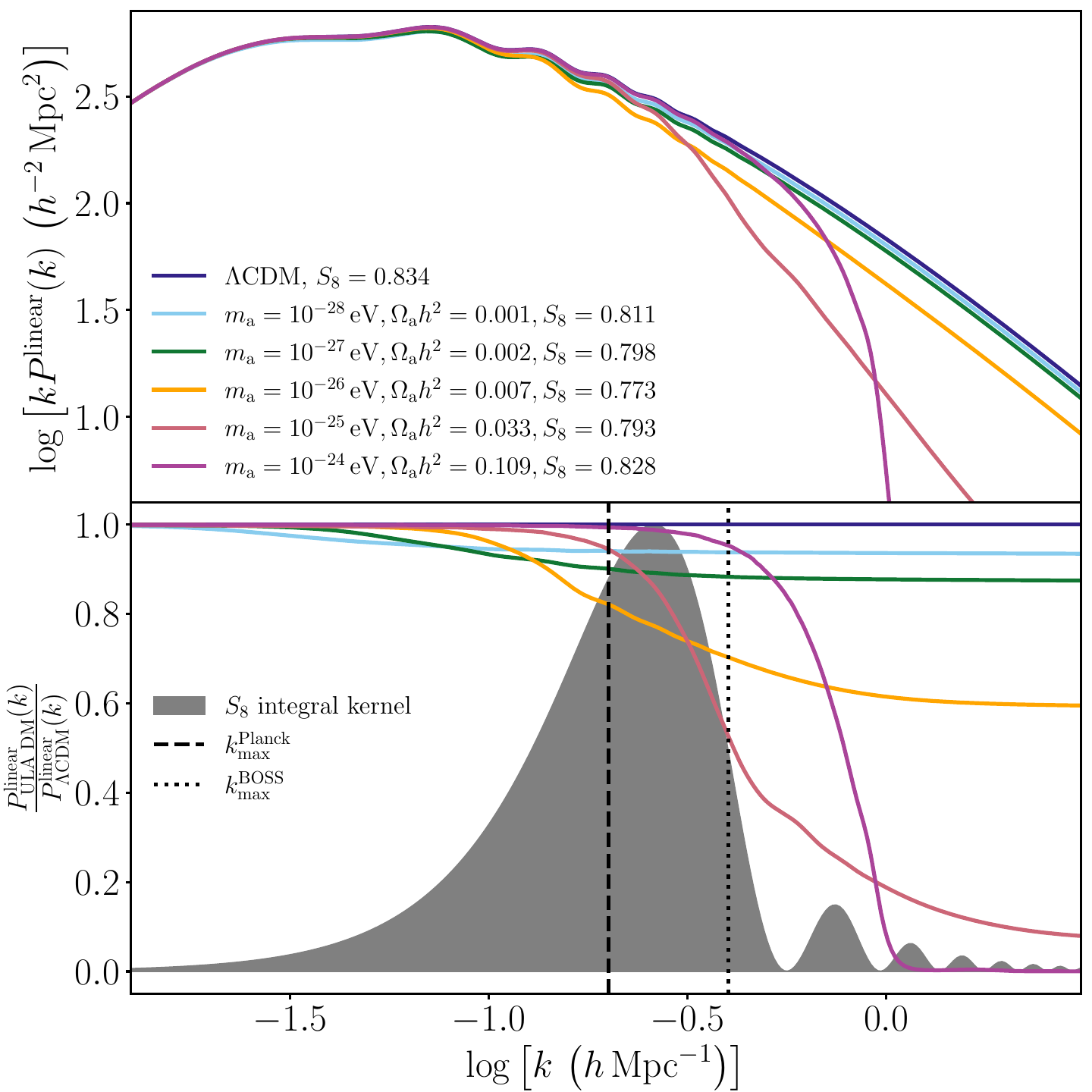} 
\caption{\label{fig:power_linear}The effect of ultra-light axions on the linear matter power spectrum (\textit{top panel}) and the ratio to the \(\Lambda\)CDM limit (\textit{bottom panel}). Power spectra are shown at the 95\% upper limit on the axion energy density \(\Omega_\mathrm{a} h^2\) given \textit{Planck} CMB and BOSS galaxy clustering data and thus reflect the tightening density constraint at lower axion mass \(m_\mathrm{a}\) (see Table \ref{tab:planck}; all parameters fixed apart from \(m_\mathrm{a}\), \(\Omega_\mathrm{a} h^2\), the cold DM density \(\Omega_\mathrm{c} h^2\) and the dark energy fraction \(\Omega_\Lambda\)). In the bottom panel, the shaded area shows the Fourier-space filter \(k^2 W^2(k)\) (in arbitrary units) of \(k P^\mathrm{linear}(k)\) in the integral calculation of the matter clumping factor \(S_8\) (see Eq.~\eqref{eq:sigma8}), where \(W(k)\) is the Fourier transform of a top-hat filter in real space with radius \(8\,h^{-1}\,\mathrm{Mpc}\) and \(\Omega_\mathrm{m}\) is the (fixed) total matter energy density. The shaded area thus indicates the wavenumbers to which \(S_8\) is most sensitive; for \(m_\mathrm{a} \leq 10^{-25}\,\mathrm{eV}\), axions suppress power and thus lower \(S_8\); for \(m_\mathrm{a} \geq 10^{-24}\,\mathrm{eV}\), the axion-induced power suppression is at too large a wavenumber to change \(S_8\) significantly. The dashed line indicates the maximum wavenumber which we probe in the \textit{Planck} likelihood (see \S~\ref{sec:planck}); the dotted line indicates the maximum wavenumber which we model in the BOSS galaxy power spectrum (see \S~\ref{sec:boss}).}
\end{figure}

Figure \ref{fig:power_linear} illustrates this Jeans scale (for wavenumbers \(k\) above the Jeans wavenumber, the linear matter power spectrum \(P^\mathrm{linear}(k)\) is suppressed relative to the \(\Lambda\)CDM limit) and how it depends on axion mass \(m_\mathrm{a}\): the lighter the axion, the larger the power suppression scale (the smaller the suppression wavenumber). In Fig.~\ref{fig:power_linear}, we show linear matter power spectra at the 95 \% upper limits on the axion energy density \(\Omega_\mathrm{a} h^2\) given a combination of \textit{Planck} CMB and BOSS galaxy clustering data (see \S~\ref{sec:boss_results}). As will be expanded later, these data set stronger constraints on the amount of axions at lower mass; Fig.~\ref{fig:power_linear} thereby illustrates how a lower \(\Omega_\mathrm{a} h^2\) reduces the strength of the power suppression. The amplitude of the linear matter power spectrum today (redshift \(z = 0\)) is often summarised by the cosmological parameter
\begin{equation}
\label{eq:sigma8}
\sigma_8 = \int \mathrm{d}\mathrm{ln}k\,\frac{k^3}{2 \pi} W^2(k) P^\mathrm{linear}(k),
\end{equation}
where \(W(k)\) is the Fourier transform of a top-hat filter in real space with radius \(8\,h^{-1}\,\mathrm{Mpc}\); large-scale structure (LSS) data are then typically used to constrain the parameter combination \(S_8 = \sqrt{\frac{\Omega_\mathrm{m}}{0.3}} \sigma_8\), where \(\Omega_\mathrm{m}\) is the total matter energy density\footnote{The parameter combination \(S_8\) was historically optimised to project away parameter degeneracies in galaxy weak lensing experiments. We typically use this parameter combination in this work with galaxy clustering since it still does a good job of projecting away degeneracies and it simplifies comparisons to the literature.}. \(S_8\) is therefore sensitive to a filtered integral of the linear matter power spectrum; the bottom panel of Fig.~\ref{fig:power_linear} shows this filter and indicates the scales to which \(S_8\) is sensitive. It can be seen that, for \(m_\mathrm{a} \geq 10^{-24}\,\mathrm{eV}\), the power suppression is on too small scales to lower significantly \(S_8\). For \(m_\mathrm{a} \leq 10^{-28}\,\mathrm{eV}\), the data constraint is too strong to leave an appreciable amount of axions that significantly lowers \(S_8\). However, there is a window for \(m_\mathrm{a} \in [10^{-27}, 10^{-25}]\,\mathrm{eV}\), where the presence of axions significantly lowers \(S_8\) and is allowed by the data we consider. \(S_8\) is lowered because the Jeans scale today for \(m_\mathrm{a} = 10^{-25} - 10^{-26}\,\mathrm{eV}\) is about \(4 - 12\,h^{-1}\,\mathrm{Mpc}\) (see Eq.~\eqref{eq:axion_Jeans}). We therefore discuss the prospects of axions resolving discrepancies in the inferred values of \(S_8\) given CMB and LSS data in \S~\ref{sec:discussion}.

\subsubsection{Cosmic microwave background}
\label{sec:linear_cmb}

\begin{figure}[tbp]
\centering 
\includegraphics[width=\textwidth]{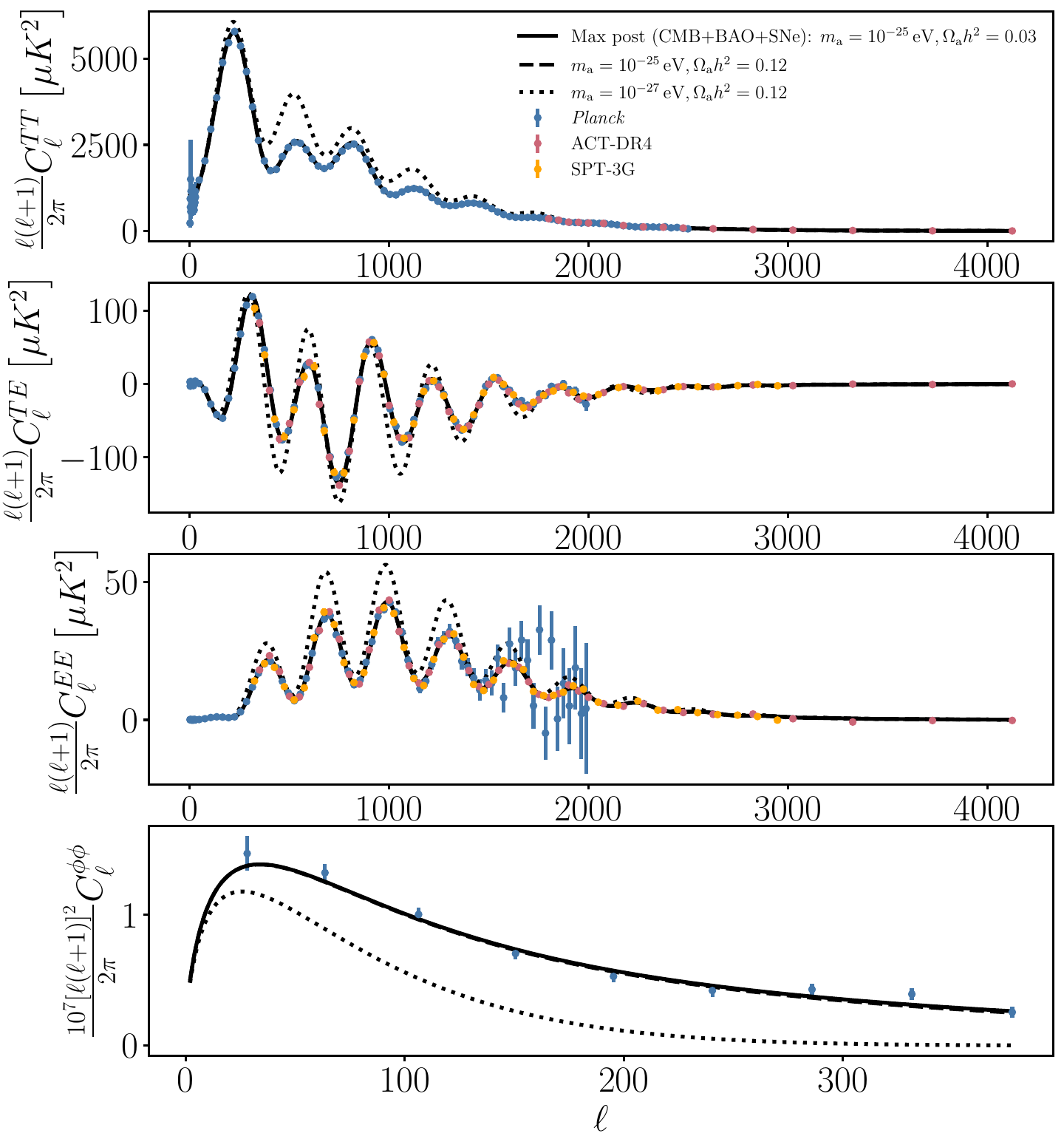} 
\caption{\label{fig:power_CMB}The effect of ultra-light axions on (\textit{from top to bottom}) the cosmic microwave background (CMB) \(TT\), \(TE\), \(EE\) and \(\phi \phi\) angular power spectra, compared to data from \textit{Planck} (\textit{blue}), the Atacama Cosmology Telescope (ACT-DR4; \textit{red}) and the South Pole Telescope (SPT-3G; \textit{orange}). We show the maximum posterior model (\textit{solid}) given \textit{Planck}, ACT and SPT CMB, galaxy baryon acoustic oscillation (BAO) and supernovae (SNe) data for axion mass \(m_\mathrm{a} = 10^{-25}\,\mathrm{eV}\). We compare this to the cases where the axion energy density \(\Omega_\mathrm{a} h^2 = 0.12\) for \(m_\mathrm{a} = 10^{-25}, 10^{-27}\,\mathrm{eV}\) (all other parameters fixed to their maximum posterior values). At the multipoles currently probed, axions for \(m_\mathrm{a} \geq 10^{-25}\,\mathrm{eV}\) are poorly constrained by these data; upcoming high-resolution CMB lensing measurements will increase sensitivity to heavier axions. Data are shown as points with \(68 \%\) c.l. errorbars.}
\end{figure}

In modelling CMB anisotropies, we consider only adiabatic initial perturbations; we defer a search for isocurvature perturbations to future work \citep[see][for consequences on the energy scale of inflation]{2013PhRvD..87l1701M,2018MNRAS.476.3063H}. Fig.~\ref{fig:power_CMB} illustrates how axions change the CMB temperature \(TT\), polarisation \(EE\), cross \(TE\) and lensing potential \(\phi \phi\) angular power spectra \(C_\ell\) as a function of multipole \(\ell\). For the multipoles probed by current data (\textit{Planck}, ACT-DR4, SPT-3G; see \S~\ref{sec:data}), the impact of axions for \(m_\mathrm{a} \geq 10^{-25}\,\mathrm{eV}\) on these data is small compared to statistical uncertainties. For axions that become a dark matter component before matter-radiation equality (\(m_\mathrm{a} \geq 10^{-27}\,\mathrm{eV}\)), much of the data constraint comes from the change in the relative heights of acoustic peaks arising from the change in the matter-to-radiation ratio (see Fig.~\ref{fig:power_CMB}). For axions that still behave like dark energy after matter-radiation equality (\(m_\mathrm{a} \leq 10^{-28}\,\mathrm{eV}\)), much of the data constraint (once the angular size of the sound horizon is constrained) comes from the change in the integrated Sachs-Wolfe effect at the smallest multipoles. The lensing power spectrum is sensitive to all axion masses through a scale (multipole)-dependent suppression arising from the matter power spectrum (see Fig.~\ref{fig:power_linear}; see Refs.~\cite{2015PhRvD..91j3512H,2017PhRvD..95l3511H,2018MNRAS.476.3063H} for summaries of the effects of axions on the CMB).

\begin{figure}
    \centering
    \includegraphics[width=\textwidth, trim={10, 30, 0, 10}, clip]{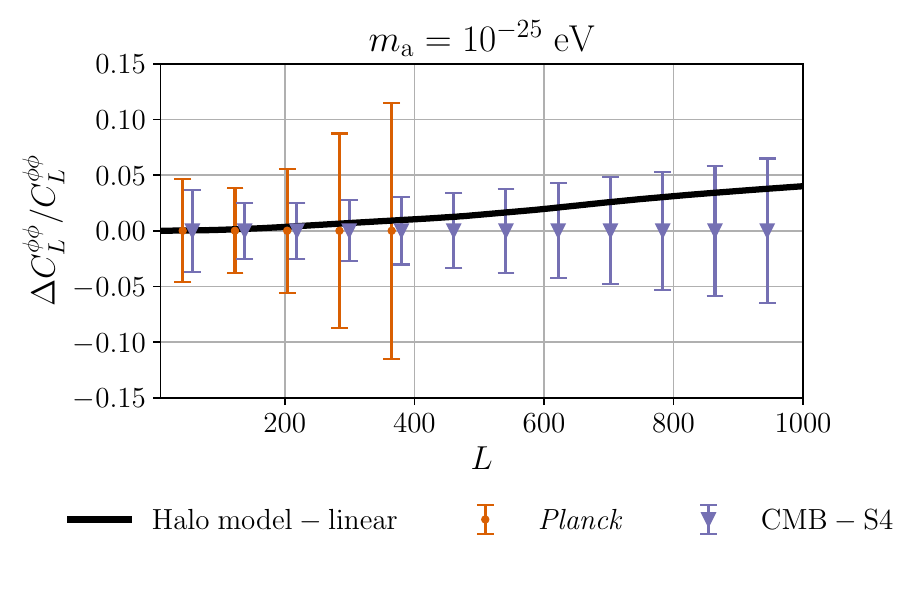}
    \caption{Fractional difference in the cosmic microwave background (CMB) lensing potential angular power spectrum \(C^{\phi \phi}_L\) as a function of multipole \(L\) between a non-linear axion halo model \citep{2022arXiv220913445V} and the linear theory prediction from \texttt{axionCAMB} \citep{2015PhRvD..91j3512H,2022ascl.soft03026G} (\textit{black line}). We show the difference in the theoretical prediction for an axion of mass $m_\mathrm{a}=10^{-25}$ eV which constitutes 50\% of the total dark matter energy density. There is a negligible difference for $L \leq 400$ compared to the error in the \textit{Planck} data that we use (\textit{orange}; see \S~\ref{sec:planck}). The inclusion of non-linearities will however be necessary for future CMB surveys such as CMB-S4 (\textit{purple}; forecasted data error from Ref.~\cite{2018MNRAS.476.3063H}).\label{fig:rel_diff_non_lin}}
\end{figure}

Since we conservatively ignore small-scale CMB lensing anisotropies (we use only multipoles \(8 \leq L \leq 400\) as recommended by the Planck Collaboration; see \S~\ref{sec:planck}), we ignore non-linear effects in the lensing power spectrum. Refs.~\cite{2017PhRvD..95l3511H,bauer/etal:2021,2022arXiv220913445V} indicated that this is a good approximation as, for axion mass \(m_\mathrm{a} < 10^{-25}\,\mathrm{eV}\), axions with observationally-allowed energy densities do not non-linearly cluster at observationally-relevant wavenumbers and redshifts. For \(m_\mathrm{a} \geq 10^{-25}\,\mathrm{eV}\), the non-linear effects are only significant for larger multipoles than we use. A halo model was developed to capture the effects of ultra-light axions on non-linear scales in CMB and galaxy lensing \citep{2022arXiv220913445V}. Using this halo model, we reconsider the impact of ignoring non-linear effects in our \textit{Planck} CMB lensing analysis (Fig.~\ref{fig:rel_diff_non_lin}). We conclude that this observable is well captured using linear theory for the $L$ range considered in this work. Forthcoming measurements of small-scale lensing anisotropies from ground-based CMB (and future galaxy weak lensing) experiments will increase sensitivity to smaller axion suppression scales and, hence, larger \(m_\mathrm{a}\). We anticipate that non-linear modelling will become necessary in this regime.

\subsection{Galaxy clustering and the effective field theory of large-scale structure}
\label{sec:eft}

\subsubsection{Galaxy power spectrum and bispectrum multipoles}
\label{sec:eft_multipoles}

In order to capture the anisotropic clustering in the galaxy distribution arising from redshift-space effects, we model galaxy power spectrum multipoles \citep[\eg][]{2020PhRvD.102f3533C}:
\begin{equation}
\label{eq:multipoles}
P_\ell (k, z) \equiv \frac{2 \ell + 1}{2} \int_{-1}^1 \mathrm{d}\mu\,\mathcal{L}_\ell (\mu) P_\mathrm{g} (k, \mu, z).
\end{equation}
Here, \(P_\mathrm{g} (k, \mu, z)\) is the full anisotropic galaxy power spectrum depending on wavenumber \(k\), the cosine of the angle between the wavenumber and the line-of-sight \(\mu\), and redshift \(z\); \(\mathcal{L}_\ell (\mu)\) are Legendre polynomials indexed by multipole \(\ell\). 
Non-linear redshift-space distortions (``fingers of God''~\cite{Jackson:1971sky}) 
are non-trivial to model with accuracy on small scales, while the power suppression effect of axions is stronger as wavenumber increases. Therefore, to increase the constraining power from galaxy data and following Ref.~\cite{2022PhRvD.105d3531I}, we estimate the reconstructed real-space\footnote{\Ie without redshift-space distortions.} galaxy power spectrum \(Q_0 (k, z) \equiv P_0 (k, z) - \frac{1}{2} P_2 (k, z) + \frac{3}{8} P_4 (k, z)\) \citep[see also][]{DAmico:2021ymi}. Ref.~\cite{2022PhRvD.105d3531I} demonstrates that this estimator effectively down-weights information in line-of-sight modes (\(\mu > 0.3\)) that are heavily contaminated by redshift-space distortions. Further, we extract information on the post-reconstructed baryon acoustic oscillation feature using the Alcock-Paczynski (AP) parameters:
\begin{equation}
\label{eq:AP}
\alpha_\parallel(z) \equiv \frac{H^\mathrm{fid}(z) r^\mathrm{fid}_\mathrm{s}(z_\mathrm{d})}{H(z) r_\mathrm{s}(z_\mathrm{d})},\,\,\,\alpha_\perp(z) \equiv \frac{D_\mathrm{A}(z) r^\mathrm{fid}_\mathrm{s}(z_\mathrm{d})}{D^\mathrm{fid}_\mathrm{A}(z) r_\mathrm{s}(z_\mathrm{d})}.
\end{equation}
Here, \(H(z)\) is the Hubble parameter, \(r_\mathrm{s}(z_\mathrm{d})\) is the sound horizon at the redshift of decoupling, \(D_\mathrm{A}(z)\) is the angular diameter distance, and ``fid'' indicates a fiducial cosmology. For the first time, in order to extract information beyond the two-point statistics described above, we model the effect of axions on the galaxy bispectrum (Fourier transform of the three-point correlation function). In this work, we consider only the angle-averaged bispectrum monopole (\(\ell = 0\)) \(B_0 (k_1, k_2, k_3)\).

\subsubsection{Introduction to the effective field theory of large-scale structure}
\label{sec:eft_intro}

In order to model the effect of ULAs on the redshift-space galaxy power spectrum and bispectrum, we calculate mildly non-linear perturbations using the effective field theory of large-scale structure \citep[EFT of LSS;][]{2012JCAP...07..051B,2018JCAP...04..030S,Cabass:2022avo,Ivanov:2022mrd} as implemented in the Einstein-Boltzmann solver~\texttt{CLASS-PT}\footnote{\url{https://github.com/Michalychforever/CLASS-PT}. Background evolution and linear matter power spectrum calculations are done in \texttt{axionCAMB}, which are passed to \texttt{CLASS-PT} for non-linear corrections.}~\citep{2020PhRvD.102f3533C}. Following the effective field theory of large-scale structure, this is a schematic view of our power spectrum model \citep{Ivanov:2019pdj,2020PhRvD.102f3533C}:
\begin{equation}
\label{eq:eft_power}
P_\ell (k, z) = P_\ell^\mathrm{tree} (k, z) + P_\ell^\mathrm{one-loop} (k, z) + P_\ell^\mathrm{counterterms} (k, z) + P_\ell^\mathrm{stochastic} (k, z).
\end{equation}
Here, \(P_\ell^\mathrm{tree} (k, z)\) captures linear bias and redshift-space distortions (\(\propto P^\mathrm{linear} (k, z)\), which is the linear matter power spectrum) \citep{1987MNRAS.227....1K}; \(P_\ell^\mathrm{one-loop} (k, z)\) captures perturbative corrections up to one loop in order (\(\propto k^2 P^\mathrm{linear} (k, z)\) on large scales); \(P_\ell^\mathrm{counterterms} (k, z)\) captures ultraviolet counterterms that consistently account for small-scale physics (\(\propto k^2 P^\mathrm{linear} (k, z)\)); and \(P_\ell^\mathrm{stochastic} (k, z)\) captures stochastic effects including shot noise and fingers of God (\(\propto\) constant, plus corrections). We also include infrared resummation to account for non-perturbative long-wavelength displacements~\citep{Senatore:2014via,Blas:2016sfa,Ivanov:2018gjr,2020PhRvD.102f3533C}, and account for the so-called Alcock-Paczynski distortion \citep{1979Natur.281..358A} (\ie the effects of an incorrect fiducial cosmology above) by wavevector rescalings \citep[see \eg][]{2022PhRvD.105d3517P,Arvanitaki:2019rax}.

The bispectrum model can be given schematically in 3D space \citep{2022PhRvD.105f3512I}:
\begin{equation}
\label{eq:eft_bispec}
B (\vec{k}_1, \vec{k}_2, z) = B^\mathrm{tree} (\vec{k}_1, \vec{k}_2, z) + B^\mathrm{counterterms} (\vec{k}_1, \vec{k}_2, z) + B^\mathrm{stochastic} (\vec{k}_1, \vec{k}_2, z),
\end{equation}
for wavevectors \(\vec{k}_1, \vec{k}_2\). Here, \(B^\mathrm{tree} (\vec{k}_1, \vec{k}_2, z)\) captures tree-level perturbations (\(\propto P^2_\mathrm{linear} (k, z)\)); \(B^\mathrm{counterterms} (\vec{k}_1, \vec{k}_2, z)\) is a proxy for ultraviolet fingers-of-God counterterms (\(\propto k_\parallel^2 P^2_\mathrm{linear} (k, z)\), where \(k_\parallel\) is the wavenumber for modes parallel to the line of sight); and \(B^\mathrm{stochastic} (\vec{k}_1, \vec{k}_2, z)\) captures stochastic effects (\(\propto P_\mathrm{linear} (k, z) + \mathrm{constant}\)). For the bispectrum, the tree-level model is sufficiently accurate for the small wavenumbers that we consider in our data (\(k < 0.08\,h\,\mathrm{Mpc}^{-1}\); see \S~\ref{sec:boss}). We again account for infrared resummation~\cite{Ivanov:2018gjr} and the Alcock-Paczynski distortion as above. We then integrate over external angles to calculate the bispectrum monopole \(B_0 (k_1, k_2, k_3)\), which can be compared to data without additional window convolutions. In comparing to BOSS data, we multiply \(B_0\) by a discreteness weight vector to account for the finite resoltuion of the Fourier grid\footnote{BOSS discreteness weight vectors can be found at \url{https://github.com/oliverphilcox/full_shape_likelihoods}.}.

\subsubsection{Axions and the effective field theory of large-scale structure}
\label{sec:eft_axions}

The EFT of LSS was originally developed with the assumption of cold, collisionless dark matter (CDM). However, we follow Ref.~\cite{2022JCAP...01..049L} in noting that the effect of ultra-light axion dark matter (not to cluster below a characteristic scale) is qualitatively the same as for free-streaming neutrinos (although the physical reason is different). Ref.~\cite{2017arXiv170704698S} found that, to first order, the additional counterterms needed to account for the effect of neutrinos have the same functional form as existing CDM counterterms. We therefore assume that the additional axion-induced counterterms will also have the same functional form, although with different constants of proportionality which must be marginalised. Further, Ref.~\cite{2022JCAP...01..049L} demonstrated that linear-order axion-wave corrections are negligible since they manifest on scales that are already heavily suppressed in the linear matter power spectrum. Ref.~\cite{2022JCAP...01..049L} also demonstrated that axion and cosmological parameters can be inferred without bias from a simulated BOSS galaxy catalogue in the presence of axions, when marginalising over the EFT of LSS model presented above. It follows that phenomenology of axions can be captured by only modifying the background evolution and linear matter power spectrum (presented in \S~\ref{sec:linear} and calculated using \texttt{axionCAMB}) as input to the EFT of LSS model presented in Eqs.~\eqref{eq:eft_power} and \eqref{eq:eft_bispec}.

We marginalise over a full set of EFT of LSS nuisance parameters: linear \(b_1\), quadratic \(b_2\), tidal \(b_{\mathcal{G}_2}\) and third-order \(b_{\Gamma_3}\) galaxy biases; monopole \(c_0\), quadrupole \(c_2\), hexadecapole \(c_4\), fingers-of-God \(\Tilde{c}\) and bispectrum \(c_1\) counterterms; power spectrum \(P_\mathrm{shot}\) and bispectrum \(B_\mathrm{shot}\) shot noise parameters; and power spectrum scale-dependent stochastic parameters \(a_0\) and \(a_2\) \citep[more details are given in][]{2022PhRvD.105f3512I}.

\begin{figure}[tbp]
\centering 
\includegraphics[width=\textwidth]{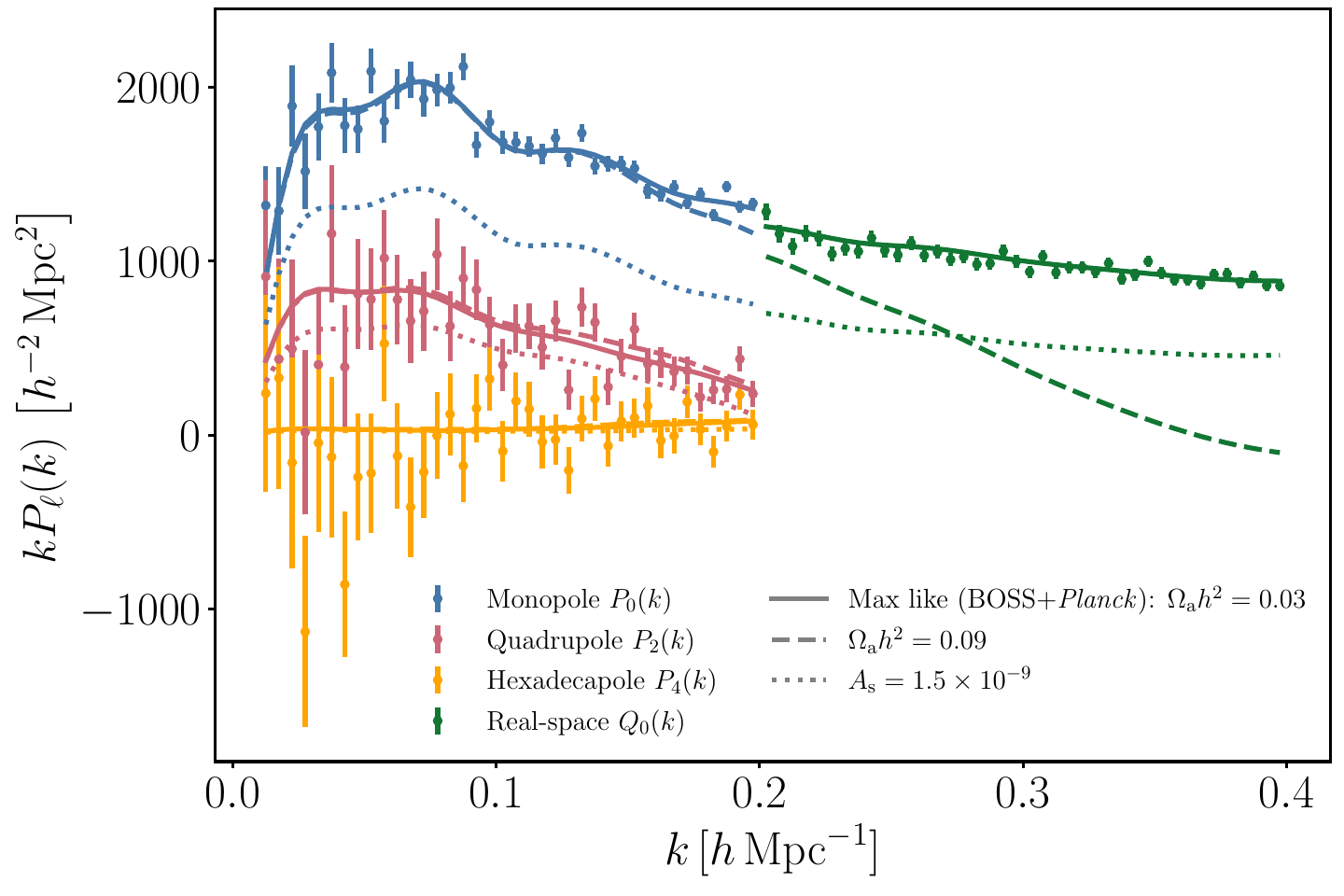} 
\caption{\label{fig:power}The effect of ultra-light axions (mass \(m_\mathrm{a} = 10^{-25}\,\mathrm{eV}\), axion energy density \(\Omega_\mathrm{a} h^2 = 0.09\), \textit{dashed lines}) on the Baryon Oscillation Spectroscopic Survey (BOSS) galaxy power spectrum, compared to maximum-likelihood model parameters (with \(\Omega_\mathrm{a} h^2 = 0.03\), \textit{solid lines}; all other parameters fixed to maximum-likelihood values). The solid line shows the maximum likelihood given BOSS galaxy power spectrum and \textit{Planck} cosmic microwave background (CMB) data; here, we maximise the likelihood with respect to all cosmological and EFT of LSS parameters, including those that are usually analytically marginalised. We also show the case (\textit{dotted lines}) where we lower the primordial power spectrum amplitude from its best-fit value given \textit{Planck} + BOSS (\(A_\mathrm{s} = 2.15 \times 10^{-9}\)) to its best-fit value given only BOSS galaxy power spectra (\(A_\mathrm{s} = 1.53 \times 10^{-9}\)). This illustrates the lack of degeneracy with heavier axions (\(m_\mathrm{a} = 10^{-25}\,\mathrm{eV}\)); nonetheless, a good fit to BOSS data is maintained given the addition of \textit{Planck} data by reducing the best-fit linear galaxy bias (see also Fig.~\ref{fig:boss_joint_axions}). We anticipate degeneracy between high \(A_\mathrm{s}\) (and also high linear galaxy bias \(b_1\)) and high \(\Omega_\mathrm{a} h^2\) for \(m_\mathrm{a} \leq 10^{-28}\,\mathrm{eV}\) since the large Jeans scale suppresses all BOSS wavenumbers in a similar way as reducing \(A_\mathrm{s}\) (or \(b_1\)). We show the galaxy power spectrum monopole \(P_0 (k)\) (\textit{blue}), quadrupole \(P_2 (k)\) (\textit{red}), hexadecapole \(P_4 (k)\) (\textit{orange}), and the reconstructed real-space galaxy power spectrum \(Q_0 (k)\) (\textit{green}), as a function of wavenumber \(k\). BOSS data are shown as points with \(68 \%\) c.l. errorbars.}
\end{figure}

Figure~\ref{fig:power} shows the effect of ultra-light axions on the galaxy power spectrum, for \(m_\mathrm{a} = 10^{-25}\,\mathrm{eV}\). The solid line shows the power spectrum for axion energy density \(\Omega_\mathrm{a} h^2 = 0.03\), while the dashed line shows the power spectrum for \(\Omega_\mathrm{a} h^2 = 0.09\) (all other parameters fixed to the same values as for the solid line). There are two main effects. The first is a scale-dependent suppression in the power spectrum, which gets stronger on smaller scales (and so is most significantly seen in the \(Q_0\) statistic). This is qualitatively similar to the effect in the linear matter power spectrum. The effect is physically caused by axions not clustering on scales below their Jeans wavelength at matter-radiation equality. The second effect is a small-scale enhancement in the galaxy quadrupole (and, to a much-lesser extent, hexadecapole). This effect is caused by a reduction in the fingers of God effect owing to lower peculiar velocities at weaker matter over-densities, although this is degenerate with the EFT of LSS counterterm parameter that controls the fingers of God amplitude. These effects are discussed in further detail in Ref.~\cite{2022JCAP...01..049L}. The dotted lines lower the primordial power spectrum amplitude \(A_\mathrm{s}\) with respect to the solid lines and thus suppress the galaxy power spectrum at all wavenumbers.

\begin{figure}[tbp]
\centering 
\includegraphics[width=\textwidth]{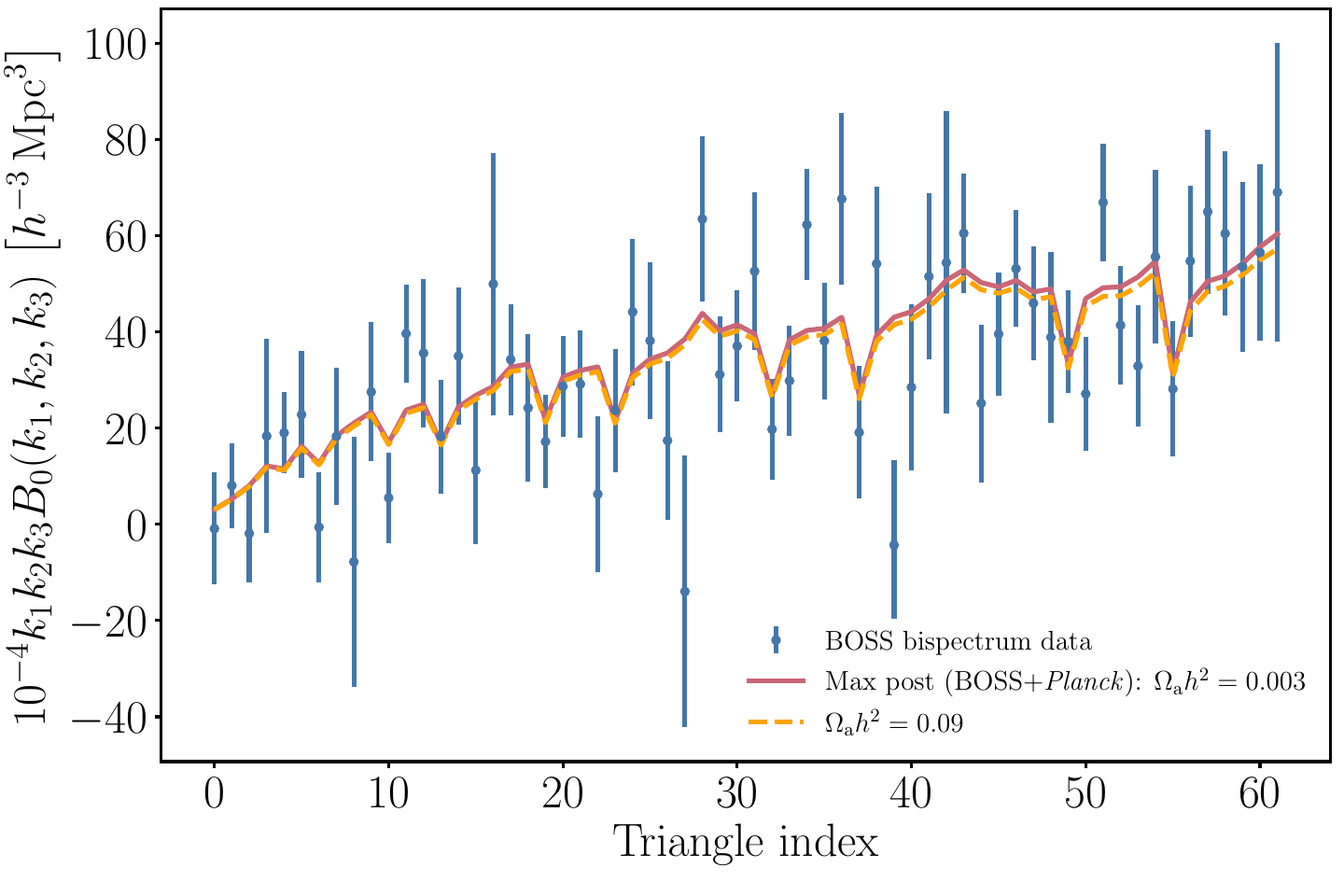} 
\caption{\label{fig:bispectrum}The effect of ultra-light axions (\(m_\mathrm{a} = 10^{-25}\,\mathrm{eV}\), \(\Omega_\mathrm{a} h^2 = 0.09\), \textit{dashed line}) on the BOSS galaxy bispectrum monopole \(B_0 (k_1, k_2, k_3)\), compared to maximum-posterior model parameters (with negligible axion densities, \textit{solid line}; all other parameters fixed to their maximum posterior values). The solid line shows the maximum posterior given BOSS power spectrum and bispectrum and \textit{Planck} CMB data. We show \(B_0\) as a function of \(k_1, k_2, k_3\) wavenumber triangles, where triangle index increases first with \(k_1\), then with \(k_2\), and then with \(k_3\), for \([k_1, k_2, k_3] \in [0.01, 0.08]\,h\,\mathrm{Mpc}^{-1}\); \ie wavenumber triangles with smaller sides on the left and larger sides on the right. BOSS data are shown as points with \(68 \%\) c.l. errorbars.}
\end{figure}

Fig.~\ref{fig:bispectrum} shows the effect of ultra-light axions on the galaxy bispectrum monopole, for \(m_\mathrm{a} = 10^{-25}\,\mathrm{eV}\). As above, the dashed line shows the effect of increasing the axion density while keeping all other parameters fixed. The effect is a small scale-dependent suppression in the bispectrum, which gets stronger for smaller-scale triangles. However, at the current statistical precision of BOSS data and on the relatively large scales modelled here in the bispectrum (\(k < 0.08\,h\,\mathrm{Mpc^{-1}}\)), the effect is negligible. We anticipate that axions will impact the smaller-scale, one-loop bispectrum \citep{Philcox:2022frc,DAmico:2022osl} more strongly; we will investigate this in future work.

\section{Data}
\label{sec:data}
\begin{table}
    \centering
    \begin{tabular}{lcccc}
        \hline
		Data & Description & Nuisance pars. & \S & Refs. \\
		\hline
		\textit{Planck} 2018 & $C_{\ell}^{TT, TE, EE, \phi \phi}$: $\ell \leq 2508$, $L \leq 400$ & $a_\mathrm{Planck}$ & \ref{sec:planck} & {\citep{2020A&A...641A...6P}} \\
		ACT-DR4 & $C_{\ell}^{TT, TE, EE}$: $326 \leq \ell \leq 4325$ & $y_\mathrm{p}$ & \ref{sec:act} & \citep{2020JCAP...12..047A} \\
		SPT-3G & $C_{\ell}^{TE, EE}$: $300 \leq \ell \leq 2999$ & Fixed & \ref{sec:spt} & \citep{2021PhRvD.104b2003D} \\
		BAO + SNe & 6dFGS, MGS, BOSS DR12, JLA & $\alpha, \beta, \delta M$ & \ref{sec:bao} & {\citep{2011MNRAS.416.3017B, 2015MNRAS.449..835R, 2017MNRAS.470.2617A, 2014A&A...568A..22B}} \\
		BOSS full-shape & $P_{0, 2, 4}(k, z)$, $Q_0(k, z)$, $B_{0}(k, z)$, AP & EFT of LSS & \ref{sec:boss} & \citep{2022PhRvD.105d3517P} \\
		\hline
    \end{tabular}
    \caption{\label{tab:data}A summary of the data used in this work.}
\end{table}

In Table \ref{tab:data}, we give a summary of the data used in this work. In \S~\ref{sec:cmb} to \ref{sec:boss}, we give more details about the data and, in \S~\ref{sec:inference}, we give details on our parameter inference method including the prior distribution.

\subsection{Cosmic microwave background}
\label{sec:cmb}

\subsubsection{\textit{Planck}}
\label{sec:planck}
\begin{sloppypar}
We consider baseline \textit{Planck} 2018 CMB temperature, polarisation and lensing angular power spectra \citep{2020A&A...641A...6P}. We use: the low-multipole \((2 \leq \ell \leq 29)\) temperature \(TT\) auto-spectrum likelihood \texttt{commander\_dx12\_v3\_2\_29}; the low-multipole \((2 \leq \ell \leq 29)\) polarisation \(EE\) auto-spectrum likelihood \texttt{simall\_100x143\_offlike5\_EE\_Aplanck\_B}; the high-multipole, nuisance-marginalised, \(TT\) \((30 \leq \ell \leq 2508)\), \(TE\) and \(EE\) \((30 \leq \ell \leq 1996)\) likelihood \texttt{plik\_lite\_v22\_TTTEEE}; and the lensing \(\phi \phi\) auto-spectrum likelihood \((8 \leq L \leq 400)\) \texttt{smicadx12\_Dec5\_ftl\_mv2\_ndlcpp\_p\_teb\_consext8}. As we use compressed, nuisance-marginalised likelihoods, we have remaining a single nuisance calibration parameter \(a_\mathrm{Planck}\). Ref.~\cite{2015PhRvD..91j3512H} demonstrated that there is no statistically-significant effect on axion parameter inference if re-marginalising nuisance foreground parameters in an axion model with \textit{Planck} data. The use of \textit{Planck} 2018 data to constrain \(\Omega_\mathrm{a} h^2\) is an update from Refs.~\cite{2018MNRAS.476.3063H, 2022JCAP...01..049L}, which used \textit{Planck} 2015 data \citep{2016A&A...594A..13P}, and from Ref.~\cite{dentler/etal:2022}, which used \textit{Planck} 2018 data to constrain only \(m_\mathrm{a}\) in the case where axions are all the dark matter. The main differences from 2015 data are a new low-\(\ell\) polarisation likelihood and a larger-scale cut in the lensing likelihood, \ie \(L_\mathrm{min}\) goes from 40 to 8. We anticipate improved bounds on the lightest axions from this additional large-scale information (see Fig.~\ref{fig:planck_test} for a breakdown of how 2018 data improves axion limits).
\end{sloppypar}

\subsubsection{Atacama Cosmology Telescope}
\label{sec:act}
We consider Atacama Cosmology Telescope (ACT) data release 4 (DR4) temperature and polarisation angular power spectra \citep{2020JCAP...12..047A}. We use the baseline nuisance-marginalised (``CMB-only'') likelihood \texttt{actpollite}. This includes \(TT\) power spectra for \(576 \leq \ell \leq 4325\) and \(TE\) and \(EE\) power spectra for \(326 \leq \ell \leq 4325\). The foreground marginalisation leaves a single nuisance parameter \(y_\mathrm{p}\), which is an overall polarisation efficiency that re-scales the \(TE\) and \(EE\) spectra. Our baseline analysis combines ACT and \textit{Planck} (see \S~\ref{sec:planck}) data. However, the cross-covariance between these data has not yet been released. Therefore, to reduce the amount of cross-covariance that we ignore, we follow Ref.~\cite{2020JCAP...12..047A} in setting the minimum multipole in the ACT \(TT\) spectrum \(\ell_\mathrm{min} = 1800\), with no cut on the \(TE\) and \(EE\) spectra. Ref.~\cite{2020JCAP...12..047A} found that these approximations are sufficient to keep the underestimation of parameter uncertainties to less than 5\%, including for one-parameter extensions of the standard cosmological model.

\subsubsection{South Pole Telescope}
\label{sec:spt}
We consider South Pole Telescope (SPT-3G) \(TE\) and \(EE\) angular power spectra for \(300 \leq \ell \leq 2999\)\footnote{In the latter stages of manuscript preparation, SPT-3G \(TT\) angular power spectra were released \citep{2022arXiv221205642B}; we will include these data in a future analysis, although we do not anticipate a significant change to our results since the multipoles contained in this release are already covered by the ACT data that we use (\S~\ref{sec:act}).} \citep{2021PhRvD.104b2003D}. We use the baseline \texttt{spt3g\_2020} likelihood that has twenty nuisance foreground and calibration parameters\footnote{We write a CosmoSIS \citep{2015A&C....12...45Z} wrapper to the Cobaya \citep{2021JCAP...05..057T} likelihood available at \url{https://github.com/xgarrido/spt_likelihoods}.}. In order to reduce the dimensionality of the parameter space, we fix the nuisance parameters to fiducial values\footnote{We take fiducial values to be the maximum prior probability values from the fiducial SPT-3G analysis: \url{https://github.com/xgarrido/spt_likelihoods/blob/master/spt3g_2020/TEEE.yaml}.}. We confirm that fixing these parameters makes no difference to the inferred cosmological posterior distribution by comparing to the case where all nuisance parameters are marginalised. Our baseline analyses combine SPT with \textit{Planck} (see \S~\ref{sec:planck}) and ACT (see \S~\ref{sec:act}) data. We follow Ref.~\cite{2021PhRvD.104b2003D} in ignoring the cross-covariance between SPT and \textit{Planck} since the survey sky overlap is small; we ignore cross-covariance between SPT and ACT for the same reason \citep[see \eg][for similar assumptions]{PhysRevD.105.083519}.

\subsection{Baryon acoustic oscillations \& supernovae}
\label{sec:bao}
We consider a compendium of galaxy baryon acoustic oscillation (BAO) data from: the 6dF Galaxy Survey (6dFGS) at \(z = 0.106\) \citep{2011MNRAS.416.3017B}; the Sloan Digital Sky Survey data release 7 Main Galaxy Sample (SDSS DR7 MGS) at \(z = 0.15\) \citep{2015MNRAS.449..835R}; and the Baryon Oscillation Spectroscopic Survey data release 12 (BOSS DR12) at \(z = [0.38, 0.51, 0.61]\) \citep{2017MNRAS.470.2617A}. These galaxy samples are largely independent. In \S~\ref{sec:boss}, we will consider instead the full-shape galaxy power spectrum and bispectrum as measured from the BOSS DR12 galaxy sample, which captures the BOSS BAO information plus additional information in the full power spectrum and bispectrum. In the data in \S~\ref{sec:boss}, the BAO information is extracted in the Alcock-Paczynski parameters defined in Eq.~\eqref{eq:AP} from the reconstructed power spectrum, taking into account their covariance with the full-shape information \citep{Philcox:2020vvt}. We never combine the full-shape BOSS likelihood in \S~\ref{sec:boss} with the 
``standard'' BOSS BAO likelihood (used in this section) as they contain identical BAO information.

We consider a compendium of type Ia supernovae (SNe) data from the Joint Light-curve Analysis (JLA) \citep{2014A&A...568A..22B}. We marginalise over the shape parameter \(\alpha\), the colour parameter \(\beta\) and the magnitude parameter \(\delta M\).

\subsection{Baryon Oscillation Spectroscopic Survey galaxy power spectrum \& bispectrum}
\label{sec:boss}

We use the twelfth data release of the Baryon Oscillation Spectrosopic Survey (BOSS DR12) \citep{2017MNRAS.470.2617A,BOSS:2012dmf}, which is part of the Sloan Digital Sky Survey (SDSS) \citep{SDSS:2011jap}. This data release contains $\sim 8 \times 10^5$ galaxies across two redshift slices (LOWZ sample: $0.2<z<0.5$; CMASS sample: $0.5<z<0.75$) and across both the north and south Galactic cap (NGC/SGC) sky cuts. We use the window-free galaxy power spectrum and bispectrum measurements described in Ref.~\cite{2022PhRvD.105d3517P}, which are measured respectively with the approaches of Refs.~\cite{Philcox:2020vbm} and \cite{Philcox:2021ukg}. We bin in \(k\) with $\Delta k = 0.005\,h\,\mathrm{Mpc}^{-1}$ for power spectra and $\Delta k = 0.01\,h\,\mathrm{Mpc}^{-1}$ for bispectra, with minimum wavenumber $k_{\rm min}=0.01\,h\,\mathrm{Mpc}^{-1}$ to avoid large-scale systematics \citep[\eg][]{Kalus:2018qsy}. As discussed in \S~\ref{sec:eft}, we fix the maximum wavenumber $k_{\rm max}=0.2\,h\,\mathrm{Mpc}^{-1}$ for the power spectrum (using the $\ell=0,2,4$ multipoles) \(P_\ell (k, z)\) \citep{Chudaykin:2020ghx,2022PhRvD.105d3517P}, $k_\mathrm{max} = 0.08\,h\,\mathrm{Mpc}^{-1}$ for the bispectrum monopole \citep{2022PhRvD.105f3512I,2022PhRvD.105d3517P} \(B_0 (k, z)\), and we include the $Q_0 (k, z)$ statistic for $k \in [0.2, 0.4]\,h\,\mathrm{Mpc}^{-1}$ following Ref.~\cite{2022PhRvD.105d3531I}. We extract from the BOSS galaxy power spectrum information on the post-reconstructed BAO feature using the AP parameters (Eq.~\eqref{eq:AP}). We use power spectrum, bispectrum and AP measurements for each of the four redshift/sky cuts (NGC and SGC, both in redshift samples with central redshifts 0.38 and 0.61). The BOSS power spectrum and bispectrum data for the North Galactic cap at \(z = 0.38\) are respectively shown in Figs.~\ref{fig:power} and \ref{fig:bispectrum}. These data are analysed with the EFT of LSS model presented in \S~\ref{sec:eft}; we use the existing public BOSS likelihood\footnote{\url{https://github.com/oliverphilcox/full_shape_likelihoods}.}, with the covariance estimated from 2048 MultiDark-Patchy simulations \citep{Kitaura:2015uqa,Rodriguez-Torres:2015vqa}. The EFT of LSS nuisance parameters (see \S~\ref{sec:eft}) are allowed to differ independently for each of the four BOSS data cuts due to their different redshifts and calibrations. Only $b_1, b_2$ and $b_{\mathcal{G}_2}$ enter the model non-linearly: the others are analytically marginalized and only this partially-marginalised likelihood is numerically sampled (see \S~\ref{sec:inference} for details about our numerical sampling approach).

\subsection{Parameter inference}
\label{sec:inference}
All the likelihoods presented in \S~\ref{sec:cmb} to \ref{sec:boss} are implemented in the cosmological parameter estimation code CosmoSIS \citep{2015A&C....12...45Z}. We infer the posterior distribution for an axion cosmological model with uniform prior distributions on: the Hubble parameter \(h\); the baryon energy density \(\Omega_\mathrm{b} h^2\); the cold dark matter energy density \(\Omega_\mathrm{b} h^2\); the axion energy density \(\Omega_\mathrm{a} h^2\); the primordial power spectrum amplitude \(A_\mathrm{s}\); the primordial power spectrum tilt \(n_\mathrm{s}\); and the reionisation optical depth \(\tau\)\footnote{When considering BOSS data alone, we do not vary \(\tau\) since large-scale structure data are insensitive to this parameter.}. We fix the neutrino energy density \(\Omega_\nu h^2 = 0.0006442\) with one massive neutrino at its minimally-allowed mass; Refs.~\cite{marsh2011b,2017PhRvD..95l3511H,2022arXiv220913445V} discuss degeneracies between axion and neutrino density parameters. We consistently calculate the helium abundance given the baryon density and number of neutrinos using the \texttt{bbn\_consistency} module \citep{2008CoPhC.178..956P}. We often show the posterior for derived cosmological parameters: the total matter energy density today \(\Omega_\mathrm{m}\) (to which axions always contribute); and the matter clumping factor \(S_8 \equiv \sqrt{\frac{\Omega_\mathrm{m}}{0.3}} \sigma_8\), where \(\sigma_8\) is the amplitude of the linear matter power spectrum averaged over 8 \(h^{-1}\) Mpc scales. We do not vary the axion mass \(m_\mathrm{a}\), but rather follow Refs.~\cite{2018MNRAS.476.3063H, 2022JCAP...01..049L} by inferring the posterior for fixed values of \(m_\mathrm{a} \in [10^{-32}\,\mathrm{eV}, 10^{-24}\,\mathrm{eV}]\). This is because the full posterior projected in the \(m_\mathrm{a}\) - \(\Omega_\mathrm{a} h^2\) plane has a highly non-trivial degeneracy, which is difficult to sample numerically in a converged manner. There is no impact on our conclusions as we are still able to determine our main results, which are constraints on \(\Omega_\mathrm{a} h^2\) and \(S_8\) as a function of \(m_\mathrm{a}\). The main limitation from our approach is that we cannot marginalise \(m_\mathrm{a}\). However, we find no need to do so in our analysis in the same way that it is not useful to marginalise dark matter parameters like mass, cross section or coupling in experimental searches. We therefore defer solving this sampling problem to future work. We consider neither lighter axions as these are indistinguishable from a cosmological constant, nor heavier axions as these are indistinguishable from cold dark matter on the scales probed by the data we use \citep[see][for more discussion and tests on axion prior choices]{dentler/etal:2022}.

We use a uniform prior on the ACT calibration parameter \(y_\mathrm{p}\) and the three supernovae standardisation parameters \([\alpha, \beta, \delta M]\). We use a Gaussian prior on the \textit{Planck} calibration parameter \(a_\mathrm{Planck} \sim \mathcal{N}(1, 0.0025)\). For the EFT of LSS nuisance parameters (see \S~\ref{sec:eft}), we use the following priors (from Ref.~\cite{2022PhRvD.105d3517P}):
\begin{equation}
\label{eq:priors}
\begin{gathered}
b_1 \sim \mathcal{U}(0,4), \quad b_2 \sim \mathcal{N}(0,1^2), \quad b_{\mathcal{G}_2} \sim \mathcal{N}(0,1^2), \quad b_{\Gamma_3} \sim \mathcal{N}\left(\frac{23}{42}(b_1-1),1^2\right)\\
\frac{c_0}{[h^{-1}\,\text{Mpc}]^2} \sim \mathcal{N}(0,30^2), \quad \frac{c_2}{[h^{-1}\,\text{Mpc}]^2} \sim \mathcal{N}(30,30^2), \quad \frac{c_4}{[h^{-1}\,\text{Mpc}]^2} \sim \mathcal{N}(0,30^2),\\
\frac{\tilde{c}}{[h^{-1}\,\text{Mpc}]^4} \sim \mathcal{N}(500,500^2), \quad \frac{c_1}{[h^{-1}\,\text{Mpc}]^2} \sim \mathcal{N}(0,5^2),\\
P_{\rm shot} \sim \mathcal{N}(0,\bar n^{-2}), \quad B_{\rm shot} \sim \mathcal{N}(1,\bar n^{-2}), \quad a_{0} \sim \mathcal{N}(0,\bar n^{-2}), \quad a_2 \sim \mathcal{N}(0,\bar n^{-2}),
\end{gathered}
\end{equation}
where the inverse galaxy number density $\bar n^{-1}=5000\,[h^{-1}\,\text{Mpc}]^3$
for the high-\(z\) samples and 
$\bar n^{-1}=3500\,[h^{-1}\,\text{Mpc}]^3$
for the low-\(z\) samples. For discussion on these priors and a comparison to the choices made in the \texttt{PyBird} implementation of the EFT of LSS model (whose parameters are a linear combination of the above and which was used in a previous axion analysis \citep{2022JCAP...01..049L}), see \S~\ref{sec:discussion} and Refs.~\cite{Nishimichi:2020tvu} and \cite{Simon:2022lde}.

We numerically sample posterior distributions using the importance nested sampling algorithm \texttt{MultiNest} \citep{2009MNRAS.398.1601F,2019OJAp....2E..10F}. We use 480 live points and we stop chains when posterior weights reach a tolerance of 1\% of their maximum, thus ensuring that we sample the bulk of the posterior weight. We check that our chains are converged with respect to the number of live points and tolerance by running test chains with 3600 live points and a tolerance of \(10^{-5}\) and determining no shift in inferred distributions.

\section{Results}
\label{sec:results}

\subsection{Cosmic microwave background, baryon acoustic oscillations \& supernovae}
\label{sec:cmb_results}

\subsubsection{\textit{Planck}}
\label{sec:cmb_results_planck}

\begin{figure}[tbp]
\centering 
\includegraphics[width=\textwidth]{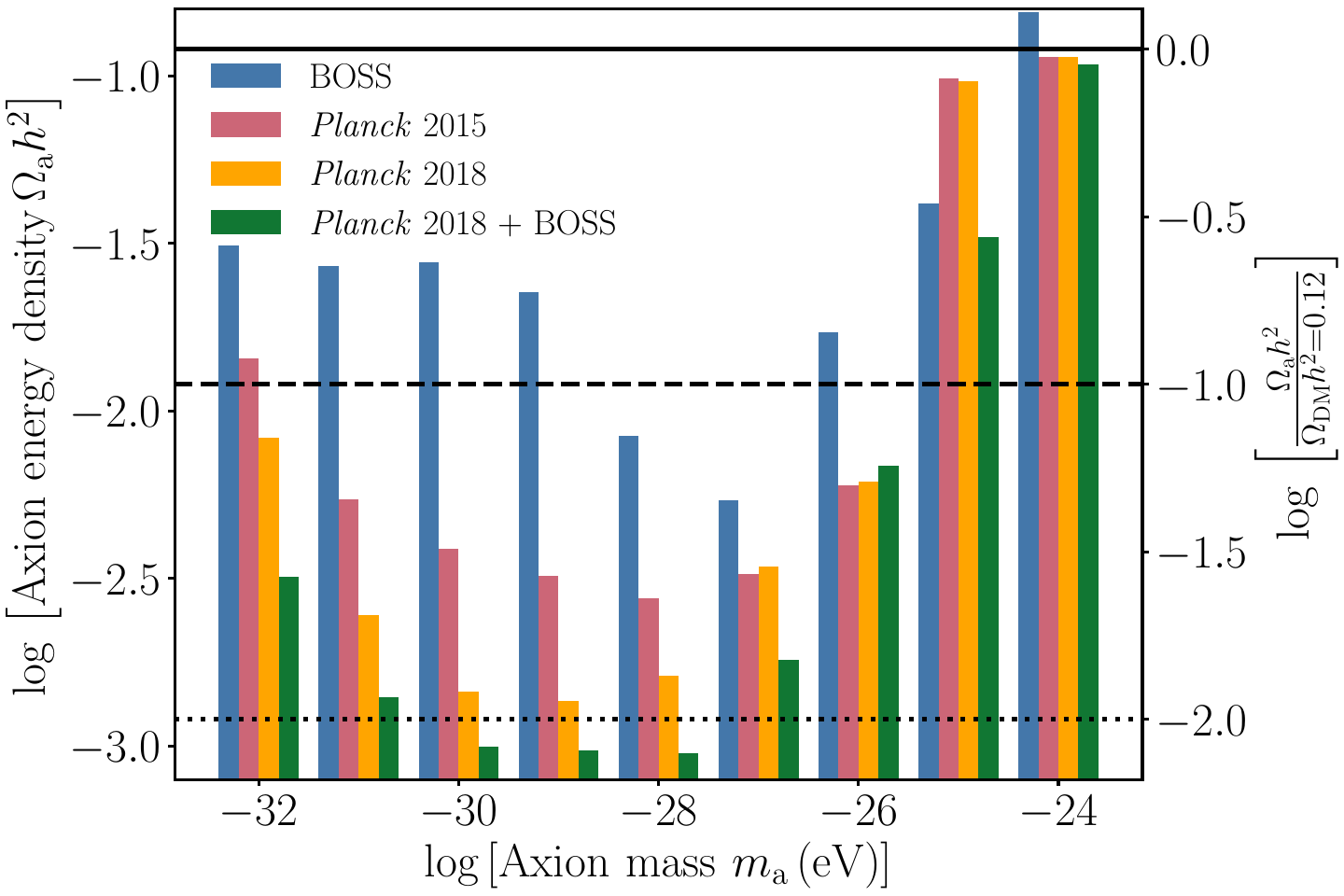} 
\caption{\label{fig:mass_comparison}95\% credible upper limits on axion energy density \(\Omega_\mathrm{a} h^2\), as a function of axion mass \(m_\mathrm{a}\), as inferred: from BOSS galaxy clustering data (\textit{blue}; see \S~\ref{sec:boss_results}); from \textit{Planck} 2015 CMB data (\textit{red}; \citep{2018MNRAS.476.3063H}); from \textit{Planck} 2018 CMB data (\textit{orange}; see \S~\ref{sec:cmb_results}); and as jointly inferred from \textit{Planck} 2018 CMB and BOSS galaxy clustering data (\textit{green}; see \S~\ref{sec:boss_results}). On the right-hand side, we show the 95\% upper limit on the ratio of the axion energy density to the best-fit dark matter (DM) energy density as inferred from \textit{Planck} in the \(\Lambda \mathrm{CDM}\) model \(\Omega_\mathrm{DM} h^2 = 0.12\). The black horizontal dashed and dotted lines respectively indicate the energy densities at which axions form 10\% and 1\% of the DM today.}
\end{figure}

\begin{table}
    \centering
    \begin{tabular}{lccccccccccc}
        \hline
		\(m_\mathrm{a}\) & \(\Omega_\mathrm{a} h^2\) (\textit{Planck}) & $S_8$ (\textit{Planck}) & $\Omega_a h^2$ (\textit{Planck}+BOSS) & $S_8$ (\textit{Planck}+BOSS) \\
		\hline
		$\Lambda$CDM & -- & $0.834^{+0.014}_{-0.013}$ & -- & $0.827\pm 0.011$ \\
		$10^{-24}\,\mathrm{eV}$ & < 0.11399 & $0.831\pm 0.014$ & < 0.10858 & $0.826^{+0.011}_{-0.012}$ \\
		$10^{-25}\,\mathrm{eV}$ & < 0.09667 & $0.811^{+0.025}_{-0.039}$ & < 0.03306 & $0.818^{+0.015}_{-0.017}$ \\
		$10^{-26}\,\mathrm{eV}$ & < 0.00615 & $0.819\pm 0.020$ & < 0.00689 & $0.804^{+0.020}_{-0.024}$ \\
		$10^{-27}\,\mathrm{eV}$ & < 0.00344 & $0.822^{+0.016}_{-0.020}$ & < 0.00181 & $0.819^{+0.013}_{-0.014}$ \\
		$10^{-28}\,\mathrm{eV}$ & < 0.00163 & $0.831^{+0.014}_{-0.012}$ & < 0.00095 & $0.824\pm 0.011$ \\
		$10^{-29}\,\mathrm{eV}$ & < 0.00136 & $0.836\pm 0.014$ & < 0.00097 & $0.826\pm 0.011$ \\
		$10^{-30}\,\mathrm{eV}$ & < 0.00145 & $0.837^{+0.014}_{-0.013}$ & < 0.00099 & $0.827\pm 0.011$ \\
		$10^{-31}\,\mathrm{eV}$ & < 0.00247 & $0.838^{+0.015}_{-0.014}$ & < 0.00140 & $0.827\pm 0.011$ \\
		$10^{-32}\,\mathrm{eV}$ & < 0.00833 & $0.843^{+0.019}_{-0.016}$ & < 0.00321 & $0.829^{+0.012}_{-0.011}$ \\
		\hline
    \end{tabular}
    \caption{\label{tab:planck}Constraints on axion energy density \(\Omega_\mathrm{a} h^2\) and the matter clumping factor \(S_8\), as a function of axion mass \(m_\mathrm{a}\) (\textit{top to bottom}), as inferred from \textit{Planck} CMB data (\textit{left}; see \S~\ref{sec:cmb_results}) and as jointly inferred from \textit{Planck} CMB and BOSS galaxy clustering data (\textit{right}; see \S~\ref{sec:boss_results}). For \(\Omega_\mathrm{a} h^2\), we give the 95\% upper c.l.; for \(S_8\), we give the maximum marginalised posterior with the asymmetric 68\% c.l.}
\end{table}

\begin{figure}[tbp]
\centering 
\includegraphics[width=\textwidth]{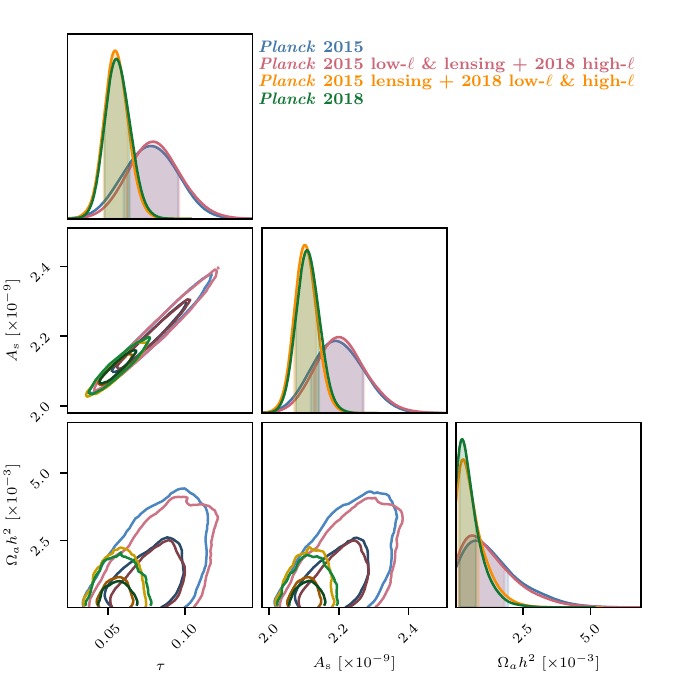} 
\caption{\label{fig:planck_test}The effect of updating from \textit{Planck} 2015 (\textit{blue}) to \textit{Planck} 2018 CMB data on axion constraints for \(m_\mathrm{a} = 10^{-30}\,\mathrm{eV}\). We systematically update parts of the 2015 data with 2018 results: first the high-multipole \(\ell\) likelihood (\textit{red}), then also the low-\(\ell\) likelihood (\textit{orange}), and then finally also the lensing likelihood (\textit{green}). We find that the vast majority of improvement in the axion energy density bound comes from 2018 low-\(\ell\) information, \ie the measurement of the large-scale reionisation bump breaks degeneracies between the reionisation optical depth \(\tau\), the primordial power spectrum amplitude \(A_\mathrm{s}\) and the axion energy density \(\Omega_\mathrm{a} h^2\). For each set, the inner and outer contours respectively indicate the 68\% and 95\% credible regions of the 2D marginalised posterior distribution, with the 1D marginalised posteriors on the diagonal, where 68\% credible regions are shaded.}
\end{figure}

We first search for ultra-light axions (ULAs) in baseline \textit{Planck} 2018 CMB temperature, polarisation and lensing data (see \S~\ref{sec:planck} for a description of the data). We note that this is an update from Refs.~\cite{2018MNRAS.476.3063H, 2022JCAP...01..049L} which considered older \textit{Planck} 2015 data. We now have access to a more robust measurement of the large-scale polarisation signal, and large-scale lensing anisotropies not previously released (\(L_\mathrm{min}\) goes from 40 to 8). We anticipate improved bounds on the lightest axions that we consider since their effect is strong on large scales. Fig.~\ref{fig:mass_comparison} shows the 95\% upper limit on the axion energy density allowed by \textit{Planck} 2018 as a function of mass (it also compares to joint constraints from \textit{Planck} CMB and BOSS galaxy clustering data and constraints from BOSS alone, which we discuss in \S~\ref{sec:boss_results}; these results are also shown in Table \ref{tab:planck}). We see the typical ``u''-shaped constraints \citep{2006PhLB..642..192A} where axions in the ``belly'' (\(10^{-30}\,\mathrm{eV} \leq m_\mathrm{a} \leq 10^{-28}\,\mathrm{eV}\)) are heavily constrained, but dark energy (DE)-like axions for \(m_\mathrm{a} < 10^{-30}\,\mathrm{eV}\) and dark matter (DM)-like axions for \(m_\mathrm{a} \geq 10^{-27}\,\mathrm{eV}\) can still be a significant cosmological component. In particular, \textit{Planck} data lose sensitivity for \(m_\mathrm{a} \geq 10^{-25}\,\mathrm{eV}\) as the scale-dependent suppression is on scales smaller than those that \(Planck\) probes and the background evolution is the same as \(\Lambda\)CDM deep into the radiation epoch.

Our \textit{Planck} 2018 results are consistent with previous \textit{Planck} 2015 limits \citep{2018MNRAS.476.3063H} for \(m_\mathrm{a} \geq 10^{-27}\,\mathrm{eV}\), but stronger for \(m_\mathrm{a} \leq 10^{-28}\,\mathrm{eV}\) (see Fig.~\ref{fig:mass_comparison}). Fig.~\ref{fig:planck_test} investigates which parts of the updated 2018 data are most important in improving axion constraints. In systematically replacing parts of the 2015 likelihood\footnote{We consider the same \textit{Planck} 2015 CMB likelihood as used in Ref.~\cite{2018MNRAS.476.3063H}: the low-\(\ell\) likelihood \texttt{lowl\_SMW\_70\_dx11d\_2014\_10\_03\_v5c\_Ap} for \(2 \leq \ell \leq 29\), the high-\(\ell\) likelihood \texttt{plik\_lite\_v18\_TTTEEE} for \(30 \leq \ell \leq 2508\) (\(TT\) power spectrum) and \(30 \leq \ell \leq 1996\) (\(TE\) and \(EE\) power spectra), and the lensing likelihood \texttt{smica\_g30\_ftl\_full\_pp} for \(40 \leq L \leq 400\).} with 2018 updates, we find that it is the inclusion of the 2018 low-\(\ell\) likelihood that accounts for the vast majority of the improvement in the axion energy density bound. The main difference at low \(\ell\) in 2018 data, arising from an analysis of high (electromagnetic) frequency polarisation modes, is a stronger and slightly lower constraint on the reionisation optical depth \(\tau\) thanks to measurement of the large-scale reionisation bump in the \(EE\) power spectrum. This \(\tau\) measurement breaks degeneracies with the primordial power spectrum amplitude \(A_\mathrm{s}\) and the axion energy density \(\Omega_\mathrm{a} h^2\). We show results for \(m_\mathrm{a} = 10^{-30}\,\mathrm{eV}\). These are indicative for all \(m_\mathrm{a} \leq 10^{-28}\,\mathrm{eV}\), since the effect of the lightest DE-like axions in CMB data is restricted to the largest scales (through the integrated Sachs-Wolfe effect) that are degenerate with the primordial amplitude \citep{2015PhRvD..91j3512H}. Hence, the improved \(\tau\) measurement does not improve axion constraints for heavier DM-like axions (\(m_\mathrm{a} \geq 10^{-27}\,\mathrm{eV}\)), whose effect is restricted to smaller scales.

\begin{figure}[tbp]
\centering 
\includegraphics[width=\textwidth]{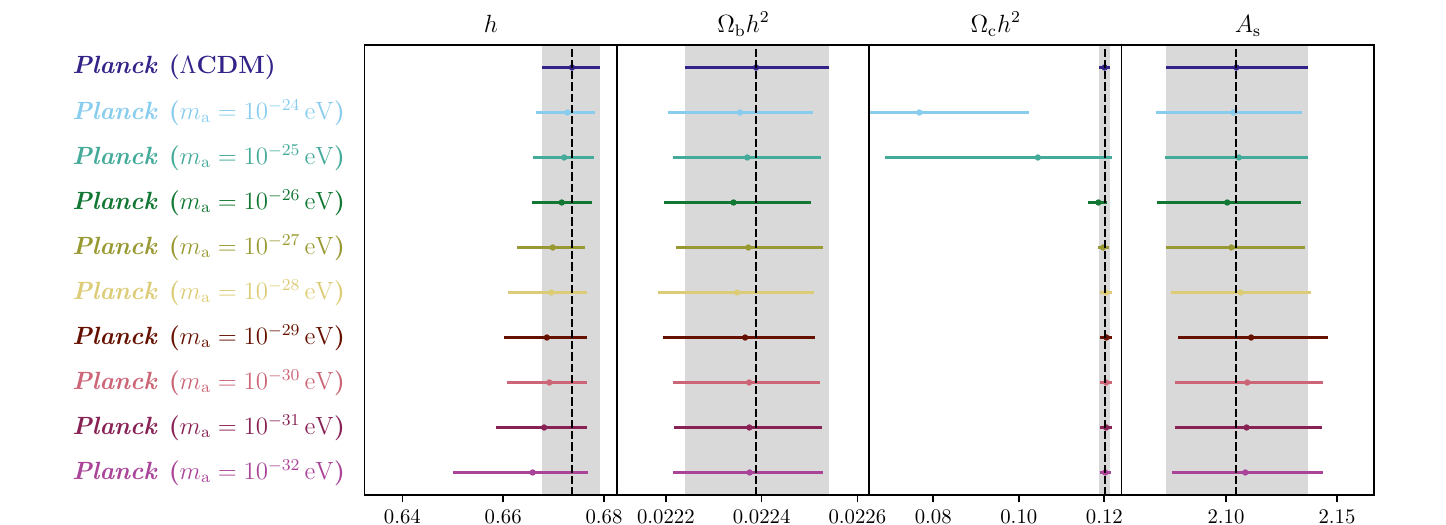}
\includegraphics[width=\textwidth]{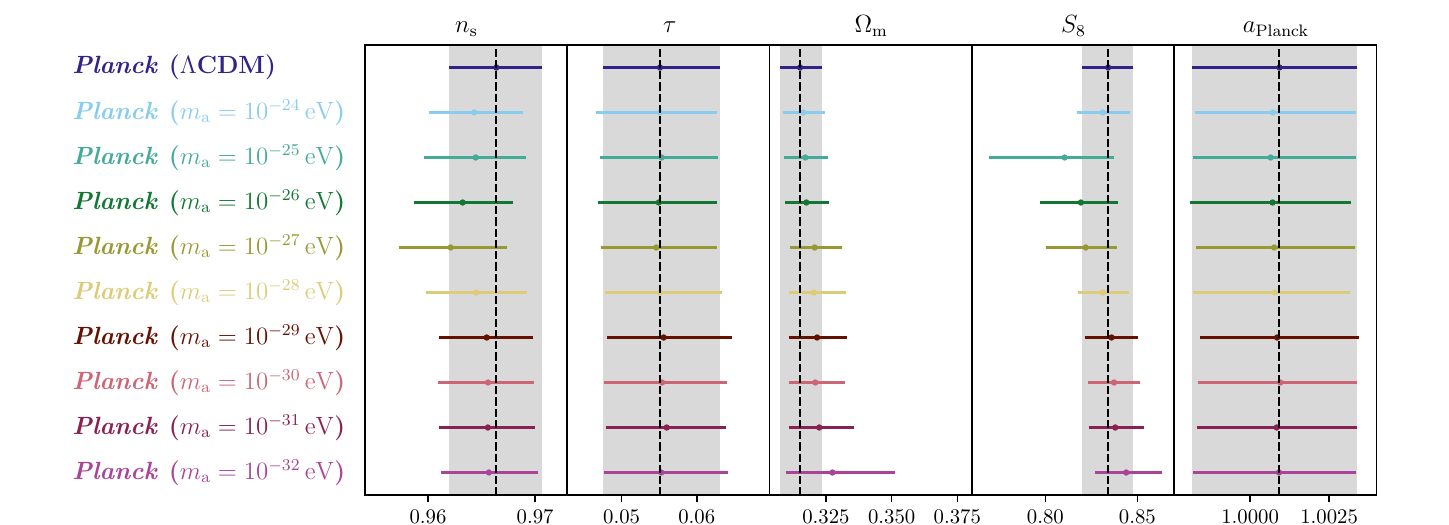}
\caption{\label{fig:planck_all}The effect of axion mass \(m_\mathrm{a}\) on cosmological parameter constraints from \textit{Planck} CMB data. We see how dark energy-like axions (\(m_\mathrm{a} < 10^{-27}\,\mathrm{eV}\)) have degeneracy with lower values of the Hubble parameter \(h\), while dark matter (DM)-like axions (\(m_\mathrm{a} \geq 10^{-27}\,\mathrm{eV}\)) have degeneracy with the cold DM density \(\Omega_\mathrm{c} h^2\) and lower values of the matter clumping factor \(S_8\). These degeneracies are explored further in Fig.~\ref{fig:mass}. Each point indicates the maximum marginalised posterior, while the errorbar indicates the marginalised 68\% c.l. \(A_\mathrm{s}\) is in units of \(10^{-9}\).}
\end{figure}

\begin{figure}[tbp]
\centering 
\includegraphics[width=\textwidth]{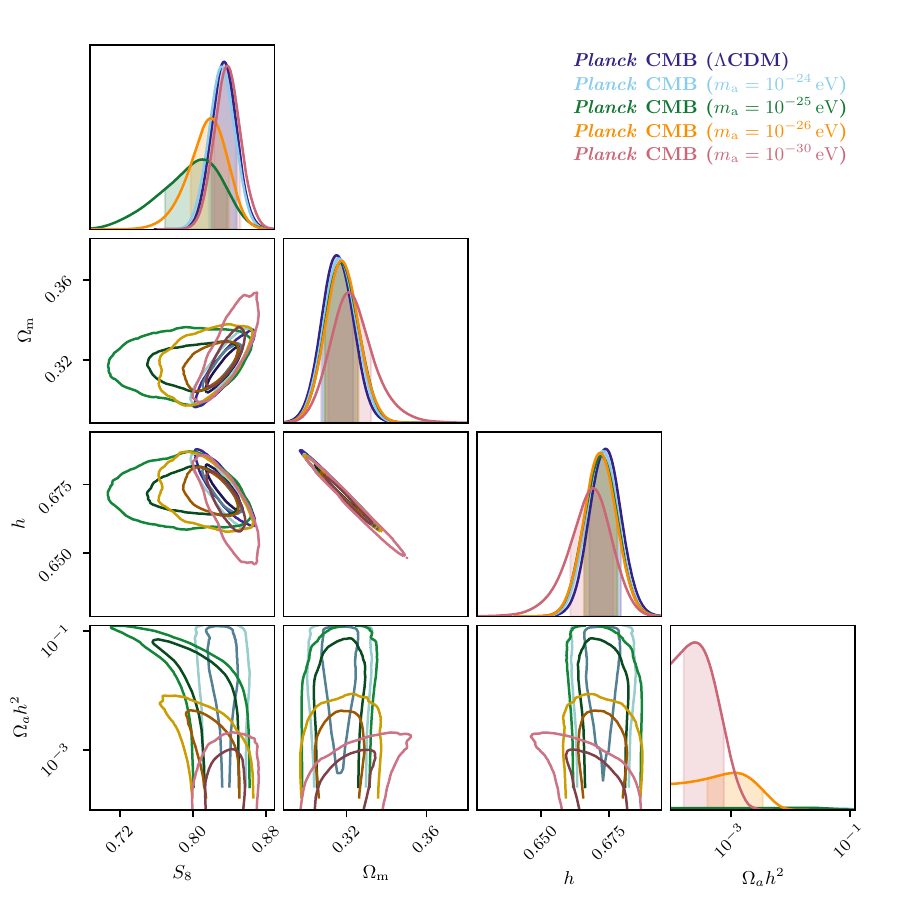} 
\caption{\label{fig:mass}The effect of ultra-light axions on the matter clumping factor \(S_8\), matter energy density \(\Omega_\mathrm{m}\) and Hubble parameter \(h\), inferred from \textit{Planck}, as a function of axion mass \(m_\mathrm{a}\). Dark matter (DM)-like axions for \(m_\mathrm{a} \in [10^{-26}, 10^{-25}]\,\mathrm{eV}\) give lower \(S_8\) values by a scale-dependent power spectrum suppression, while dark energy-like axions (\eg \(m_\mathrm{a} = 10^{-30}\,\mathrm{eV}\)) give lower \(h\) values by causing accelerated expansion after matter-radiation equality. DM-like axions with \(m_\mathrm{a} = 10^{-24}\,\mathrm{eV}\) have a negligible effect on \(S_8\) as the power spectrum suppression is on scales smaller than those to which \(S_8\) is sensitive. For each set, the inner and outer contours respectively indicate the 68\% and 95\% credible regions of the 2D marginalised posterior distribution, with the 1D marginalised posteriors on the diagonal, where 68\% credible regions are shaded.}
\end{figure}

Figure \ref{fig:planck_all} shows the \textit{Planck} constraints on other cosmological parameters. DE-like axions \((m_\mathrm{a} < 10^{-27}\,\mathrm{eV})\) are consistent with lower values of \(h\) as they drive accelerated expansion after matter-radiation equality \citep{2015PhRvD..91j3512H}\footnote{We are considering different axion models than those that are typically invoked to \textit{increase} \(h\) (and so address the Hubble parameter tension). These so-called ``early dark energy'' axions are contrived to induce a burst of accelerated expansion \textit{before} recombination and typically require non-trivial axion potentials \cite[see \eg][]{2019PhRvL.122v1301P}.}. As DE-like axions have all started oscillating (and so behave like DM) by today, they count towards the total matter energy density \(\Omega_\mathrm{m}\), but are not degenerate with cold DM in the CMB. Thus, for \(m_\mathrm{a} \leq 10^{-27}\,\mathrm{eV}\), larger values of \(\Omega_\mathrm{m}\) are allowed. This also drives compatibility with larger values of the matter clumping factor since \(S_8 \propto \sqrt{\Omega_\mathrm{m}}\). Conversely, DM-like axions \((m_\mathrm{a} > 10^{-27}\,\mathrm{eV})\) are degenerate with cold DM (CDM) in the CMB and so, in the high-mass limit where axions are poorly constrained, the CDM density is also poorly constrained. Further, DM-like axions suppress the matter power spectrum on scales below their de Broglie wavelength. Thus, when DM-like axions can comprise a significant fraction (\(\gtrsim 2\%\)) of the total DM budget (\(10^{-27}\,\mathrm{eV} \leq m_\mathrm{a} \leq 10^{-25}\,\mathrm{eV}\)), \textit{Planck} data are compatible with lower values of \(S_8\) than in the \(\Lambda\)CDM model, since \(S_8\) integrates over lower-amplitude modes. For \(m_\mathrm{a} > 10^{-25}\,\mathrm{eV}\), the power spectrum suppression is on scales smaller than those to which the \(S_8\) parameter is most sensitive and so \(\Lambda\)CDM values of \(S_8\) are returned. This suggests that axions with \(m_\mathrm{a} \in [10^{-27}, 10^{-25}]\,\mathrm{eV}\) could help to resolve the so-called \(S_8\) tension by bringing CMB data into compatibility with the lower \(S_8\) values inferred from large-scale structure data. Fig.~\ref{fig:mass} explicitly illustrates that it is degeneracy with the axion energy density that allows lower values of \(h\) and higher values of \(\Omega_\mathrm{m}\) for DE-like axions (\(m_\mathrm{a} \sim 10^{-30}\,\mathrm{eV}\)), and lower values of \(S_8\) for DM-like axions (\(m_\mathrm{a} \sim [10^{-26} - 10^{-25}]\,\mathrm{eV}\)).

\begin{table}
    \centering
    \begin{tabular}{lcc}
        \hline
		Data & \(m_\mathrm{a}\) & Bayes factor relative to $\Lambda$CDM \\
		\hline
		\textit{Planck} & \multirow{4}{*}{$10^{-25}\,\mathrm{eV}$} & -1.8 \\
		  \textit{Planck} + ACT-DR4 & & -0.6 \\
		\textit{Planck} + SPT-3G & & -1.8 \\
		All CMB + BAO + SNe & & -0.4 \\
        \hline
        & $10^{-24}\,\mathrm{eV}$ & -1.5 \\
		& $10^{-25}\,\mathrm{eV}$ & -2.6 \\
		& $10^{-26}\,\mathrm{eV}$ & -1.6 \\
		& $10^{-27}\,\mathrm{eV}$ & -4.0 \\
		\textit{Planck} + BOSS & $10^{-28}\,\mathrm{eV}$ & -3.1 \\
		& $10^{-29}\,\mathrm{eV}$ & -3.1 \\
		& $10^{-30}\,\mathrm{eV}$ & -3.2 \\
		& $10^{-31}\,\mathrm{eV}$ & -2.5 \\
		& $10^{-32}\,\mathrm{eV}$ & -2.6 \\
		\hline
    \end{tabular}
    \caption{\label{tab:evidence}Bayes factor (log-ratio of model evidences; \textit{right column}) for the indicated data (\textit{left column}) given the indicated axion model (\textit{middle column}) relative to the \(\Lambda\)CDM model. For all the data combinations shown, the Bayesian evidence favours the \(\Lambda\)CDM model; although we find that axions can improve consistency between datasets (see \S~\ref{sec:boss_results}), there is no preference for an extension beyond \(\Lambda\)CDM given these data.}
\end{table}
Although axions with \(m_\mathrm{a} = 10^{-25}\,\mathrm{eV}\) can comprise the dark matter according to \textit{Planck} data, there is no preference for such a model compared to \(\Lambda\)CDM according to the Bayesian evidence. The log-ratio of model evidences (or Bayes factor) given \textit{Planck} data is 1.8 in favour of \(\Lambda\)CDM (see Table \ref{tab:evidence}). This amounts to ``positive'' evidence in favour of \(\Lambda\)CDM according to the Jeffreys scale as given by Ref.~\cite{doi:10.1080/01621459.1995.10476572}. This lack of preference for extended cosmological models 
is consistent with previous studies \citep[\eg][]{2017PhRvL.119j1301H}, and there is no improvement in the maximum likelihood (or minimum chi-squared).

\subsubsection{All CMB, BAO \& supernovae}
\label{sec:cmb_results_all_cmb}

\begin{figure}[tbp]
\centering 
\includegraphics[width=\textwidth]{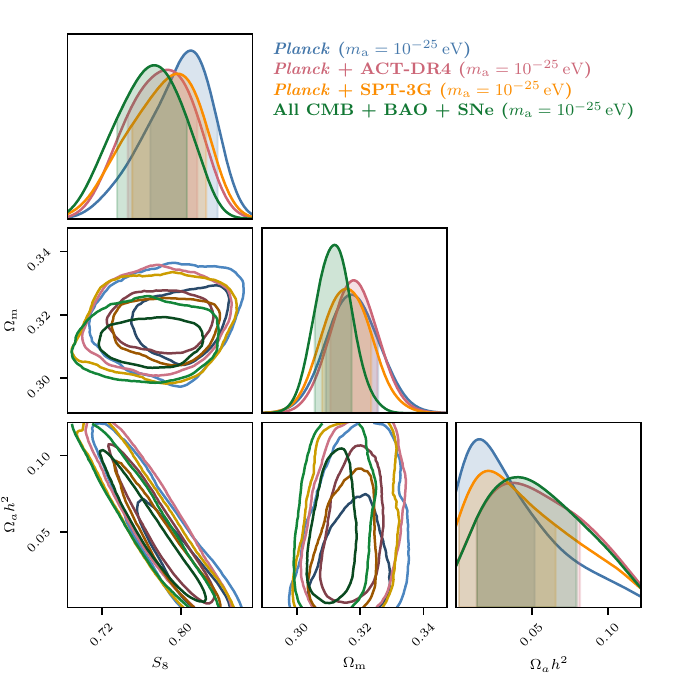} 
\caption{\label{fig:cmb}The effect of current higher-resolution CMB data (\textit{Planck} and ACT-DR4 in \textit{red}; \textit{Planck} and SPT-3G in \textit{orange}), galaxy baryon acoustic oscillations (BAO) and supernovae (SNe) (all combined with \textit{Planck} in \textit{green}; \textit{Planck} only in blue) on axion constraints for \(m_\mathrm{a} = 10^{-25}\,\mathrm{eV}\). For each set, the inner and outer contours respectively indicate the 68\% and 95\% credible regions of the 2D marginalised posterior distribution, with the 1D marginalised posteriors on the diagonal, where 68\% credible regions are shaded. \textit{From left to right}, \(S_8\) is the matter clumping factor, \(\Omega_\mathrm{m}\) is the matter energy density and \(\Omega_\mathrm{a} h^2\) is the physical axion energy density.}
\end{figure}

\begin{table}
    \centering
    \begin{tabular}{lccc}
        \hline
		Data & $S_8$ ($\Lambda$CDM) & $\Omega_a h^2$ (\(m_\mathrm{a} = 10^{-25}\,\mathrm{eV}\)) & $S_8$ (\(m_\mathrm{a} = 10^{-25}\,\mathrm{eV}\)) \\
		\hline
		\textit{Planck} & $0.834^{+0.014}_{-0.013}$ & < 0.09667 & $0.811^{+0.025}_{-0.039}$ \\
		\textit{Planck} + ACT-DR4 & $0.835^{+0.013}_{-0.012}$ & < 0.10745 & $0.789^{+0.027}_{-0.041}$ \\
		\textit{Planck} + SPT-3G & $0.828^{+0.014}_{-0.011}$ & < 0.10580 & $0.799^{+0.027}_{-0.046}$ \\
		All CMB + BAO + SNe & $0.827\pm 0.010$ & < 0.10610 & $0.774^{+0.032}_{-0.037}$ \\
		\hline
    \end{tabular}
    \caption{\label{tab:cmb_bao}Constraints on axion energy density \(\Omega_\mathrm{a} h^2\) and the matter clumping factor \(S_8\), for different CMB, galaxy BAO and supernovae data combinations (see \S~\ref{sec:cmb} for a description of the data). For \(\Omega_\mathrm{a} h^2\), we give the 95\% upper c.l.; for \(S_8\), we give the maximum marginalised posterior with the asymmetric 68\% c.l.}
\end{table}

\begin{figure}[tbp]
\centering 
\includegraphics[width=\textwidth]{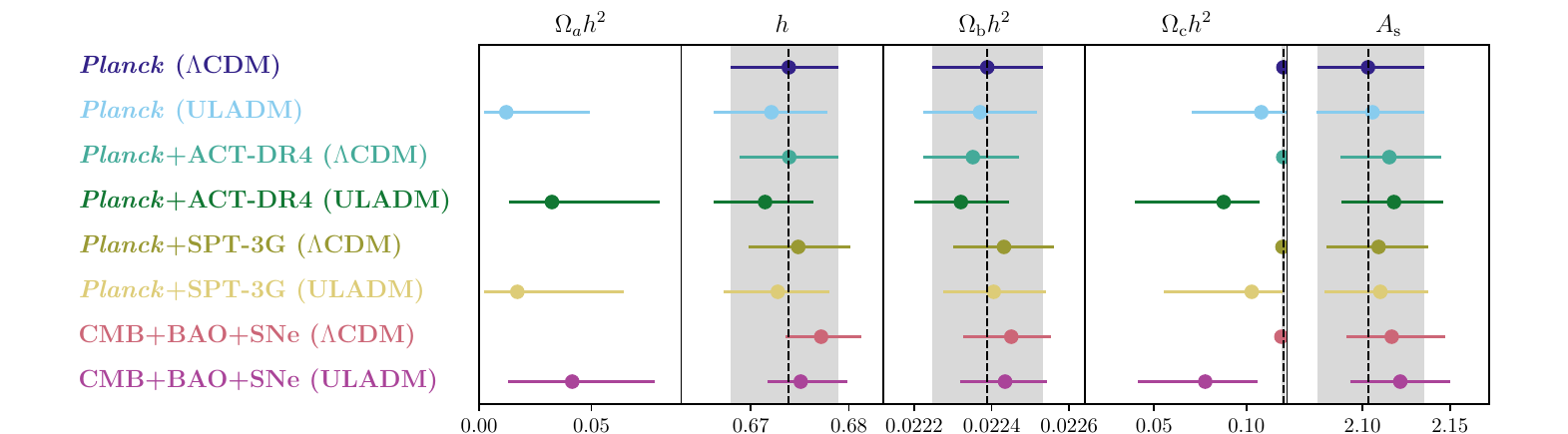}
\includegraphics[width=\textwidth]{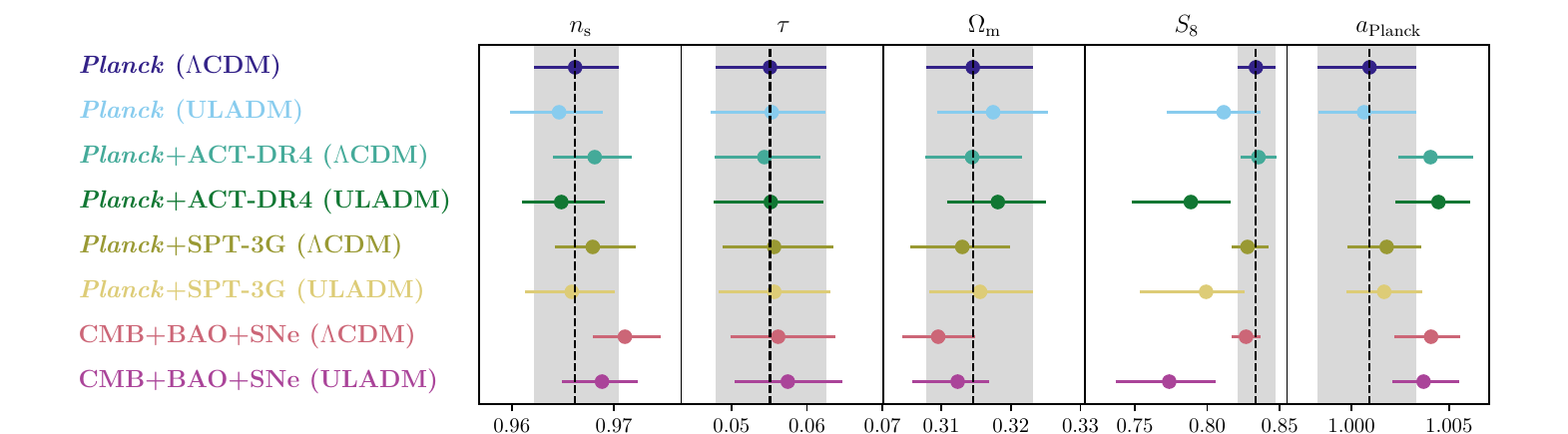}
\caption{\label{fig:cmb_all}The effect of current higher-resolution CMB data (ACT-DR4, SPT-3G), galaxy BAO and supernovae (SNe) on ultra-light axion and cosmological constraints for \(m_\mathrm{a} = 10^{-25}\,\mathrm{eV}\) (ULA DM), also comparing to \(\Lambda\)CDM and \textit{Planck}-only constraints. Each point indicates the marginalised mean, while the errorbar indicates the marginalised 68\% c.l. \(A_\mathrm{s}\) is in units of \(10^{-9}\).}
\end{figure}

For the first time in a ULA search, we consider the addition of higher-resolution CMB data from the ACT and SPT experiments. We defer a systematic search of the axion mass parameter space to future work, in anticipation of upcoming high-resolution lensing data. In this study, we focus on the impact of current ACT (see \S~\ref{sec:act}) and SPT (see \S~\ref{sec:spt}) data (and a compendium of low-\(z\) galaxy BAO and supernovae; see \S~\ref{sec:bao}) on DM-like axions that most significantly increase compatibility with low values of \(S_8\): \(m_\mathrm{a} = 10^{-25}\,\mathrm{eV}\). Fig.~\ref{fig:cmb} illustrates the effect on the \(S_8\) - \(\Omega_\mathrm{m}\) - \(\Omega_\mathrm{a} h^2\) planes from adding these data to the \textit{Planck} data considered above. The posterior shifts with respect to \textit{Planck} alone are small. There is a \(\sim 0.5 \sigma\) decrease in \(\Omega_\mathrm{m}\) when adding BAO and SNe data (\(\sim 1 \sigma\) decrease seen in the \(\Lambda\)CDM case; see Fig.~\ref{fig:cmb_all}). In particular, the axion energy density bounds at \(m_\mathrm{a} = 10^{-25}\,\mathrm{eV}\) are slightly weakened with the addition of these data (see also Table \ref{tab:cmb_bao}). Correspondingly, there is a shift to even lower values of \(S_8\) driven by its parameter degeneracy with \(\Omega_\mathrm{a} h^2\). This weakening of constraints is consistent with previous searches for massive neutrinos in high-resolution CMB data \citep[\eg][]{2020JCAP...12..047A}. Similarly to neutrinos, DM-like axions are constrained in primary CMB anisotropy power spectra through the lensing-induced smoothing of acoustic peaks. Here, gravitational lensing by lower-redshift (mostly \(z < 2\)) large-scale structure dampens the amplitude of peaks in angular power spectra. It follows that the amount of lensing-induced smoothing is sensitive to the presence of ultra-light axions or neutrinos which suppress the growth of structure and thus reduce the amount of smoothing. The amount of lensing relative to the best-fit \(\Lambda\)CDM expectation is quantified by the multiplicative correction to the theoretical expectation \(A_\mathrm{L}\). In particular, both ACT-DR4 (\(A_\mathrm{L} = 1.01 \pm 0.11\)) \citep{2020JCAP...12..047A} and SPT-3G (\(A_\mathrm{L} = 0.98 \pm 0.12\)) \citep{2021PhRvD.104b2003D} prefer lower values of \(A_\mathrm{L}\) compared to \textit{Planck} (\(A_\mathrm{L} = 1.180 \pm 0.065\)) \citep{2020A&A...641A...6P}. This means that when adding ACT or SPT data to \textit{Planck}, constraints on models that suppress structure and lower the lensing signal are weakened, \eg massive neutrinos \citep{2020JCAP...12..047A} or ultra-light axions. Fig.~\ref{fig:cmb_all} shows marginalised constraints on all other cosmological parameters, also comparing to the \(\Lambda\)CDM case. We see the typical degeneracy for DM-like axions with standard cold DM meaning that weakened constraints on \(\Omega_\mathrm{a} h^2\) lead to correspondingly-weakened constraints on \(\Omega_\mathrm{c} h^2\). We note a \(\sim 1 \sigma\) increase in the \textit{Planck} calibration parameter \(a_\mathrm{Planck}\) when adding ACT data which is seen in both \(\Lambda\)CDM and axion models. Similarly as for \textit{Planck} data, there is no preference given these combined datasets for an axion model compared to \(\Lambda\)CDM according to the Bayesian evidence (see Table \ref{tab:evidence}). The Bayes factors amount to evidence in favour of \(\Lambda\)CDM that ranges from ``positive'' to ``not worth more than a bare mention'' according to the Jeffreys scale \citep{doi:10.1080/01621459.1995.10476572}.

\begin{figure}[tbp]
\centering 
\includegraphics[width=\textwidth]{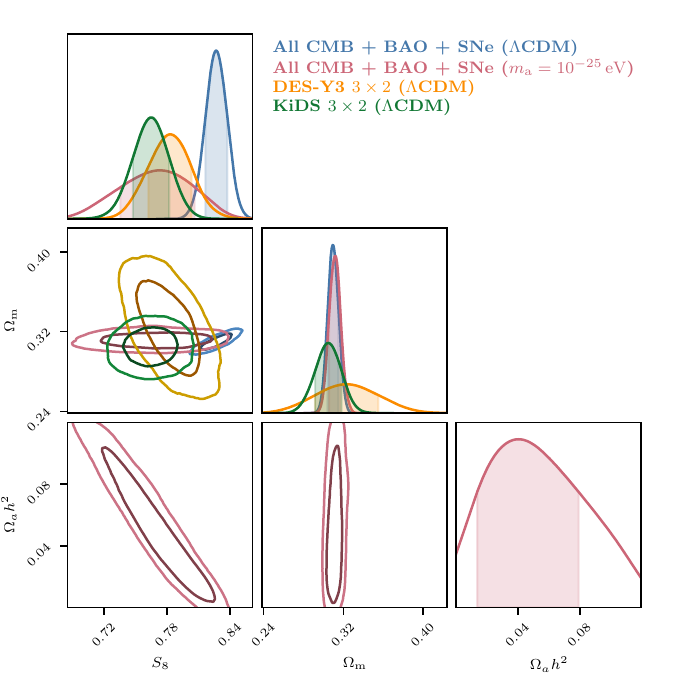} 
\caption{\label{fig:s8}Comparison of CMB (\textit{Planck}, ACT-DR4, SPT-3G), galaxy BAO and supernovae (SNe) constraints with fiducial galaxy weak lensing and clustering (\(3 \times 2\)) \(\Lambda\)CDM constraints from the Dark Energy Survey (DES) and the Kilo-Degree Survey (KiDS) (all with fixed neutrino mass). In \(\Lambda\)CDM, CMB, BAO and SNe data prefer systematically higher values of the matter clumping factor \(S_8\) than is inferred from fiducial \(3 \times 2\) analyses. When axions of \(m_\mathrm{a} = 10^{-25}\,\mathrm{eV}\) contribute to the energy budget with energy density \(\Omega_\mathrm{a} h^2\), CMB, BAO and SNe data are consistent with lower values of \(S_8\). In order to assess consistency between data in an axion model, it is necessary to re-analyse the \(3 \times 2\) data in the presence of axions; in \S~\ref{sec:boss_results}, we consider the first part with galaxy clustering from BOSS. For each set, the inner and outer contours respectively indicate the 68\% and 95\% credible regions of the 2D marginalised posterior distribution, with the 1D marginalised posteriors on the diagonal, where 68\% credible regions are shaded. \textit{From left to right}, \(S_8\) is the matter clumping factor, \(\Omega_\mathrm{m}\) is the matter energy density and \(\Omega_\mathrm{a} h^2\) is the physical axion energy density.}
\end{figure}

We demonstrate above that, in an axion model with \(m_\mathrm{a} = 10^{-25}\,\mathrm{eV}\), the combination of \textit{Planck}, ACT-DR4 and SPT-3G CMB, galaxy BAO and supernovae data are compatible with lower values of the matter clumping factor (\(S_8 = 0.774^{+0.032}_{-0.037}\)) than in \(\Lambda\)CDM (\(S_8 = 0.827 \pm 0.010\)). Fig.~\ref{fig:s8} compares this result to fiducial \(\Lambda\)CDM constraints from combined galaxy weak lensing and clustering (\(3 \times 2\)) data. We consider \(\Lambda\)CDM constraints (with fixed neutrino energy density) from the combination of galaxy clustering, galaxy lensing shear and galaxy -- galaxy lensing two-point correlation functions (\(3 \times 2\)) as measured by the Dark Energy Survey (DES) \citep{2022PhRvD.105b3520A}\footnote{This is the publicly-released posterior chain \texttt{chain\_3x2pt\_fixednu\_lcdm}.}. We also consider \(\Lambda\)CDM constraints (with fixed neutrino energy density) from the same combination of three \(\times\) two-point correlation functions as measured by the Kilo-Degree Survey (KiDS) \citep{2021A&A...646A.140H}, which includes redshift-space galaxy clustering data from BOSS \citep{2017MNRAS.464.1640S} and galaxy -- galaxy lensing data from the survey overlap between KiDS, BOSS and the spectroscopic 2-degree Field Lensing Survey (2dFLenS) \citep{10.1093/mnras/stw1990}\footnote{This is the publicly-released posterior chain \texttt{samples\_multinest\_blindC\_EE\_nE\_w}.}. We note that the KiDS \(3 \times 2\) constraints are therefore not entirely independent of the CMB + BAO + SNe compendium we consider, since part of the BAO measurements we use is derived from the same BOSS data (see \S~\ref{sec:bao}) as goes into the KiDS \(3 \times 2\) measurement. However, we note that the addition of BAO and SNe data makes only a small difference to the \(S_8\) constraint from \textit{Planck} + ACT and \textit{Planck} + SPT (see \eg Fig.~\ref{fig:cmb}). We anticipate that, in \S~\ref{sec:boss_results}, we will consider the full-shape galaxy clustering power spectrum from BOSS, which will be much more significantly correlated with the KiDS \(3 \times 2\) analysis. Despite this proviso, Fig.~\ref{fig:s8} illustrates how, in the \(\Lambda\)CDM model, \(3 \times 2\) analyses from both DES (\(S_8 = 0.783 \pm 0.020\)) and KiDS (\(S_8 = 0.765 \pm 0.017\)) prefer systematically lower values of \(S_8\) than the compendium of CMB, BAO and SNe data (\(S_8 = 0.827 \pm 0.010\)). This is a manifestation of the so-called ``\(S_8\) tension'', where many galaxy clustering, weak lensing and galaxy cluster observations prefer lower values of \(S_8\) than is inferred from CMB observations, with statistical significance ranging from 2 to 3 \(\sigma\) depending on the data comparison \citep[see \eg][for a recent review]{2022JHEAp..34...49A}. However, when axions of \(m_\mathrm{a} = 10^{-25}\,\mathrm{eV}\) contribute to the energy budget, the CMB, BAO and SNe compendium is compatible with the low \(S_8\) values preferred by DES and KiDS in the \(\Lambda\)CDM model. We therefore hypothesise that axions could resolve the \(S_8\) tension. In order to assess this, we must reanalyse the \(3 \times 2\) data in the axion model. In this work, we consider the first part of this in analysing full-shape galaxy clustering information from BOSS. We present these results in \S~\ref{sec:boss_results}.

\subsection{Baryon Oscillation Spectroscopic Survey galaxy power spectrum \& bispectrum}
\label{sec:boss_results}
We now consider the effect on ultra-light axion constraints from the galaxy power spectrum and bispectrum as measured from the Baryon Oscillation Spectroscopic Survey (BOSS; see \S~\ref{sec:boss} for a description of the data).

\subsubsection{\(\Lambda\)CDM}
\label{sec:boss_results_lcdm}

\begin{figure}[tbp]
\centering 
\includegraphics[width=\textwidth]{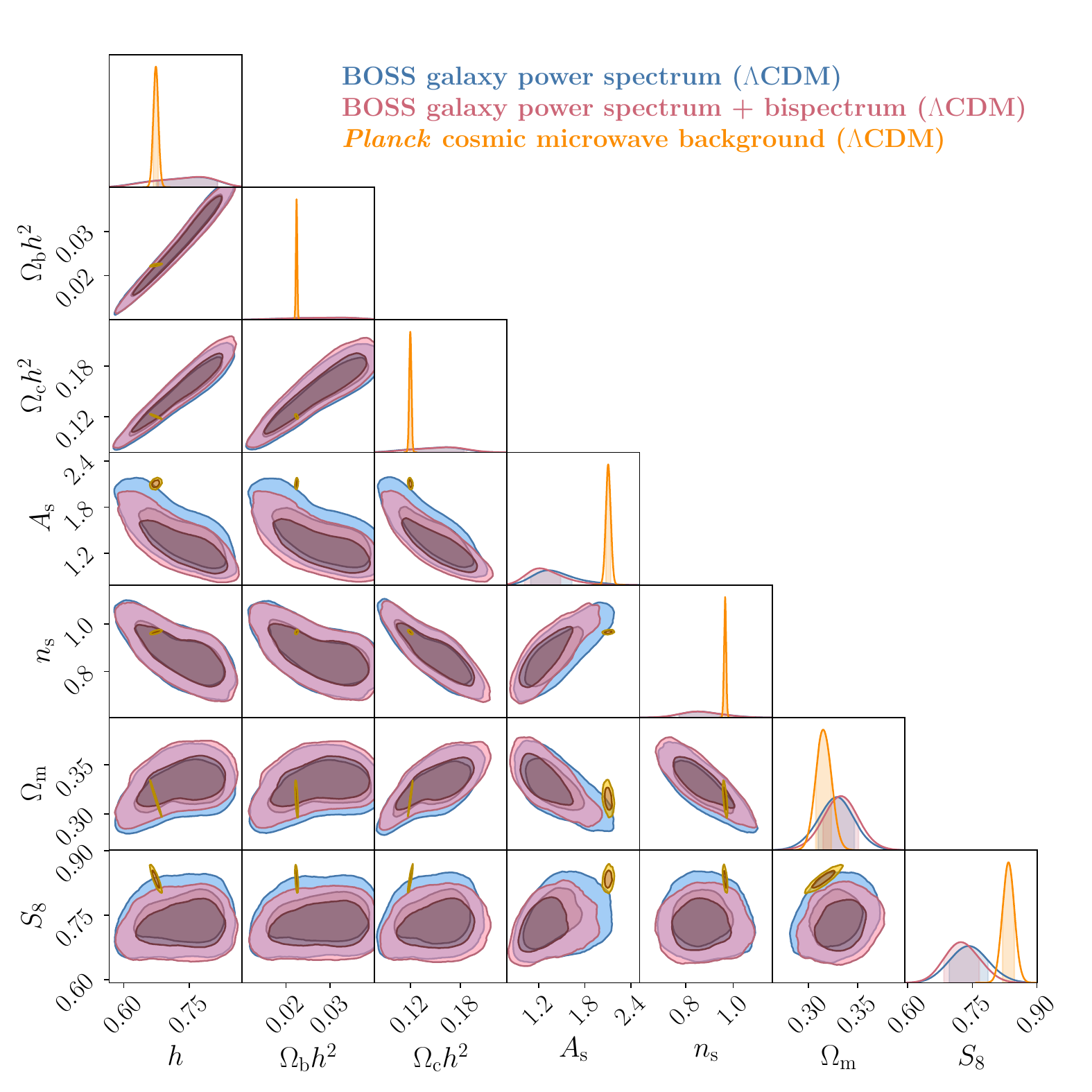} 
\caption{\label{fig:boss_lcdm}Comparison of BOSS galaxy clustering and \textit{Planck} CMB constraints on \(\Lambda\)CDM cosmological parameters. BOSS data alone (in particular without an \(\Omega_\mathrm{b} h^2\) prior) are much less constraining than \textit{Planck} data on the standard cosmological model. For each set, the darker and lighter shaded contours respectively indicate the 68\% and 95\% credible regions of the 2D marginalised posterior distribution, with the 1D marginalised posteriors on the diagonal, where 68\% credible regions are shaded. \(A_\mathrm{s}\) is in units of \(10^{-9}\).}
\end{figure}

Before studying the combination of \textit{Planck} CMB and BOSS galaxy clustering data, we assess constraints independently from each dataset. Fig.~\ref{fig:boss_lcdm} shows \(\Lambda\)CDM cosmological constraints from the BOSS galaxy power spectrum only (\(P_0, P_2, P_4, Q_0\) and the post-reconstructed BAO  Alcock-Paczynski parameters), the BOSS galaxy power spectrum and bispectrum monopole (additionally \(B_0\)), and \textit{Planck} CMB data (previously shown in \S~\ref{sec:cmb_results}). In particular, we consider BOSS constraints without a prior on the baryon energy density \(\Omega_\mathrm{b} h^2\) or any other cosmological parameters. It is striking how much more constraining is \textit{Planck} data on the full cosmological model than BOSS data alone. However, as is typical of large-scale structure experiments, BOSS provides more competitive constraints when projected onto the plane of derived parameters \(\Omega_\mathrm{m}\) and \(S_8\).

\subsubsection{Parameter tension metrics}
\label{sec:boss_results_metrics}

In order to assess consistency between datasets in their cosmological constraints, we consider three metrics of parameter tension (the difference in \(S_8\) only, the difference in the \(S_8\) - \(\Omega_\mathrm{m}\) plane, and the difference in the full posterior distribution). We now describe these metrics in more detail. The first metric, which is most widely quoted in the literature, is the discrepancy in the marginalised \(S_8\) constraint from two datasets (labelled 1 and 2), defined as
\begin{equation}
\label{eq:tension_s8}
\frac{\Delta S_8}{\sigma_{S_8}} = \frac{\mu_1 - \mu_2}{\sqrt{\sigma^2_1 + \sigma^2_2}}.
\end{equation}
Here, \(\mu_i\) and \(\sigma_i\) are respectively the parameter posterior mean and standard deviation given experiment \(i\). This metric is given in the third column of Table \ref{tab:tension}. We also consider a second metric, which is an extension of Eq.~\eqref{eq:tension_s8} to higher dimensions:
\begin{equation}
\label{eq:tension_s8_om}
\chi^2 = (\vec{\mu}_1 - \vec{\mu}_2)^\mathrm{T} (C_1 + C_2)^{-1} (\vec{\mu}_1 - \vec{\mu}_2).
\end{equation}
Here, \(\vec{\mu}_i\) is now the vector of parameter posterior means and \(C_i\) is the posterior covariance, both given experiment \(i\). We then calculate the probability \(p\) to exceed \(\chi^2\) (for a \(\chi^2\) distribution with degrees of freedom equal to the number of parameters) and convert this to a number \(N\) of \(\sigma\) using the standard Gaussian interpretation\footnote{\(N = \sqrt{2}\,\mathrm{erf}^{-1}(p)\).}.

Both Eqs.~\eqref{eq:tension_s8} and \eqref{eq:tension_s8_om} are good measures of parameter discrepancy in the limit of Gaussian posterior distributions. We therefore give, in the fourth column of Table \ref{tab:tension}, the metric defined in Eq.~\eqref{eq:tension_s8_om} evaluated for the marginalised posterior in the \(S_8 - \Omega_\mathrm{m}\) plane. In this plane, the BOSS data are most constraining and the distribution is reasonably Gaussian. It is important nonetheless also to consider consistency in the full set of parameters constrained by both datasets. However, the full BOSS posterior distribution appears highly non-Gaussian and so the metrics defined above will not be a good measure of consistency in the full parameter space. We therefore elect to calculate the full posterior distribution of the parameter difference \(\Delta \vec{\theta}\) (marginalised over the parameters \(\vec{\theta}\)) \citep{2021PhRvD.104d3504R}:
\begin{equation}
\label{eq:tension_full}
\mathcal{P}(\Delta \vec{\theta}) = \int \mathrm{d}\vec{\theta} \mathcal{P}_1(\vec{\theta}) \mathcal{P}_2(\vec{\theta} - \Delta \vec{\theta}).
\end{equation}
Here, \(\mathcal{P}_i (\vec{\theta})\) is the posterior distribution given experiment \(i\). We can then calculate the significance of the inferred parameter shift (relative to none) by integrating \(\mathcal{P}(\Delta \vec{\theta})\) above the iso-probability contour that goes through \(\Delta \vec{\theta} = 0\) (this probability to exceed can be converted to a number of \(\sigma\) as above)\footnote{We numerically evaluate this integral using the \texttt{tensiometer} package: \url{https://github.com/mraveri/tensiometer}.}. In this way, this third tension metric accounts for non-Gaussianities in the parameter posterior distribution. We therefore give, in the final column of Table \ref{tab:tension}, the metric derived from Eq.~\eqref{eq:tension_full} as evaluated in the volume of all parameters constrained by both \textit{Planck} and BOSS [\(h, \Omega_\mathrm{b} h^2, \Omega_\mathrm{c} h^2, A_\mathrm{s}, n_\mathrm{s}, \Omega_\mathrm{m}, S_8\) and \(\Omega_\mathrm{a} h^2\) when part of the model].

There are many proposed approaches to evaluating parameter consistency in high-dimensional and non-Gaussian distributions. Although these different approaches tend to agree in terms of trend (\ie they typically agree with respect to an increasing or decreasing tension) \citep[see \eg][]{2021MNRAS.505.6179L}, they typically disagree as to the particular value of tension. We therefore urge caution when interpreting Table \ref{tab:tension} that it is most useful as a measure of relative tension given different models. All three metrics considered in Table \ref{tab:tension} (and Fig.~\ref{fig:boss_lcdm}) illustrate that the addition of BOSS bispectrum data \(B_0\) increases the discrepancy with respect to \textit{Planck} mostly by preferring slightly lower values of the primordial power spectrum amplitude \(A_\mathrm{s}\). This in turn pushes \(S_8\) to slightly lower values.

\subsubsection{BOSS-only axion constraints}
\label{sec:boss_results_boss_only}

\begin{figure}[tbp]
\centering 
\includegraphics[width=\textwidth]{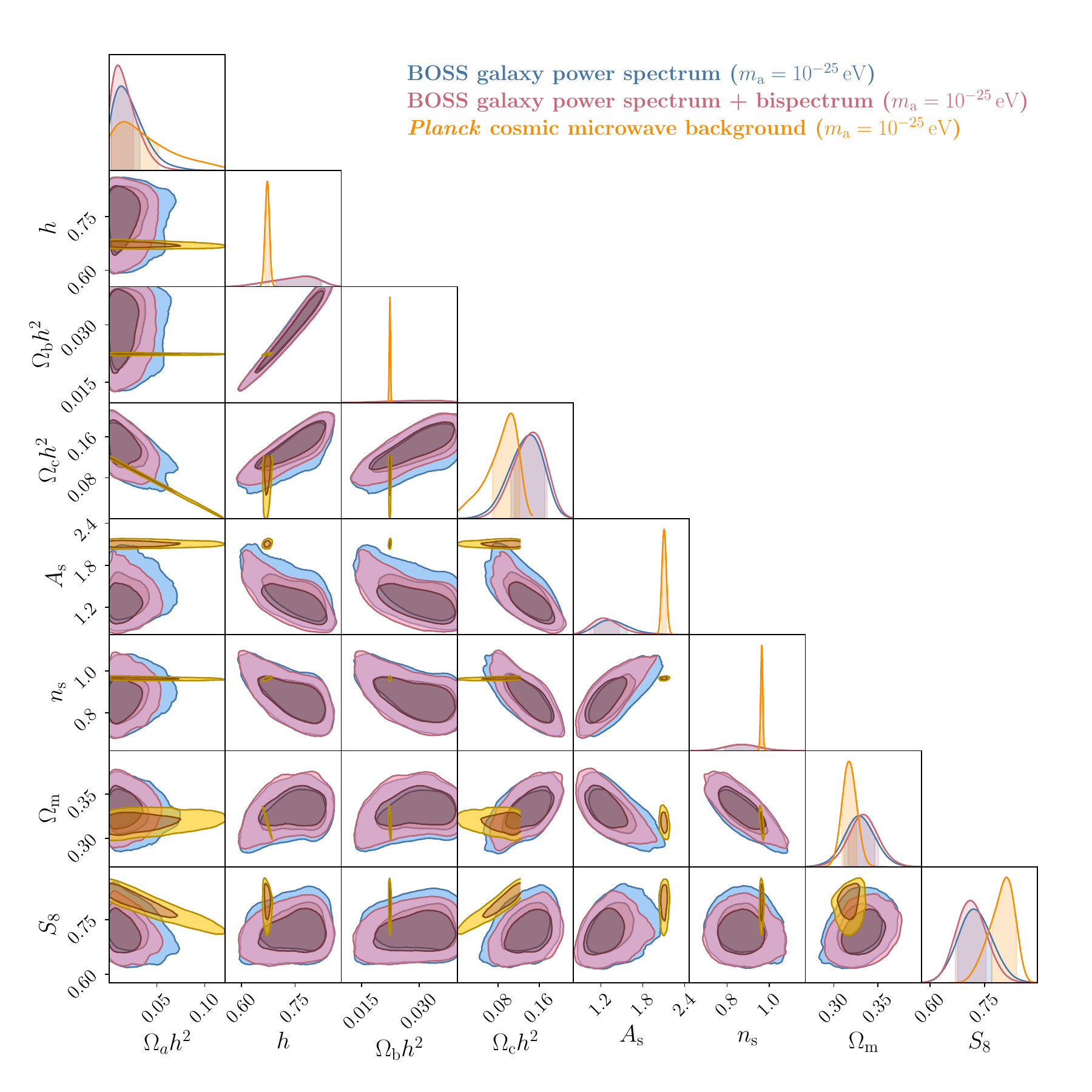} 
\caption{\label{fig:boss_axions}Comparison of BOSS galaxy clustering and \textit{Planck} CMB constraints on axion and cosmological parameters, for axion mass \(m_\mathrm{a} = 10^{-25}\,\mathrm{eV}\). BOSS data alone are more constraining than \textit{Planck} data on axion energy density \(\Omega_\mathrm{a} h^2\) since BOSS probes smaller scales (\(k < 0.4\,h\,\mathrm{Mpc}^{-1}\)); in the extended axion model, there is more posterior overlap in the \(S_8\) - \(\Omega_\mathrm{m}\) plane than in \(\Lambda\)CDM (see Fig.~\ref{fig:boss_lcdm}). For each set, the darker and lighter shaded contours respectively indicate the 68\% and 95\% credible regions of the 2D marginalised posterior distribution, with the 1D marginalised posteriors on the diagonal, where 68\% credible regions are shaded. \(A_\mathrm{s}\) is in units of \(10^{-9}\).}
\end{figure}

\begin{table}
    \centering
    \begin{tabular}{llccc}
        \hline
		Data & Model & \(S_8\) (\(\sigma\)) & $S_8 - \Omega_\mathrm{m}$ (\(\sigma\)) & All parameters (\(\sigma\)) \\
		\hline
		\textit{Planck}, BOSS [no \(B_0\)] & $\Lambda$CDM & 2.1 & 2.02 & 1.77 \\
		 & $m_\mathrm{a} = 10^{-25}\,\mathrm{eV}$ & 1.32 & 1.14 & 2.14 \\
		\hline
		\textit{Planck}, BOSS & $\Lambda$CDM & 2.70 & 2.82 & 4.36 \\
		 & $m_\mathrm{a} = 10^{-25}\,\mathrm{eV}$ & 1.63 & 1.57 & 3.70 \\
		 & $m_\mathrm{a} = 10^{-26}\,\mathrm{eV}$ & 3.63 & 3.81 & 5.38 \\
		 & $m_\mathrm{a} = 10^{-27}\,\mathrm{eV}$ & 2.28 & 2.11 & 3.63 \\
		 & $m_\mathrm{a} = 10^{-28}\,\mathrm{eV}$ & 1.78 & 1.76 & 3.31 \\
		 & $m_\mathrm{a} = 10^{-29}\,\mathrm{eV}$ & 1.74 & 2.44 & 3.19 \\
		 & $m_\mathrm{a} = 10^{-30}\,\mathrm{eV}$ & 2.22 & 2.82 & 4.11 \\
		 & $m_\mathrm{a} = 10^{-31}\,\mathrm{eV}$ & 2.24 & 2.73 & 2.95 \\
      & $m_\mathrm{a} = 10^{-32}\,\mathrm{eV}$ & 2.58 & 2.78 & 3.19 \\
		\hline
    \end{tabular}
    \caption{\label{tab:tension}Discrepancy in parameters (given in the top row) as inferred from the two datasets given in the first column, for the model given in the second column. The third column is the discrepancy in the marginalised \(S_8\) constraint, and the fourth column is the discrepancy in the marginalised \(S_8 - \Omega_\mathrm{m}\) plane, both with the reasonable approximation of a Gaussian posterior distribution. The final column is the discrepancy in the marginalised constraint on all cosmological (and axion) parameters, where we account for non-Gaussianity by calculating the full parameter difference posterior. The full details of the tension metrics that we use are given in \S~\ref{sec:boss_results}.}
\end{table}

\begin{table}
    \centering
    \begin{tabular}{lcc}
        \hline
		\(m_\mathrm{a}\) & \(\Omega_\mathrm{a} h^2\) (BOSS) & $S_8$ (BOSS) \\
		\hline
		$\Lambda$CDM & -- & $0.723^{+0.041}_{-0.037}$ \\
		$10^{-24}\,\mathrm{eV}$ & < 0.15539 & $0.718^{+0.038}_{-0.039}$ \\
		$10^{-25}\,\mathrm{eV}$ & < 0.04174 & $0.709^{+0.043}_{-0.037}$ \\
		$10^{-26}\,\mathrm{eV}$ & < 0.01717 & $0.653\pm 0.040$ \\
		$10^{-27}\,\mathrm{eV}$ & < 0.00542 & $0.719^{+0.040}_{-0.038}$ \\
		$10^{-28}\,\mathrm{eV}$ & < 0.00842 & $0.742^{+0.050}_{-0.040}$ \\
		$10^{-29}\,\mathrm{eV}$ & < 0.02259 & $0.759^{+0.044}_{-0.043}$ \\
		$10^{-30}\,\mathrm{eV}$ & < 0.02771 & $0.745^{+0.041}_{-0.040}$ \\
		$10^{-31}\,\mathrm{eV}$ & < 0.02706 & $0.744^{+0.040}_{-0.042}$ \\
		$10^{-32}\,\mathrm{eV}$ & < 0.03126 & $0.737^{+0.040}_{-0.038}$ \\
		\hline
    \end{tabular}
    \caption{\label{tab:boss}Constraints on axion energy density \(\Omega_\mathrm{a} h^2\) and the matter clumping factor \(S_8\), as a function of axion mass \(m_\mathrm{a}\) (\textit{top to bottom}), as inferred from BOSS galaxy clustering data. For \(\Omega_\mathrm{a} h^2\), we give the 95\% upper c.l.; for \(S_8\), we give the maximum marginalised posterior with the asymmetric 68\% c.l. For consistency with other masses, at \(m_\mathrm{a} = 10^{-26}\,\mathrm{eV}\), we give the upper limit on the axion density; nonetheless, \(\Omega_\mathrm{a} h^2 = 0\) is disfavoured at \(\sim 2.7 \sigma\), \ie the maximum marginalised posterior \(\Omega_\mathrm{a} h^2 = 0.0100^{+0.0048}_{-0.0037}\).}
\end{table}

Figure \ref{fig:boss_axions} shows the same set of posterior contours as in Fig.~\ref{fig:boss_lcdm} but for an axion model with \(m_\mathrm{a} = 10^{-25}\,\mathrm{eV}\). Although BOSS is less constraining than \textit{Planck} on \(\Lambda\)CDM parameters, it is significantly more constraining on the axion energy density. This is driven by the addition of smaller-scale data in the reconstructed real-space galaxy power spectrum \(Q_0\) (see Fig.~\ref{fig:boss_test} and discussion below). However, since \textit{Planck} alone is unconstraining on the axion energy density at this mass (see also \S~\ref{sec:cmb_results}), it is more consistent with the lower values of \(S_8\) that BOSS (and indeed other large-scale structure experiments) prefer. This means that there is more posterior overlap in the \(S_8\) - \(\Omega_\mathrm{m}\) plane and this is reflected in the improved tension metrics in Table \ref{tab:tension}. Notably, the tension in \(S_8\) when comparing \textit{Planck} to full BOSS data is reduced from \(2.70\,\sigma\) (\(\Lambda\)CDM) to \(1.63\,\sigma\) (for \(m_\mathrm{a} = 10^{-25}\,\mathrm{eV}\)). However, there is no degeneracy between \(\Omega_\mathrm{a} h^2\) and \(A_\mathrm{s}\) at \(m_\mathrm{a} = 10^{-25}\,\mathrm{eV}\). This is because \(\Omega_\mathrm{a} h^2\) largely affects the small-scale power spectrum, while \(A_\mathrm{s}\) is constrained by the overall normalisation at all wavenumbers (see Fig.~\ref{fig:power}). This means there is no improvement in the \(A_\mathrm{s}\) discrepancy between \textit{Planck} and BOSS even in the presence of axions at \(m_\mathrm{a} = 10^{-25}\,\mathrm{eV}\) and this is reflected in the full parameter space tension metric given in Table \ref{tab:tension}. Indeed, when not including bispectrum data, the full parameter tension increases slightly.

\begin{figure}[tbp]
\centering 
\includegraphics[width=\textwidth]{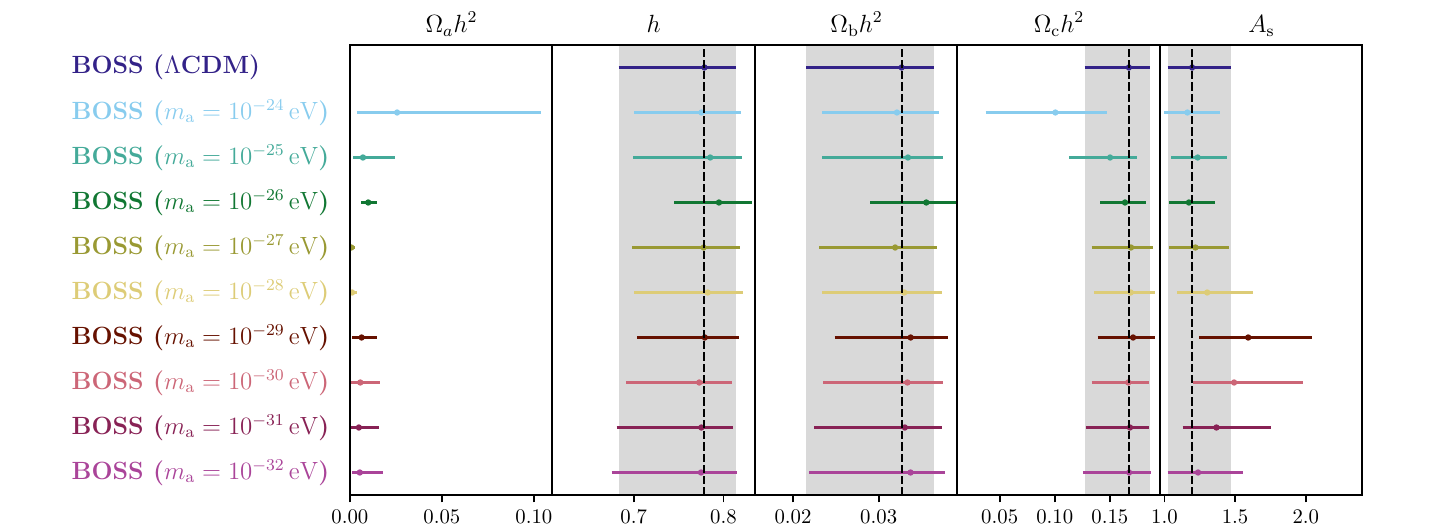}
\includegraphics[width=\textwidth]{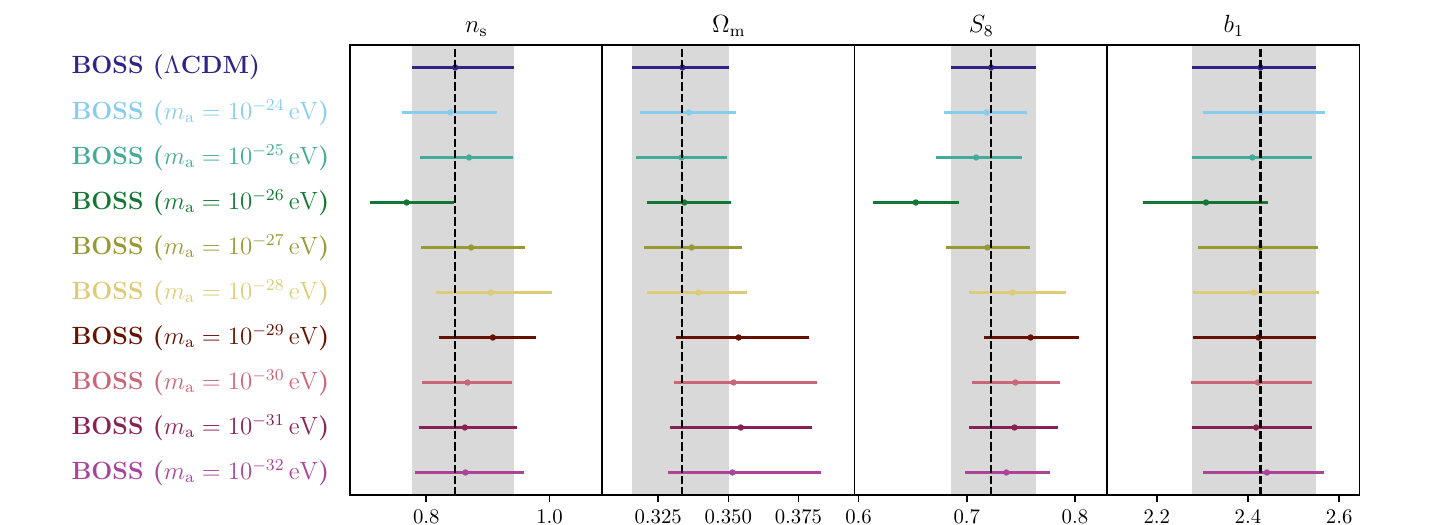}
\caption{\label{fig:boss_only_all}The effect of axion mass \(m_\mathrm{a}\) on cosmological parameter constraints from BOSS galaxy clustering data. We stress that even the \(\Lambda\)CDM constraints differ from those reported in Ref.~\cite{2022PhRvD.105d3517P} as, in this work, we do not use a Big Bang nucleosynthesis (BBN) prior on the baryon energy density \(\Omega_\mathrm{b} h^2\). Each point indicates the marginalised mean, while the errorbar indicates the marginalised 68\% c.l. \(A_\mathrm{s}\) is in units of \(10^{-9}\); \(b_1\) is the linear galaxy bias at \(z = 0.61\) in the north Galactic cap (NGC; similar values are found in all four redshift/sky samples).}
\end{figure}

In Fig.~\ref{fig:mass_comparison}, we show the 95 \% upper limits on \(\Omega_\mathrm{a} h^2\) derived from BOSS data across the full mass range that we consider (see also Table~\ref{tab:boss}). For \(m_\mathrm{a} < 10^{-25}\,\mathrm{eV}\), the BOSS-only constraints are weaker than \textit{Planck}, although the BOSS data are crucial in strengthening the combined CMB and galaxy clustering limit at nearly all masses (see below). Nonetheless, we see the typical ``u''-shaped constraints (that we see with CMB data) also given BOSS alone: at higher mass, BOSS loses sensitivity since the scale-dependent suppression manifests at larger wavenumbers than those we model in BOSS (crucially, BOSS probes smaller scales than \textit{Planck} and so we have improved sensitivity for \(m_\mathrm{a} = 10^{-25}\,\mathrm{eV}\)); at lower mass, BOSS loses sensitivity owing to degeneracy with \(A_\mathrm{s}\). The degeneracy with \(A_\mathrm{s}\) for \(m_\mathrm{a} \leq 10^{-28}\,\mathrm{eV}\) is illustrated in Fig.~\ref{fig:boss_only_all}, where we show how axions impact BOSS constraints on all cosmological parameters. This degeneracy arises at low mass since the axion Jeans wavenumber is then smaller than the smallest wavenumber that we model in BOSS. This means that axions suppress all BOSS wavenumbers, which is degenerate with lowering \(A_\mathrm{s}\) and so lowering the overall power amplitude. BOSS data are therefore compatible with higher values of \(S_8\) (driven by higher \(A_\mathrm{s}\) and also higher \(\Omega_\mathrm{m}\)) than for \(\Lambda\)CDM for \(m_\mathrm{a} \leq 10^{-28}\,\mathrm{eV}\) (the effects of higher \(A_\mathrm{s}\) and \(\Omega_\mathrm{a} h^2\) do not cancel perfectly at the scales to which \(S_8\) is sensitive). This drives an increase in compatibility between BOSS and \textit{Planck} around \(m_\mathrm{a} \sim 10^{-29}\,\mathrm{eV}\), including (unlike with heavier axions) with regards to the \(A_\mathrm{s}\) discrepancy (see Table~\ref{tab:tension}). Beyond degeneracy with \(A_\mathrm{s}\), we also see the typical degeneracy with \(\Omega_\mathrm{c} h^2\) for (heavier) DM-like axions and degeneracy with \(\Omega_\mathrm{m}\) for (lighter) DE-like axions since they additionally count as matter by today. Fig.~\ref{fig:boss_only_all} also reveals that at \(m_\mathrm{a} = 10^{-26}\,\mathrm{eV}\), rather than an upper limit on the axion density, BOSS data alone disfavour no axions at \(\sim 2.7 \sigma\) significance; we discuss this in more detail below (see Fig.~\ref{fig:boss_joint_axions_m26} and surrounding discussion).

\subsubsection{Joint \textit{Planck} and BOSS axion constraints}
\label{sec:boss_results_joint}

\begin{figure}[tbp]
\centering 
\includegraphics[width=\textwidth]{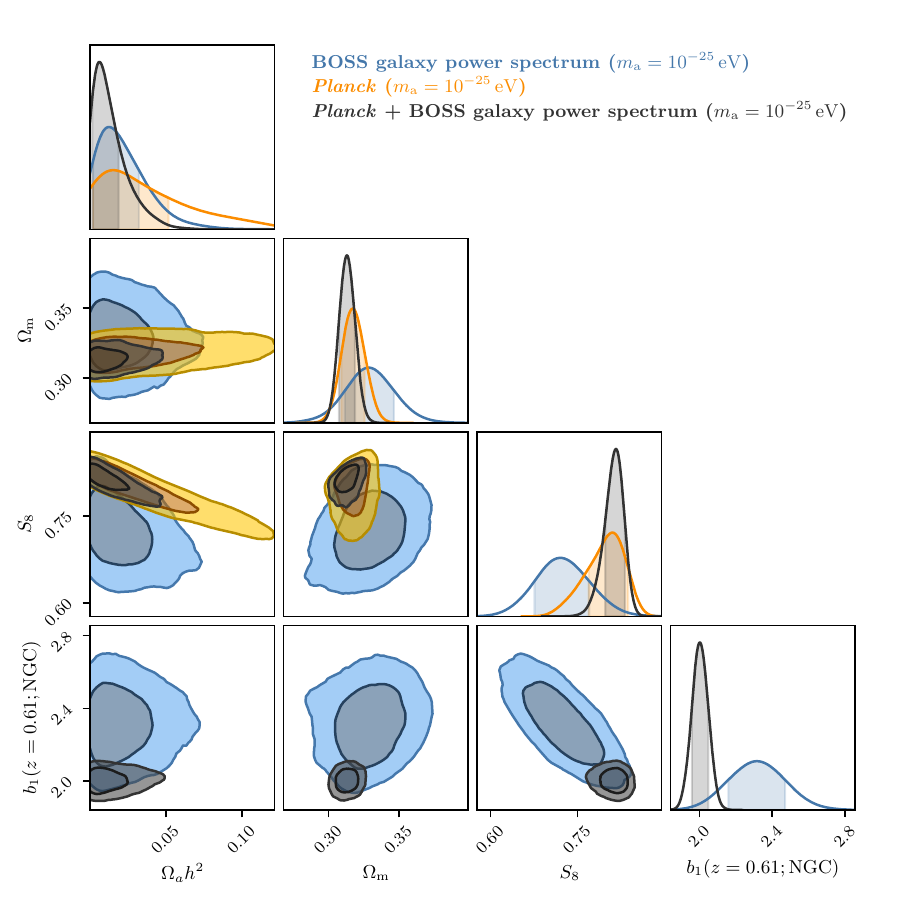} 
\caption{\label{fig:boss_joint_axions}Comparison of BOSS galaxy power spectrum (\textit{blue}), \textit{Planck} CMB (\textit{orange}) and joint (\textit{black}) constraints on axion and cosmological parameters, for axion mass \(m_\mathrm{a} = 10^{-25}\,\mathrm{eV}\). The strongest bound on the axion energy density \(\Omega_\mathrm{a} h^2\) comes from combining the datasets; in order to maintain a good fit to the galaxy data in the joint constraint, lower (though still physically plausible) values of the linear galaxy bias \(b_1\) are preferred. For each set, the darker and lighter shaded contours respectively indicate the 68\% and 95\% credible regions of the 2D marginalised posterior distribution, with the 1D marginalised posteriors on the diagonal, where 68\% credible regions are shaded. \textit{From left to right}, \(\Omega_\mathrm{a} h^2\) is the physical axion energy density, \(\Omega_\mathrm{m}\) is the matter energy density, \(S_8\) is the matter clumping factor and \(b_1 (z = 0.61; \mathrm{NGC})\) is the linear galaxy bias at redshift \(z = 0.61\) in the north Galactic cap (NGC; similar values are found in all four redshift/sky samples).}
\end{figure}

\begin{figure}[tbp]
\centering 
\includegraphics[width=\textwidth]{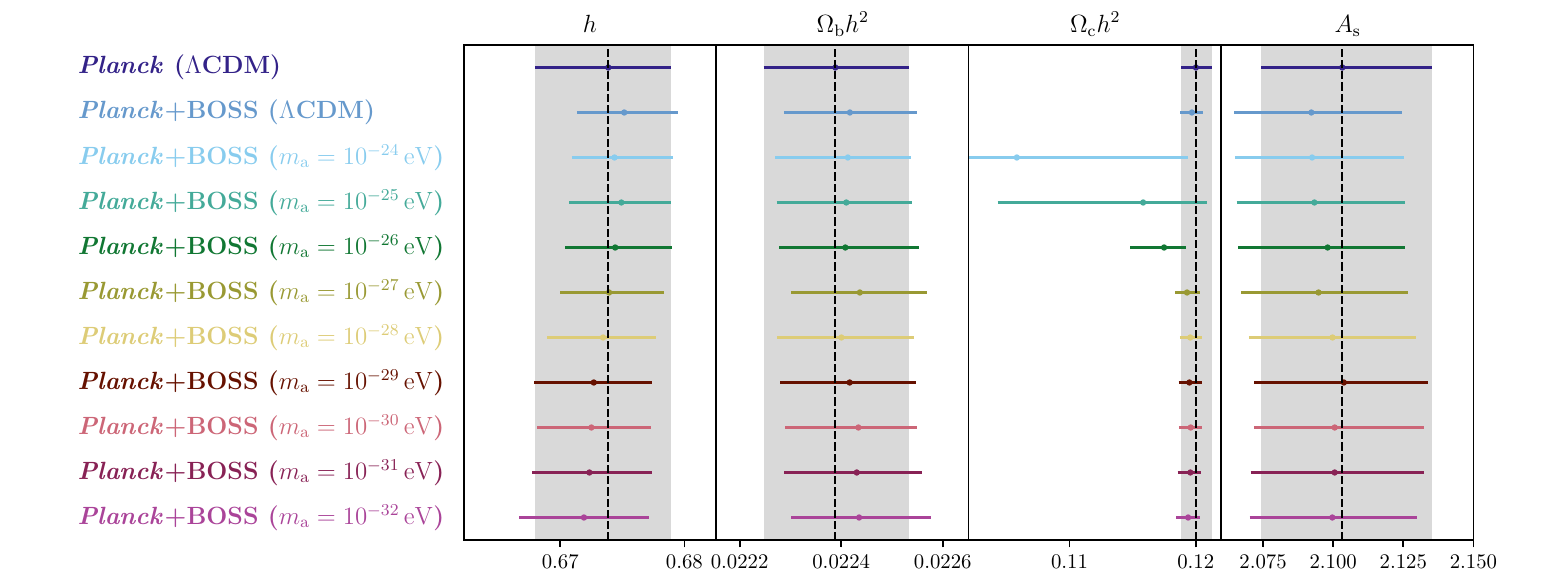}
\includegraphics[width=\textwidth]{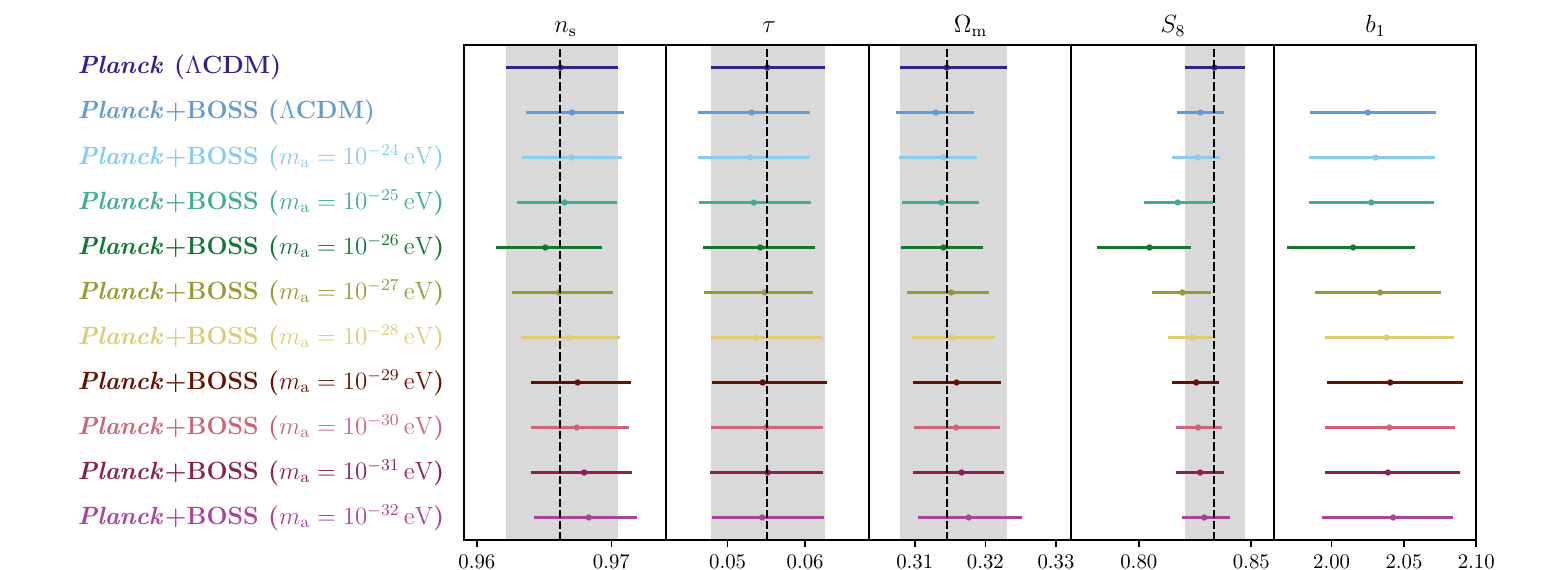}
\caption{\label{fig:boss_all}The effect of axion mass \(m_\mathrm{a}\) on cosmological parameter constraints from the joint inference of \textit{Planck} CMB and BOSS galaxy clustering data, and a comparison to the \textit{Planck} \(\Lambda\)CDM inference. Since \textit{Planck} is much more constraining than BOSS alone on \(\Lambda\)CDM cosmological parameters, the joint constraints on these parameters are broadly consistent with the \textit{Planck} \(\Lambda\)CDM case. Each point indicates the marginalised mean, while the errorbar indicates the marginalised 68\% c.l. \(A_\mathrm{s}\) is in units of \(10^{-9}\); \(b_1\) is the linear galaxy bias at \(z = 0.61\) in the north Galactic cap (NGC; similar values are found in all four redshift/sky samples). The \(\Omega_\mathrm{c} h^2\) constraint at \(m_\mathrm{a} = 10^{-24}\,\mathrm{eV}\) extends to 0.05; we zoom-in for clarity at other masses.}
\end{figure}

In Fig.~\ref{fig:boss_joint_axions}, we show the joint constraint from \textit{Planck} and the BOSS galaxy power spectrum on axions for \(m_\mathrm{a} = 10^{-25}\,\mathrm{eV}\). The strongest limit on the axion energy density comes from combining the datasets. Since \textit{Planck} is significantly more constraining than BOSS alone on \(\Lambda\)CDM parameters, the joint constraint on those parameters is largely driven by \textit{Planck} (see also Fig.~\ref{fig:boss_all}). BOSS (and galaxy clustering data in general) are constraining on a degenerate combination \(b_1 S_8\) of the power spectrum amplitude \(S_8\) and the linear galaxy bias \(b_1\), since this combination scales the large-scale galaxy power spectrum (see \S~\ref{sec:eft}; although this degeneracy is partly broken by the quadrupole's sensitivity to \(f \sigma_8\), where \(f\) is the growth rate). It follows that, in the joint constraint, since \textit{Planck} drives higher values of the power spectrum amplitude (even in the presence of axions) that a good fit to BOSS data is maintained by preferring a lower value of \(b_1\). This is illustrated in Fig.~\ref{fig:boss_joint_axions}, where the joint constraint on \(b_1\) is lower than for BOSS alone (moving along the \(b_1 S_8\) degeneracy), but still has a value \(b_1 \sim 2\) that is consistent with previous findings. This behaviour is observed at other axion masses and in the \(\Lambda\)CDM case (see Fig.~\ref{fig:boss_all}).

\begin{figure}[tbp]
\centering 
\includegraphics[width=\textwidth]{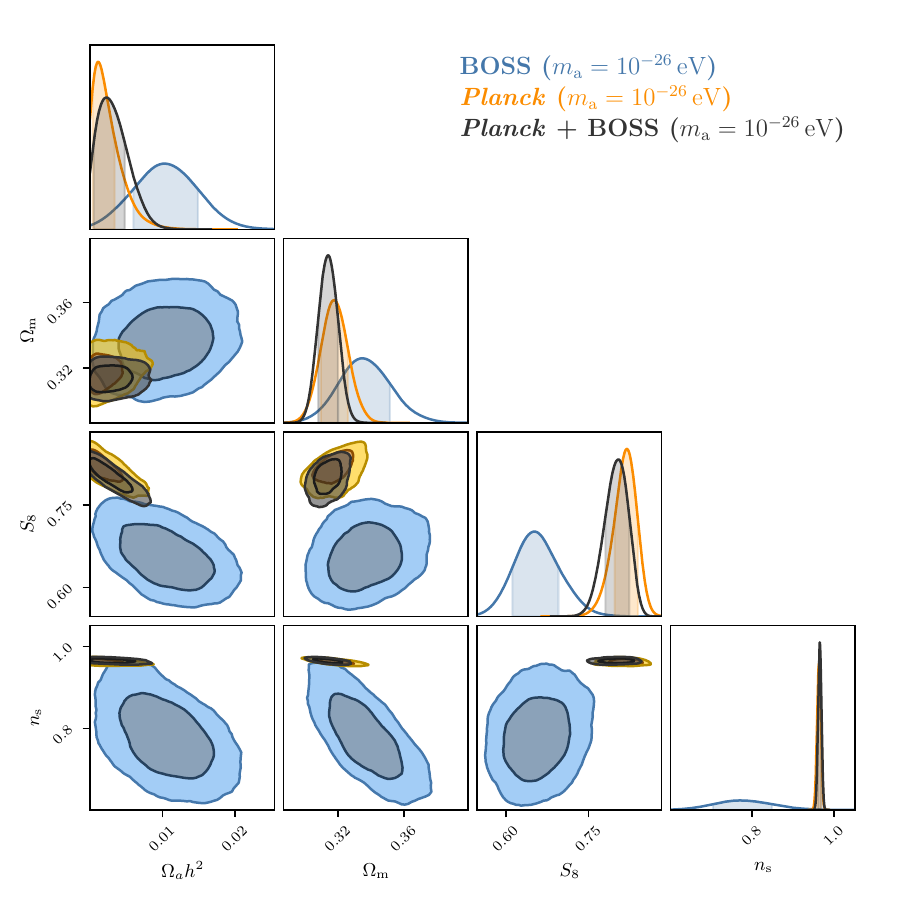} 
\caption{\label{fig:boss_joint_axions_m26}Comparison of BOSS galaxy clustering (all data; \textit{blue}), \textit{Planck} CMB (\textit{orange}) and joint (\textit{black}) constraints on axion and cosmological parameters, for axion mass \(m_\mathrm{a} = 10^{-26}\,\mathrm{eV}\). Although BOSS data give a hint of a significant axion energy density at this mass, \textit{Planck} data disfavour this scenario. The consequence is that the joint axion constraint is weaker than for \textit{Planck} data alone. However, we note that, unlike axions at other masses, axions with \(m_\mathrm{a} = 10^{-26}\,\mathrm{eV}\) increase the discrepancy between \textit{Planck} and BOSS data with respect to \(\Lambda\)CDM (see Table \ref{tab:tension}) and so the joint constraint should be considered with caution. For each set, the darker and lighter shaded contours respectively indicate the 68\% and 95\% credible regions of the 2D marginalised posterior distribution, with the 1D marginalised posteriors on the diagonal, where 68\% credible regions are shaded. \textit{From left to right}, \(\Omega_\mathrm{a} h^2\) is the physical axion energy density, \(\Omega_\mathrm{m}\) is the matter energy density, \(S_8\) is the matter clumping factor and \(n_\mathrm{s}\) is the primordial power spectrum tilt.}
\end{figure}

\begin{figure}[tbp]
\centering 
\includegraphics[width=0.9\textwidth]{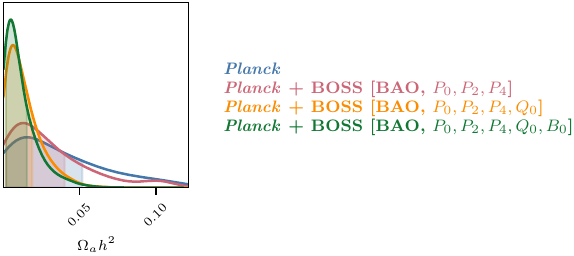} 
\caption{\label{fig:boss_test}The effect of adding different parts of the BOSS galaxy clustering data on axion energy density \(\Omega_\mathrm{a} h^2\) constraints for \(m_\mathrm{a} = 10^{-25}\,\mathrm{eV}\). We systematically add to the \textit{Planck} CMB likelihood (\textit{blue}) different parts of the BOSS likelihood: first, BAO and power spectrum multipoles \([P_0, P_2, P_4]\) up to maximum wavenumber \(k_\mathrm{max} = 0.2\,h\,\mathrm{Mpc}^{-1}\) (\textit{red}); then, also the reconstructed real-space power spectrum \(Q_0\) for \(k \in [0.2, 0.4]\,h\,\mathrm{Mpc}^{-1}\) (\textit{orange}); and finally, also the bispectrum monopole \(B_0\) (\textit{green}). We find that the vast majority of the improvement in the bound comes from the addition of smaller-scale information in the \(Q_0\) likelihood, since the suppression effect of axions is stronger on smaller scales (see Fig.~\ref{fig:power}). For each data cut, we show the 1D marginalised posterior for \(\Omega_\mathrm{a} h^2\), where the 68\% credible region is shaded.}
\end{figure}

\begin{figure}[tbp]
\centering 
\includegraphics[width=\textwidth]{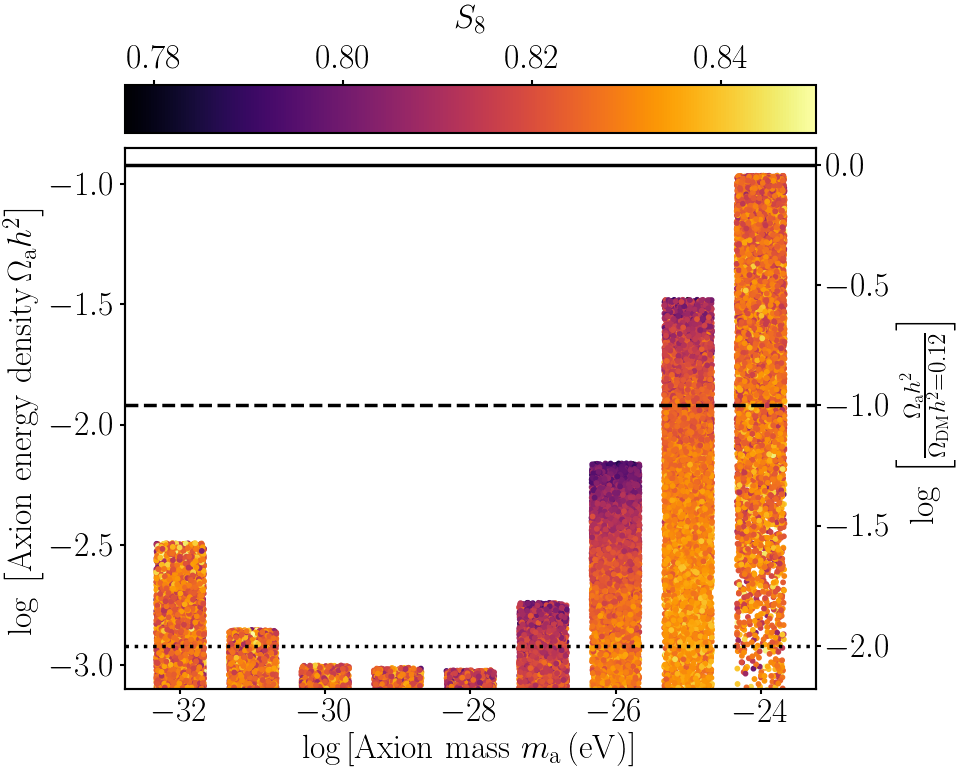} 
\caption{\label{fig:mass_s8}95\% credible upper limits on axion energy density \(\Omega_\mathrm{a} h^2\), as a function of axion mass \(m_\mathrm{a}\), as jointly inferred from \textit{Planck} CMB and BOSS galaxy clustering data. We illustrate the degeneracy with the matter clumping factor \(S_8\) by colouring (unweighted) posterior samples according to their \(S_8\) value. The lowest values of \(S_8\) are allowed for dark matter-like axions with \(m_\mathrm{a} \in [10^{-27}, 10^{-25}]\,\mathrm{eV}\). On the right-hand side, we show the 95\% upper limit on the ratio of the axion energy density to the best-fit dark matter (DM) energy density as inferred from \textit{Planck} \(\Omega_\mathrm{DM} h^2 = 0.12\).}
\end{figure}

\begin{figure}[tbp]
\centering 
\includegraphics[width=\textwidth]{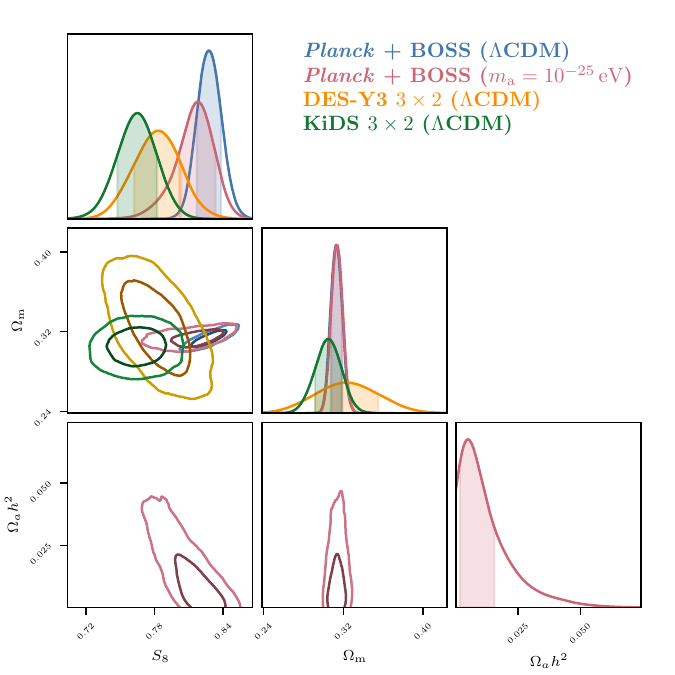} 
\caption{\label{fig:boss_s8}Comparison of joint \textit{Planck} CMB and BOSS galaxy clustering constraints (in both axion and \(\Lambda\)CDM models) with fiducial galaxy weak lensing and clustering (\(3 \times 2\)) \(\Lambda\)CDM constraints from the Dark Energy Survey (DES) and the Kilo-Degree Survey (KiDS) (all with fixed neutrino mass). \textit{Planck} and BOSS data are consistent with lower values of \(S_8\) in the presence of axions with mass \(m_\mathrm{a} = 10^{-25}\,\mathrm{eV}\) compared to the \(\Lambda\)CDM case. In order to assess consistency between all data in an axion model, it is necessary to re-analyse the \(3 \times 2\) data in the presence of axions; we discuss the future analysis of cosmic shear data in \S~\ref{sec:discussion}. We note caution in assessing parameter tension by eye, especially as the \textit{Planck} + BOSS and KiDS datasets are not independent, since KiDS uses BOSS clustering information in their \(3 \times 2\) measurement. For each set, the inner and outer contours respectively indicate the 68\% and 95\% credible regions of the 2D marginalised posterior distribution, with the 1D marginalised posteriors on the diagonal, where 68\% credible regions are shaded. \textit{From left to right}, \(S_8\) is the matter clumping factor, \(\Omega_\mathrm{m}\) is the matter energy density and \(\Omega_\mathrm{a} h^2\) is the physical axion energy density.}
\end{figure}

Figure \ref{fig:mass_comparison} shows the joint limit from \textit{Planck} and BOSS on the axion energy density across the full axion mass range to which we are sensitive (\(10^{-32}\,\mathrm{eV} \leq m_\mathrm{a} \leq 10^{-24}\,\mathrm{eV}\); see also Table \ref{tab:planck}). At nearly all masses, the strongest bound comes from combining the datasets. Fig.~\ref{fig:boss_all} shows the joint constraints on the other cosmological parameters and the linear galaxy bias. As discussed above (Fig.~\ref{fig:boss_joint_axions}), since \textit{Planck} is much more constraining on \(\Lambda\)CDM parameters, the joint \textit{Planck} + BOSS constraints on these parameters is largely driven by \textit{Planck}. Nonetheless, we note the typical degeneracy of \(\Omega_\mathrm{a} h^2\) with, for DE-like axions, lower values of \(h\) and higher values of \(\Omega_\mathrm{m}\), and for DM-like axions, with lower values of \(\Omega_\mathrm{c} h^2\) (see also \textit{Planck} data in \S~\ref{sec:cmb_results}). BOSS, in general, strengthens the limit on the amount of axions. However, in the DM-like mass range to which we are sensitive (\(10^{-27}\,\mathrm{eV} \leq m_\mathrm{a} \leq 10^{-25}\,\mathrm{eV}\)), the joint bound leaves enough axions still to drive consistency with lower values of \(S_8\) (see also Table \ref{tab:planck}). Below, we consider which parts of the BOSS data are most responsible for improving constraints (Fig.~\ref{fig:boss_test}) and discuss further the implications of these results for the \(S_8\) tension (Figs.~\ref{fig:mass_s8} and \ref{fig:boss_s8}). Similarly as for the CMB data considered in \S~\ref{sec:cmb_results}, with the addition of BOSS data, there remains no preference for axion models according to the Bayesian evidence (see Table \ref{tab:evidence}). The Bayes factors amount to evidence in favour of \(\Lambda\)CDM ranging from ``positive'' to ``strong'' \citep{doi:10.1080/01621459.1995.10476572}.

It is striking that the addition of BOSS data strengthens axion bounds at all masses apart from \(m_\mathrm{a} = 10^{-26}\,\mathrm{eV}\), where in fact the bound is weakened. Fig.~\ref{fig:boss_joint_axions_m26} breaks down the constraint at this mass into its constituent parts. While \textit{Planck} alone sets a 95\% credible upper limit \(\Omega_\mathrm{a} h^2 < 0.00615\), BOSS alone actually favours a contribution of axions \(\Omega_\mathrm{a} h^2 = 0.0100^{+0.0048}_{-0.0037}\), which excludes no axions at \(\sim 2.7 \sigma\) (the best-fit model with respect to BOSS data has a chi-squared reduced by \(\Delta \chi^2 = -7.7\)). This discrepancy in the axion constraint, however, increases the tension between all parameters inferred from \textit{Planck} and BOSS, as seen in all three tension metrics shown in Table \ref{tab:tension}. In particular, the preference in BOSS data for axions of \(m_\mathrm{a} = 10^{-26}\,\mathrm{eV}\) increases the discrepancy in \(S_8\) from \(2.70\,\sigma\) in the \(\Lambda\)CDM case to \(3.63\,\sigma\). This is because the power suppression of axions combines with the already-low value of \(A_\mathrm{s}\) (\(\Omega_\mathrm{a} h^2\) and \(A_\mathrm{s}\) are constrained from different parts of the galaxy power spectrum; see, \eg Fig.~\ref{fig:power}) to lower further the power spectrum amplitude \(S_8\) that is inferred from BOSS. For completeness, we show the joint constraint although we caution that it derives from two datasets that are in more discrepancy than in the \(\Lambda\)CDM case. As before, \textit{Planck} dominates the constraint on \(\Lambda\)CDM parameters, while the joint limit on \(\Omega_\mathrm{a} h^2\) is slightly weaker than for \textit{Planck} alone. Ref.~\cite{2022JCAP...01..049L} in their analysis of previous BOSS data do not report a preference for axions at this mass. There are a number of differences with respect to this study (summarised at the start of \S~\ref{sec:boss_results}). However, in particular, previously, the primordial power spectrum tilt was fixed: \(n_\mathrm{s} = 0.9611\). Fig.~\ref{fig:boss_joint_axions_m26} illustrates that fixing \(n_\mathrm{s}\) at this value will break degeneracy with \(\Omega_\mathrm{a} h^2\) such that the preference for a non-zero contribution is removed (this degeneracy with \(n_\mathrm{s}\) in this mass range is also seen in \textit{Planck} data; see Fig.~\ref{fig:planck_all})\footnote{Using a Big Bang nucleosynthesis (BBN) prior on the baryon energy density \(\Omega_\mathrm{b} h^2 \sim \mathcal{N}(0.02268, 0.00038)\) \citep{Schoneberg_2019} reduces the significance for an axion component at \(m_\mathrm{a} = 10^{-26}\,\mathrm{eV}\) to \(2.1 \sigma\); further adding a \textit{Planck}-motivated prior \(n_\mathrm{s} \sim \mathcal{N}(0.9649, 0.0042)\) \citep{2020A&A...641A...6P} reduces the significance to \(1.7 \sigma\). Weakening the prior on the EFT of LSS bias and counterterm parameters (see \S~\ref{sec:boss} and \ref{sec:inference}) (by doubling the standard deviation in Gaussian prior distributions and doubling the width in uniform prior distributions) increases the significance to \(3.2 \sigma\).}. This preference for axions is not seen at any other mass. At all other masses, BOSS data strengthen the axion limit and also increase consistency between \textit{Planck} and BOSS datasets (see Table \ref{tab:tension}). The fixed axion masses which we consider are arbitrary. Thus, this result means that there is a preference in BOSS data alone for a contribution of axions with a mass in a window \(m_\mathrm{a} \in [10^{-27}, 10^{-25}]\,\mathrm{eV}\), which motivates future work where we additionally sample \(m_\mathrm{a}\).

Notwithstanding \(m_\mathrm{a} = 10^{-26}\,\mathrm{eV}\), BOSS data otherwise always improve axion limits with respect to \textit{Planck} alone, and axions improve consistency between the datasets. Fig.~\ref{fig:boss_test} illustrates which parts of the BOSS data are most constraining at \(m_\mathrm{a} = 10^{-25}\,\mathrm{eV}\). We systematically add different parts of the BOSS data to a joint constraint with \textit{Planck}. We find that it is the addition of the small-scale reconstructed real-space galaxy power spectrum \(Q_0\) for \(0.2\,h\,\mathrm{Mpc}^{-1} < k < 0.4\,h\,\mathrm{Mpc}^{-1}\) which drives the vast majority of the improvement in the bound. This arises because the power suppression effect of axions is always stronger on smaller scales.

Figure \ref{fig:mass_s8} illustrates the degeneracy between \(\Omega_\mathrm{a} h^2\) and \(S_8\) within the 95\% credible upper limits on \(\Omega_\mathrm{a} h^2\) that are allowed by the joint analysis of \textit{Planck} and BOSS. As we saw above, there is no such degeneracy for DE-like axions (\(m_\mathrm{a} < 10^{-28}\,\mathrm{eV}\)) or for DM-like axions where the power suppression scale is too small (\(m_\mathrm{a} \geq 10^{-24}\,\mathrm{eV}\)). In the mass window (\(10^{-28}\,\mathrm{eV} \leq m_\mathrm{a} \leq 10^{-25}\,\mathrm{eV}\)) however, the joint constraint still allows enough axions to drive consistency with lower values of \(S_8\). Although more axions are allowed at higher masses (in the DM-like regime), since the suppression scale is smaller at higher mass, there is less total power suppression at the wavenumbers to which \(S_8\) is sensitive. The lowest values of \(S_8\) are in fact found at \(m_\mathrm{a} = 10^{-26}\,\mathrm{eV}\) (\(S_8 = 0.804^{+0.020}_{-0.024}\); see also Table \ref{tab:planck} and Fig.~\ref{fig:boss_all}). However we caution that this constraint arises from two datasets that are in stronger tension than the \(\Lambda\)CDM case. Nonetheless, at \(m_\mathrm{a} = 10^{-25}\,\mathrm{eV}\) (\(S_8 = 0.818^{+0.015}_{-0.017}\)) and \(m_\mathrm{a} = 10^{-27}\,\mathrm{eV}\) (\(S_8 = 0.819^{+0.013}_{-0.014}\)), the parameter discrepancy between \textit{Planck} and BOSS is reduced and the joint constraint on \(S_8\) is shifted to lower values than the \(\Lambda\)CDM case (\(S_8 = 0.827 \pm 0.011\)). Fig.~\ref{fig:boss_s8} updates Fig.~\ref{fig:s8} with the joint \textit{Planck} + BOSS constraints (see \S~\ref{sec:cmb_results} for details about the DES and KiDS \(\Lambda\)CDM contours that we show). In comparison to Fig.~\ref{fig:s8}, we note how the addition of BOSS data more strongly constrains the axion energy density and in turn reduces the extent to which low values of \(S_8\) are allowed. Nonetheless, there remains a tail in the posterior to lower values of \(S_8\) in the presence of axions with \(m_\mathrm{a} = 10^{-25}\,\mathrm{eV}\). Fully assessing the consistency with galaxy weak lensing experiments like DES and KiDS requires re-analysing these data in the axion models we consider here. We discuss the prospects for this in \S~\ref{sec:discussion}.

\section{Discussion}
\label{sec:discussion}

In \S~\ref{sec:results}, we present several new results in searching for ultra-light axions in a compendium of CMB and large-scale structure data. In \S~\ref{sec:cmb_results}, we present legacy constraints on the axion energy density from \textit{Planck} 2018 CMB temperature, polarisation and lensing anisotropies. We find that, compared to previous \textit{Planck} 2015 results \citep{2018MNRAS.476.3063H}, a new measurement of the optical depth to reionisation (through large-scale polarisation) breaks parameter degeneracies and improves energy density bounds for DE-like axions (\(m_\mathrm{a} \leq 10^{-28}\,\mathrm{eV}\); see Fig.~\ref{fig:planck_test}). Further, we search for axions in a compendium of higher-resolution CMB data (ACT-DR4, SPT-3G), galaxy BAO and supernovae data. We find that the addition of these data marginally weakens the axion energy density bound for \(m_\mathrm{a} = 10^{-25}\,\mathrm{eV}\) (see Table \ref{tab:cmb_bao}).

In \S~\ref{sec:boss_results}, we present axion constraints from BOSS galaxy clustering data. We find that the addition of BOSS to \textit{Planck} improves axion energy density bounds at nearly all axion masses that we consider (\(10^{-32}\,\mathrm{eV} \leq m_\mathrm{a} \leq 10^{-25}\,\mathrm{eV}\)). Crucially, we find that the inclusion of new small-scale modes (\(Q_0\) for \(k \in [0.2, 0.4]\,h\,\mathrm{Mpc}^{-1}\)) strengthens the constraint at \(m_\mathrm{a} = 10^{-25}\,\mathrm{eV}\) with respect to \textit{Planck} only (see Fig.~\ref{fig:boss_test}). This is driven by gaining sensitivity to larger wavenumbers where the power suppression of heavier axions manifests. Gains in sensitivity from BOSS data to lighter, DE-like axions (\(m_\mathrm{a} \leq 10^{-28}\,\mathrm{eV}\)) are limited by degeneracy between \(\Omega_\mathrm{a} h^2\) and \(A_\mathrm{s}\) at those masses. This arises since the axion-induced power suppression occurs at wavenumbers smaller than we model in BOSS data and so the axion effect is degenerate with an overall re-scaling of the galaxy power spectrum amplitude (\eg see Fig.~\ref{fig:boss_only_all}). This suggests that robustly modelling larger-volume galaxy surveys can improve sensitivity to DE-like axions. Robustly modelling smaller-scale correlations in galaxy positions will be extremely challenging owing to the non-trivial way that galaxies trace dark matter on small scales (\ie non-linear galaxy bias). We therefore suggest alternative probes like galaxy and CMB weak lensing (that are insensitive to galaxy bias) to increase sensitivity at \(m_\mathrm{a} \geq 10^{-24}\,\mathrm{eV}\) (see above and below for more discussion about probes complementary to galaxy correlations). Nonetheless, our results demonstrate the power in combining CMB and large-scale structure data when constraining dark matter models beyond standard CDM.

\subsection{Comparison to previous work}
\label{sec:discussion_previous}

There are a number of differences between this study and a previous BOSS analysis presented in Ref.~\cite{2022JCAP...01..049L}. First, as discussed above, we model more of the BOSS data, in particular, additionally, the galaxy power spectrum hexadecapole \(P_4\), the small-scale real-space galaxy power spectrum \(Q_0\) (where we conservatively project away hard-to-model non-linear redshift-space distortions) and the galaxy bispectrum monopole \(B_0\). Further, we choose less informative priors on cosmological parameters, \ie we do not use BBN information to place a prior on the baryon energy density \(\Omega_\mathrm{b} h^2\) and, importantly, we do not fix the primordial power spectrum tilt \(n_\mathrm{s}\). The latter is important as we do in general observe degeneracy between \(\Omega_\mathrm{a} h^2\) and \(n_\mathrm{s}\) (\eg see Fig.~\ref{fig:boss_only_all}) and this degeneracy will be broken by fixing \(n_\mathrm{s}\). We thus find that our bounds from BOSS alone are weaker than those reported in Ref.~\cite{2022JCAP...01..049L}. Ref.~\cite{2022JCAP...01..049L} combined \textit{Planck} and BOSS through a \textit{Planck}-motivated prior on cosmological and axion parameters (except \(n_\mathrm{s}\) which remained fixed) combined with the BOSS likelihood. Instead, in this work, for the first time, we jointly sample the \textit{Planck} and BOSS likelihoods in a full axion and cosmological model in setting axion constraints. We find in general that our combined constraints are stronger than those reported in Ref.~\cite{2022JCAP...01..049L}. We attribute a large degree of this to the information gained by updating to \textit{Planck} 2018 data (\textit{Planck} 2015 data was previously considered) for low masses (see above) and using the small-scale \(Q_0\) statistic for higher masses. The results in Ref.~\cite{2022JCAP...01..049L} are affected by an error in the BOSS data weights, which has since been corrected and does not affect the results presented here.

There are further pipeline differences between the two analyses. In particular, beyond the different and more complete compression of the BOSS data discussed above, we use different implementations of the BOSS likelihood and EFT of LSS theory calculations (namely, \texttt{CLASS-PT}/\texttt{full\_shape\_likelihoods} and, previously, \texttt{PyBird}) and, correspondingly, different EFT of LSS parameter priors (namely, so-called ``East Coast'' and, previously, ``West Coast'' priors). In general, the different prior choices will lead to differences in parameter inference given the same set of BOSS data (in \(\Lambda\)CDM, the cosmological constraints are consistent within \(\sim 1 \sigma\); see Ref.~\cite{Simon:2022lde}). Importantly, Refs.~\cite{Nishimichi:2020tvu} and \cite{Simon:2022lde} demonstrated that, with external CMB information from \textit{Planck}, the prior sensitivity is significantly reduced, while future larger-volume galaxy surveys will have sufficient constraining power also to lose prior sensitivity. We defer to future work a detailed study of the effect of EFT of LSS priors on BOSS axion constraints since, in this work, it is non-trivial to disentangle the other analysis differences.

\subsection{\(m_\mathrm{a} = 10^{-26}\,\mathrm{eV}\)}
\label{sec:discussion_m26}

A striking difference between this work and Ref.~\cite{2022JCAP...01..049L} is the axion constraint at \(m_\mathrm{a} = 10^{-26}\,\mathrm{eV}\). Unlike at other axion masses that we consider, at \(m_\mathrm{a} = 10^{-26}\,\mathrm{eV}\), rather than setting an upper limit on the axion energy density, we find, given BOSS data only, \(\Omega_\mathrm{a} h^2 = 0.0100^{+0.0048}_{-0.0037}\) that excludes no axions at \(\sim 2.7 \sigma\) significance. However, such a large contribution of axions at this mass is disfavoured by \textit{Planck} (\(\Omega_\mathrm{a} h^2 < 0.00615\)) and so the tension in parameter inference between these datasets is increased at this axion mass with respect to \(\Lambda\)CDM (at all other masses, the tension is reduced; see more discussion below). For completeness, we consider the joint constraint (which is thus weaker than \textit{Planck} alone) although we caution that this derives from two datasets in greater discrepancy than in the standard CDM model. We find that the preference for axions at this mass opens up degeneracy with other cosmological parameters, in particular \(n_\mathrm{s}\) and \(b_1\) (although in an opposite sense as at other masses; \eg see Figs.~\ref{fig:boss_only_all} and \ref{fig:boss_joint_axions_m26}). This explains why this preference was not observed in Ref.~\cite{2022JCAP...01..049L} where \(n_\mathrm{s}\) was fixed; indeed, when giving a BBN prior on \(\Omega_\mathrm{b} h^2\) and a \textit{Planck} prior on \(n_\mathrm{s}\), the significance of the axion preference is reduced to only \(1.7 \sigma\). Further, if an inflation-motivated prior that excluded low values of \(n_\mathrm{s}\) was imposed, we anticipate that the significance would also be reduced.

The anomalous (with respect to other masses) degeneracy between higher \(\Omega_\mathrm{a} h^2\) and lower \(n_\mathrm{s}\) and \(b_1\) suggests an effect from marginalisation over other EFT of LSS parameters; indeed, weakening the prior on EFT of LSS bias and counterterm parameters increases the axion preference to \(3.2 \sigma\). As discussed in \S~\ref{sec:boss}, in order to reduce the dimensions of the sampling task, we analytically marginalise over a number of bias and counterterm parameters. We therefore leave for future work a detailed study of the effect of nuisance parameter marginalisation by numerically sampling the full joint cosmological and EFT of LSS posterior distribution. We stress, nonetheless, owing to the way that we consider axion masses only at a number of fixed values, that there is an element of the look-elsewhere effect where it is not surprising to find one of the nine axion masses has a mild preference unlike the others. Future galaxy data (\eg from the Dark Energy Spectroscopic Instrument \citep{2016arXiv161100036D} or the \textit{Rubin} Observatory \citep{2012arXiv1211.0310L}) will be crucial in determining if this preference is only a statistical anomaly or otherwise. Reconciliation with the \textit{Planck} bound may be connected to the \(A_\mathrm{L}\) anomaly. As discussed above, we find that CMB datasets with lower (and theoretically-consistent) amounts of lensing weaken axion bounds and so we hypothesise that the \(A_\mathrm{L}\) anomaly in \textit{Planck} is strengthening the bound and increasing the discrepancy with BOSS at \(m_\mathrm{a} = 10^{-26}\,\mathrm{eV}\). We will investigate this hypothesis in future work.

\subsection{Comparison to other axion probes and future prospects}
\label{sec:discussion_other}

\begin{figure}[tbp]
\centering 
\includegraphics[width=\textwidth]{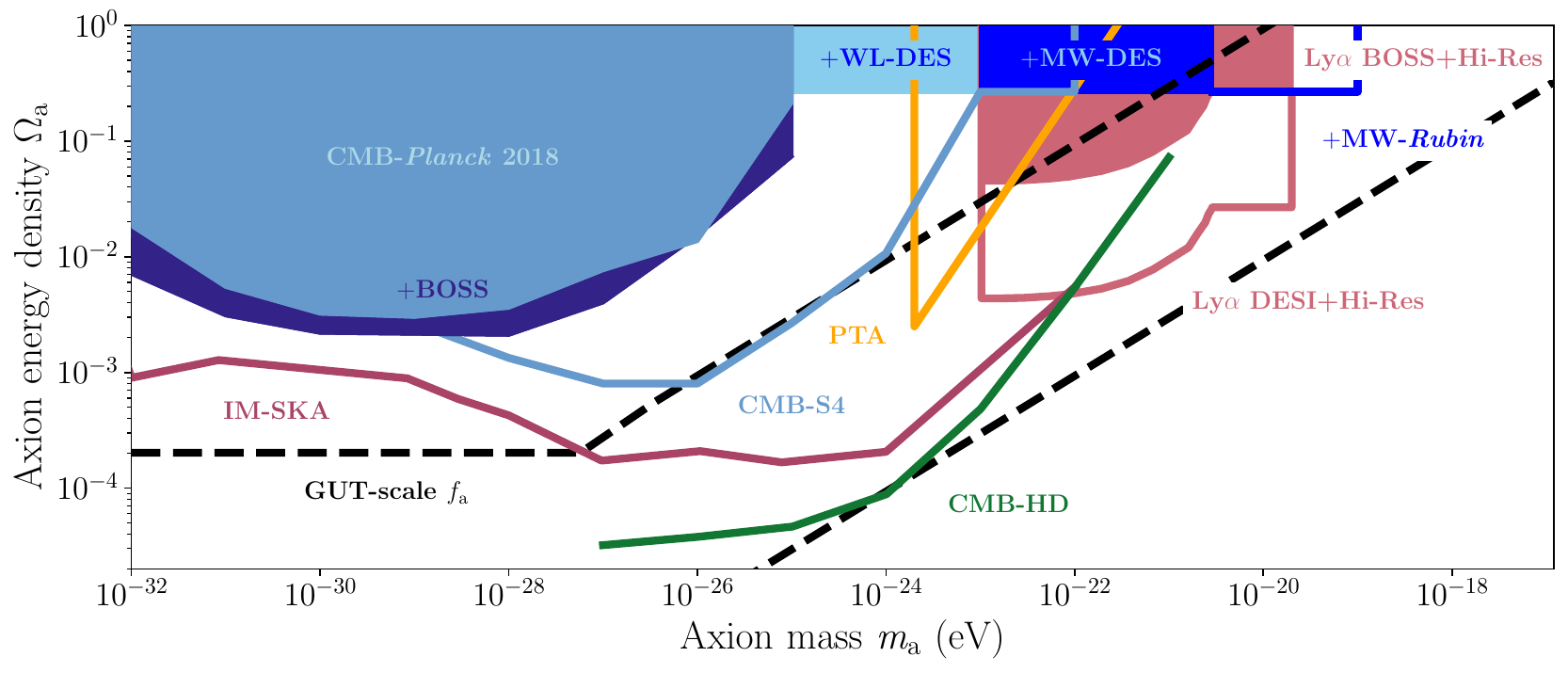} 
\caption{\label{fig:all_comparison}95\% c.l. axion energy density \(\Omega_\mathrm{a}\) bounds presented in this work from \textit{Planck} 2018 CMB data (\textit{top left}) and from a joint analysis of \textit{Planck} CMB and BOSS galaxy clustering data (+BOSS) compared to other cosmological bounds (\textit{shaded solid}) and projected bounds (\textit{thick lines}). Our results are complementary to existing bounds at higher masses derived from probes of smaller-scale structure: a joint analysis of \textit{Planck} CMB and galaxy weak lensing data from the Dark Energy Survey (+WL-DES) \citep{dentler/etal:2022}; the Milky Way sub-halo mass function from DES (+MW-DES) \citep{2021PhRvL.126i1101N}; the strongest lower limit for axions being all the DM (\(m_\mathrm{a} > 2 \times 10^{-20}\,\mathrm{eV}\)) comes from high-resolution Lyman-alpha forest data (Ly\(\alpha\) Hi-Res) \citep{rogers/peiris:2021,2021PhRvD.103d3526R}, while Ref.~\cite{2017arXiv170800015K} considered a sub-dominant axion contribution (see also Ref.~\cite{2017arXiv170309126A} for a BOSS Lyman-alpha forest analysis). We show projected bounds for: the CMB-S4 experiment \citep{2017PhRvD..95l3511H}; the CMB-HD experiment using the Ostriker-Vishniac signal \citep{Farren_2022}; the \textit{Rubin} Observatory using the MW sub-halo mass function \citep{2022arXiv220307354B}; future Lyman-alpha forest data from the Dark Energy Spectroscopic Instrument (DESI) and high-resolution quasar spectra (Hi-Res) \citep{2019BAAS...51c.567G}; intensity mapping from the Square Kilometre Array (IM-SKA) \citep{bauer/etal:2021}; and pulsar timing array (PTA) residuals \citep{2019BAAS...51c.567G}. We indicate (\textit{between the black dotted lines}) the parameter space where the axion decay constant \(f_\mathrm{a}\) is at the Grand Unified Theory (GUT) scale (see Eq.~\eqref{eq:axion_density}). Where bounds exclude axions being all the DM, we additionally exclude higher energy densities (up to \(\Omega_\mathrm{a} = 1\)) by enforcing that the Universe is not over-closed. The projections rely on different assumptions and have varying degrees of rigour and so are only indicative of future progress.}
\end{figure}

Notwithstanding the mild preference for axions at \(m_\mathrm{a} = 10^{-26}\,\mathrm{eV}\) given BOSS data only, our \textit{Planck} and joint \textit{Planck} and BOSS analyses set strong limits on the axion energy density for \(m_\mathrm{a} \leq 10^{-25}\,\mathrm{eV}\). Axions are well-motivated in a range of particle masses and can be produced in a mixture with other axions (the so-called ``axiverse'' \citep[\eg][]{Arvanitaki:2009fg}) and/or with other DM and DE particle candidates. This motivates a search for axions across a wide range of masses and for sub-dominant energy densities so that an axion of a particular mass is not prematurely excluded by assuming that it constitutes the entirety of the DM or DE. Fig.~\ref{fig:all_comparison}\footnote{An up-to-date version of Fig.~\ref{fig:all_comparison} is maintained at \url{https://keirkwame.github.io/DM_limits}.} compares our new bounds for \(m_\mathrm{a} \leq 10^{-25}\,\mathrm{eV}\) to other cosmological bounds across the mass range where the gravitational effect of axions is distinguishable in the large-scale structure from standard cold DM (\(10^{-32}\,\mathrm{eV} \leq m_\mathrm{a} \leq 10^{-18}\,\mathrm{eV}\))\footnote{There are many ongoing and proposed experimental efforts to probe axions at and above this mass range; see Refs.~\cite{2022arXiv220314915A,2022arXiv220314923J} for recent reviews; Fig.~\ref{fig:all_comparison} shows only cosmological probes.}. The details of each experiment and current and projected bounds are given in the caption. We stress that, to probe across the parameter space, it is necessary to use complementary probes of large- and small-scale structure to search for, respectively, lighter and heavier axions. We anticipate progress in this regard from ongoing, upcoming and proposed CMB (\eg ACT, SPT, Simons Observatory, CMB-S4 \citep{Dvorkin:2022bsc}, CMB-HD), large-scale structure (\eg \textit{Rubin}, DESI), intensity mapping (\eg SKA) and pulsar timing array observations.

\subsection{Axions as a resolution to the \(S_8\) parameter tension}
\label{sec:discussion_s8}

A key aim of this study is not only to search for axions as a DM and DE candidate, but also to consider the extent to which axions can improve consistency between CMB and large-scale structure datasets in their parameter inference, in particular with respect to the so-called \(S_8\) tension. The cosmological parameter \(S_8\), through its dependence on the matter power spectrum amplitude \(\sigma_8\), is a measure of the clustering of matter at \(z = 0\) when averaged over \(8\,h^{-1}\,\mathrm{Mpc}\) scales. CMB experiments prefer higher values of \(S_8\) than various large-scale structure analyses with statistical significance ranging from 2 to 3 \(\sigma\) depending on the data comparison \citep[see \eg][for a recent review]{2022JHEAp..34...49A}. Since CMB experiments generally probe structure at higher redshift (even CMB lensing is more sensitive to structure at earlier times than current galaxy surveys), most concrete model solutions to the \(S_8\) tension invoke a redshift-dependent suppression in the growth of structure, \eg decaying dark matter \citep[\eg][]{2015JCAP...09..067E,2020arXiv200809615A}. In this way, it is argued that this explains why probes of later-time structure have lower amplitude. In this work, we investigate the hypothesis that the \(S_8\) tension is a discrepancy between probes of larger- and smaller-scale structure, with axions as a concrete model, and with no need to invoke a late-time decay in the nature of DM.

Fig.~\ref{fig:power_linear} illustrates our hypothesis by showing how \textit{Planck} CMB data lose sensitivity to small-scale modes to which \(S_8\) is sensitive. It follows that it is possible to invoke a scale-dependent suppression in the matter power spectrum that is consistent with current CMB data on large scales, while lowering the value of \(S_8\) to improve compatibility with galaxy surveys (in particular, galaxy weak lensing) as a more direct probe of the scales to which \(S_8\) is sensitive. Indeed, we find (in \S~\ref{sec:cmb_results}) axions with \(m_\mathrm{a} \in [10^{-28}, 10^{-25}]\,\mathrm{eV}\) as a good candidate where they are compatible with a compendium of ``large-scale'' probes (CMB, galaxy BAO and supernovae) and the lower values of \(S_8\) that are inferred from fiducial galaxy clustering and weak lensing (\(3 \times 2\)) analyses (see Fig.~\ref{fig:s8}). Lighter axions are largely incompatible with \textit{Planck} data (except as a highly sub-dominant contribution that does little to \(S_8\)), while heavier axions are unconstrained by these data but suppress wavenumbers larger than those to which \(S_8\) is sensitive.

In order to assess whether axions can resolve parameter tensions between CMB and large-scale structure data, it is necessary also to analyse large-scale structure data in an axion model. In \S~\ref{sec:boss_results}, we analyse BOSS galaxy clustering data and model the effect of axions in the mildly non-linear regime using the effective field theory of large-scale structure (see \S~\ref{sec:eft} for details). We find two regimes in which the \(S_8\) discrepancy between \textit{Planck} and BOSS is reduced. The first is for \(m_\mathrm{a} \sim 10^{-25}\,\mathrm{eV}\), where, as discussed above, large axion contributions are unconstrained by CMB data and so bring CMB data into compatibility with lower values of \(S_8\): the \(S_8\) discrepancy is reduced from \(2.70 \sigma\) in \(\Lambda\)CDM to \(1.63 \sigma\). The second regime is for \(m_\mathrm{a} \sim 10^{-28}\,\mathrm{eV}\), where, instead, the effect of axions in BOSS data is partly degenerate with the overall amplitude of the galaxy power spectrum since all BOSS wavenumbers are suppressed by axions of this mass. This weakens BOSS constraints on the lightest axions, while allowing higher values of the primordial power spectrum amplitude \(A_\mathrm{s}\) and also \(S_8\) (the effects of higher \(A_\mathrm{s}\) and higher \(\Omega_\mathrm{a} h^2\) not cancelling exactly). Thus, the \(S_8\) discrepancy is reduced to \(1.78 \sigma\), but, importantly, values of \(A_\mathrm{s}\) (which are low in this BOSS analysis; see Ref.~\cite{Simon:2022lde} for a discussion on the effect of EFT of LSS priors on cosmological parameter inference) are brought into greater compatibility with the higher values inferred from \textit{Planck}.

Indeed, although \(S_8\) is a reasonably good compression of the information contained in large-scale structure data (though not necessarily optimal for all experiments), it is necessary to assess tension in the full posterior, in particular accounting for non-Gaussianity in the distribution (which one-parameter tension metrics do not capture). In this work, in order better to capture tension in the full parameter space, we estimate the posterior distribution of the parameter difference \citep{2021PhRvD.104d3504R} inferred given the two experiments (\textit{Planck} and BOSS). If the two experiments are in perfect agreement, the parameter difference posterior will peak at zero; we assign the significance of the discrepancy between experiments to the amount of shift from perfect agreement (see \S~\ref{sec:boss_results} for more details). We find that this measure of tension improves from \(\Lambda\)CDM at nearly all axion masses apart from \(m_\mathrm{a} = 10^{-26}\,\mathrm{eV}\) (as discussed above). There is no single tension metric on which the community has converged; different metrics tend to disagree in terms of absolute value though they agree with regards to increasing or decreasing tension \citep[see \eg][]{2021MNRAS.505.6179L}. The metrics we use in this work (and quite generally) depend on the parameterisation of the model. Since \(S_8\) and \(\sigma_8\) may not be optimal measures of the matter clustering information directly probed by CMB and even many large-scale structure experiments, we defer to future work studies of the agreement between datasets directly in the linear matter power spectrum using Bayesian metrics like the posterior predictive distribution.

Nonetheless, our results suggest that axions with masses in a window \([10^{-28}, 10^{-25}]\,\mathrm{eV}\) can be a promising candidate to improve consistency between CMB and large-scale structure observations, in particular by bringing \textit{Planck} CMB and BOSS galaxy clustering data into consistency with lower values of \(S_8\) that are preferred by galaxy weak lensing data (see \eg Figs.~\ref{fig:mass_s8} and \ref{fig:boss_s8}). We stress, though, that this is achieved through only upper limits on the axion energy density and there is no preference for model extensions beyond \(\Lambda\)CDM given these data according to the Bayesian evidence in any of our analysis (see Table \ref{tab:evidence}). A more stringent test of the ability for axions to address cosmological parameter tension is the inclusion of galaxy weak lensing data. It is common in the literature to include the effect of weak lensing through a prior on \(S_8\) derived from \(\Lambda\)CDM \(S_8\) constraints, see e.g.~\cite{Hill:2020osr,Ivanov:2020ril}. This is a good measure of the information content in the \(\Lambda\)CDM model, but we caution that this may not be the case in extended models like axions which affect in a non-trivial way the non-linear modes probed by galaxy shear (indeed, we see with BOSS how the \(S_8\) constraint changes with axions). We therefore leave for future work an analysis of galaxy shear and \(3 \times 2\) clustering and shear data using a fully non-linear halo model of axion structure formation \citep[\eg][]{2022arXiv220913445V}. This will build on initial studies of DES cosmic shear in the limited case that axions comprise the entirety of the DM \citep{dentler/etal:2022}.

\section{Conclusions}
\label{sec:conclusions}
We present a comprehensive search for ultra-light axions as a well-motivated dark matter and dark energy particle candidate using a compendium of CMB and large-scale structure data. We set the strongest bounds to date on the axion energy density for axion masses \(m_\mathrm{a} \in [10^{-32}, 10^{-25}]\,\mathrm{eV}\) through a joint analysis of \textit{Planck} 2018 CMB and BOSS full-shape galaxy power spectrum and bispectrum data, modelling the effect of axions in the mildly non-linear regime using the effective field theory of large-scale structure. We exclude axions being more than 10\% of the DM today for \(m_\mathrm{a} \leq 10^{-26}\,\mathrm{eV}\) and more than 1\% for \(m_\mathrm{a} \in [10^{-30}, 10^{-28}]\,\mathrm{eV}\). We give legacy constraints from \textit{Planck} 2018 CMB data and find that measurements of the optical depth to reionisation break parameter degeneracies and improve bounds for DE-like axions (\(m_\mathrm{a} \leq 10^{-28}\,\mathrm{eV}\)). For the first time, we consider high-resolution CMB data from the Atacama Cosmology Telescope and the South Pole Telescope (in combination with galaxy BAO and supernovae data), which we find to weaken marginally axion bounds at \(m_\mathrm{a} = 10^{-25}\,\mathrm{eV}\). Similarly to the effect of massive neutrinos, we attribute this weakening to the lower (and theoretically-consistent) amounts of lensing observed in ACT and SPT angular power spectra, which allow more structure suppression arising from axions. In the first full joint analysis of \textit{Planck} 2018 and BOSS full-shape data, we find that galaxy clustering information strengthens axion energy density limits at nearly all masses that we consider. The exception is at \(m_\mathrm{a} = 10^{-26}\,\mathrm{eV}\), where BOSS data alone have a mild preference for a non-zero axion contribution, excluding no axions at \(\sim 2.7 \sigma\). The significance is reduced to only \(1.7 \sigma\) when including Big Bang nucleosynthesis constraints on the baryon energy density \(\Omega_\mathrm{b} h^2\) and \textit{Planck} constraints on the primordial power spectrum tilt \(n_\mathrm{s}\). Such an axion contribution is, further, disfavoured by \textit{Planck} and we caution that the look-elsewhere effect applies owing to the large number of axion masses that we consider. Future galaxy data (\eg DESI, \textit{Rubin}) will be crucial in assessing the significance of this result.

We propose axions as a candidate to address the so-called ``\(S_8\) tension'', where CMB experiments infer systematically higher values of \(S_8\) (which is sensitive to the matter power spectrum amplitude at \(z = 0\)) than various large-scale structure datasets, with significance ranging from 2 to 3 \(\sigma\) \citep[\eg][]{2022JHEAp..34...49A}. We hypothesise that the scale-dependent power spectrum suppression (relative to standard cold DM) arising from axion DM can reconcile current CMB data (which probe larger scales and prefer higher amplitude) with the more direct probes of smaller-scale structure in galaxy clustering and weak lensing that prefer lower amplitude. We indeed find that a compendium of ``large-scale'' data (CMB, galaxy BAO and supernovae) are compatible with lower values of \(S_8 = 0.774^{+0.032}_{-0.037}\) for \(m_\mathrm{a} = 10^{-25}\,\mathrm{eV}\) than in \(\Lambda\)CDM (\(S_8 = 0.827 \pm 0.010\)). This is achieved since this data combination is not sensitive to the small-scale suppression arising from axions of this mass, while the axion suppression still occurs at wavenumbers to which \(S_8\) is sensitive, thus lowering its value. Although BOSS full-shape data, in general, strengthen axion density bounds (apart from at \(m_\mathrm{a} = 10^{-26}\,\mathrm{eV}\)), we find that axions can improve inferred parameter consistency between \textit{Planck} and BOSS and that the joint \textit{Planck} and BOSS constraint is still consistent with lower values of \(S_8\) than \(\Lambda\)CDM in a window of masses \([10^{-28}, 10^{-25}]\,\mathrm{eV}\). In future work, we will assess consistency with upcoming CMB and galaxy weak lensing data using a fully non-linear (halo) model of axion structure formation \citep[\eg][]{2022arXiv220913445V,dentler/etal:2022}.

\acknowledgments

The authors thank Daniel Grin for valuable discussions. The Dunlap Institute is funded through an endowment established by the David Dunlap family and the University of Toronto. RH is a CIFAR Azrieli Global Scholar (Gravity \& the Extreme Universe Program 2019) and a 2020 Alfred P. Sloan Research Fellow; and is supported by the Natural Sciences and Engineering Research Council of Canada Discovery Grant Program and the Connaught Fund. MMI is supported by the National Aeronautics and Space Administration (NASA) through the NASA Hubble Fellowship grant \#HST-HF2-51483.001-A awarded by the Space Telescope Science Institute, which is operated by the Association of Universities for Research in Astronomy, Incorporated, under NASA contract NAS5-26555. OHEP is a Junior Fellow of the Simons Society of Fellows and thanks the Institute for Advanced Study for their hospitality and abundance of baked goods. KA is supported by Japan Society for the Promotion of Science (JSPS) Overseas Research Fellowships. DJEM is supported by an Ernest Rutherford Fellowship from the Science and Technologies Facilities Council (ST/T004037/1).


\bibliographystyle{JHEP}
\bibliography{axion_s8_paper}







\end{document}